\documentclass[twocolumn]{aastex631}

\usepackage{mathtools,hyperref}
\usepackage{amsmath,amsfonts,amsthm,bm}
\usepackage{array,multirow,graphicx}
\usepackage{graphicx}
\usepackage[figuresright]{rotating}
\usepackage{natbib}
\usepackage[all]{hypcap} 
\usepackage{fancyhdr}
\usepackage{booktabs}
\usepackage{orcidlink}
\usepackage{array}
\usepackage[hang,flushmargin]{footmisc}

\usepackage{needspace}

\def\q{\mathbf{q}}
\def\p{\mathbf{p}}
\def\x{\mathbf{x}}
\def\R{\mathbf{R}}
\def\W{\mathbf{W}}
\def\F{\mathbf{F}}
\def\E{\mathbf{E}}
\def\H{\mathbf{H}}
\def\G{\mathbf{G}}
\def\C{\mathbf{C}}

\def\N{\mathbf{N}}

\def\I{\mathbf{I}}
\def\Q{\mathbf{Q}}
\def\M{\mathbf{M}}

\def\kpara{k_{\parallel}}
\def\kperp{k_{\perp}}

\def\bd{\mathbf{d}}

\newcolumntype{L}{>{$}l<{$}}
\newcolumntype{C}{>{$}c<{$}}
\newcolumntype{R}{>{$}r<{$}}

\newcommand{\highzlimit}{3,496\,mK$^2$}
\newcommand{\highzk}{0.36\,$h$Mpc$^{-1}$}
\newcommand{\lowzlimit}{457\,mK$^2$}
\newcommand{\lowzk}{0.34\,$h$Mpc$^{-1}$}

\def\Ha{\hyperlink{cite.h1c_idr2_limits}{H22a}}
\def\Hb{\hyperlink{cite.h1c_idr2_theory}{H22b}}

\newcommand{\edited}[1]{#1}

\shorttitle{Improved Constraints from HERA Phase I}
\shortauthors{The HERA Collaboration}

\begin{document}

\title{Improved Constraints on the 21\,cm EoR Power Spectrum and the X-Ray Heating of the IGM with HERA Phase I Observations} 

\collaboration{94}{The HERA Collaboration:} 

\correspondingauthor{Joshua~S.~Dillon}
\email{jsdillon@berkeley.edu}

\author{Zara~Abdurashidova}
\affiliation{Department of Astronomy, University of California, Berkeley, CA}

\author{Tyrone~Adams}
\affiliation{South African Radio Astronomy Observatory, Cape Town, South Africa}

\author{James~E.~Aguirre\,\orcidlink{0000-0002-4810-666X}\!\!}
\affiliation{Department of Physics and Astronomy, University of Pennsylvania, Philadelphia, PA}

\author{Paul~Alexander}
\affiliation{Cavendish Astrophysics, University of Cambridge, Cambridge, UK}

\author{Zaki~S.~Ali}    
\affiliation{Department of Astronomy, University of California, Berkeley, CA}

\author{Rushelle~Baartman}
\affiliation{South African Radio Astronomy Observatory, Cape Town, South Africa}

\author{Yanga~Balfour}
\affiliation{South African Radio Astronomy Observatory, Cape Town, South Africa}

\author{Rennan~Barkana\,\orcidlink{0000-0002-1557-693X}\!\!}
\affiliation{School of Physics and Astronomy, Tel-Aviv University, Tel-Aviv, 69978, Israel}

\author{Adam~P.~Beardsley\,\orcidlink{0000-0001-9428-8233}\!\!}
\affiliation{Department of Physics, Winona State University, Winona, MN}
\affiliation{School of Earth and Space Exploration, Arizona State University, Tempe, AZ}

\author{Gianni~Bernardi\,\orcidlink{0000-0002-0916-7443}\!\!}
\affiliation{INAF-Istituto di Radioastronomia, via Gobetti 101, 40129 Bologna, Italy}
\affiliation{Department of Physics and Electronics, Rhodes University, PO Box 94, Grahamstown, 6140, South Africa}
\affiliation{South African Radio Astronomy Observatory, Cape Town, South Africa}

\author{Tashalee~S.~Billings}
\affiliation{Department of Physics and Astronomy, University of Pennsylvania, Philadelphia, PA}

\author{Judd~D.~Bowman\,\orcidlink{0000-0002-8475-2036}\!\!}
\affiliation{School of Earth and Space Exploration, Arizona State University, Tempe, AZ}

\author{Richard~F.~Bradley}
\affiliation{National Radio Astronomy Observatory, Charlottesville, VA}

\author{Daniela~Breitman\,\orcidlink{0000-0002-2349-3341}\!\!}
\affiliation{Scuola Normale Superiore, 56126 Pisa, PI, Italy}

\author{Philip~Bull\,\orcidlink{0000-0001-5668-3101}\!\!}
\affiliation{Jodrell Bank Centre for Astrophysics, University of Manchester, Manchester, M13 9PL, UK}
\affiliation{Department of Physics \& Astronomy, Queen Mary University of London, London, UK}
\affiliation{Department of Physics and Astronomy, University of Western Cape, Cape Town 7535, South Africa}

\author{Jacob~Burba}
\affiliation{Department of Physics, Brown University, Providence, RI}

\author{Steve~Carey}
\affiliation{Cavendish Astrophysics, University of Cambridge, Cambridge, UK}

\author{Chris~L.~Carilli\,\orcidlink{0000-0001-6647-3861}\!\!}
\affiliation{National Radio Astronomy Observatory, Socorro, NM}

\author{Carina~Cheng}
\affiliation{Department of Astronomy, University of California, Berkeley, CA}

\author{Samir~Choudhuri\,\orcidlink{0000-0002-2338-935X}\!\!}
\affiliation{Centre for Strings, Gravitation and Cosmology, Department of Physics, Indian Institute of Technology Madras, Chennai 600036, India}
\affiliation{Department of Physics \& Astronomy, Queen Mary University of London, London, UK}

\author{David~R.~DeBoer\,\orcidlink{0000-0003-3197-2294}\!\!}
\affiliation{Department of Astronomy, University of California, Berkeley, CA}

\author{Eloy~de~Lera~Acedo\,\orcidlink{0000-0001-8530-6989}\!\!}
\affiliation{Cavendish Astrophysics, University of Cambridge, Cambridge, UK}

\author{Matt~Dexter}
\affiliation{Department of Astronomy, University of California, Berkeley, CA}

\author{Joshua~S.~Dillon\,\orcidlink{0000-0003-3336-9958}\!\!}
\affiliation{Department of Astronomy, University of California, Berkeley, CA}

\author{John~Ely}
\affiliation{Cavendish Astrophysics, University of Cambridge, Cambridge, UK}

\author{Aaron~Ewall-Wice\,\orcidlink{0000-0002-0086-7363}\!\!}
\affiliation{Department of Astronomy, University of California, Berkeley, CA}

\author{Nicolas~Fagnoni\,\orcidlink{0000-0001-5300-3166}\!\!}
\affiliation{Cavendish Astrophysics, University of Cambridge, Cambridge, UK}

\author{Anastasia~Fialkov}
\affiliation{Institute of Astronomy, University of Cambridge, Madingley Road, Cambridge CB3 0HA, United Kingdom}
\affiliation{Kavli Institute for Cosmology, Madingley Road, Cambridge CB3 0HA, UK}

\author{Randall~Fritz}
\affiliation{South African Radio Astronomy Observatory, Cape Town, South Africa}

\author{Steven~R.~Furlanetto\,\orcidlink{0000-0002-0658-1243}\!\!}
\affiliation{Department of Physics and Astronomy, University of California, Los Angeles, CA}

\author{Kingsley~Gale-Sides}
\affiliation{Cavendish Astrophysics, University of Cambridge, Cambridge, UK}

\author{Hugh~Garsden}
\affiliation{Jodrell Bank Centre for Astrophysics, University of Manchester, Manchester, M13 9PL, UK}
\affiliation{Department of Physics \& Astronomy, Queen Mary University of London, London, UK}

\author{Brian~Glendenning}
\affiliation{National Radio Astronomy Observatory, Socorro, NM}

\author{Adélie~Gorce\,\orcidlink{0000-0002-1712-737X}\!\!}
\affiliation{Department of Physics and McGill Space Institute, McGill University, 3600 University Street, Montreal, QC H3A 2T8, Canada}

\author{Deepthi~Gorthi\,\orcidlink{0000-0002-0829-167X}\!\!}
\affiliation{Department of Astronomy, University of California, Berkeley, CA}

\author{Bradley~Greig\,\orcidlink{0000-0002-4085-2094}\!\!}
\affiliation{School of Physics, University of Melbourne, Parkville, VIC 3010, Australia}

\author{Jasper~Grobbelaar}
\affiliation{South African Radio Astronomy Observatory, Cape Town, South Africa}

\author{Ziyaad~Halday}
\affiliation{South African Radio Astronomy Observatory, Cape Town, South Africa}

\author{Bryna~J.~Hazelton\,\orcidlink{0000-0001-7532-645X}\!\!}
\affiliation{Department of Physics, University of Washington, Seattle, WA}
\affiliation{eScience Institute, University of Washington, Seattle, WA}

\author{Stefan~Heimersheim}
\affiliation{Institute of Astronomy, University of Cambridge, Madingley Road, Cambridge CB3 0HA, UK}

\author{Jacqueline~N.~Hewitt\,\orcidlink{0000-0002-4117-570X}\!\!}
\affiliation{Department of Physics and MIT Kavli Institute, Massachusetts Institute of Technology, Cambridge, MA}

\author{Jack~Hickish}
\affiliation{Department of Astronomy, University of California, Berkeley, CA}

\author{Daniel~C.~Jacobs}
\affiliation{School of Earth and Space Exploration, Arizona State University, Tempe, AZ}

\author{Austin~Julius}
\affiliation{South African Radio Astronomy Observatory, Cape Town, South Africa}

\author{Nicholas~S.~Kern\,\orcidlink{0000-0002-8211-1892}\!\!}
\affiliation{Department of Physics and MIT Kavli Institute, Massachusetts Institute of Technology, Cambridge, MA}

\author{Joshua~Kerrigan\,\orcidlink{0000-0002-1876-272X}\!\!}
\affiliation{Department of Physics, Brown University, Providence, RI}

\author{Piyanat~Kittiwisit\,\orcidlink{0000-0003-0953-313X}\!\!}
\affiliation{Department of Physics and Astronomy, University of Western Cape, Cape Town 7535, South Africa}

\author{Saul~A.~Kohn\,\orcidlink{0000-0001-6744-5328}\!\!}
\affiliation{Department of Physics and Astronomy, University of Pennsylvania, Philadelphia, PA}

\author{Matthew~Kolopanis\,\orcidlink{0000-0002-2950-2974}\!\!}
\affiliation{School of Earth and Space Exploration, Arizona State University, Tempe, AZ}

\author{Adam~Lanman\,\orcidlink{0000-0003-2116-3573}\!\!}
\affiliation{Department of Physics, Brown University, Providence, RI}
\affiliation{Department of Physics and McGill Space Institute, McGill University, 3600 University Street, Montreal, QC H3A 2T8, Canada}

\author{Paul~La~Plante\,\orcidlink{0000-0002-4693-0102}\!\!}
\affiliation{Department of Astronomy, University of California, Berkeley, CA}
\affiliation{Department of Physics and Astronomy, University of Pennsylvania, Philadelphia, PA}

\author{David~Lewis}
\affiliation{School of Earth and Space Exploration, Arizona State University, Tempe, AZ}

\author{Adrian~Liu\,\orcidlink{0000-0001-6876-0928}\!\!}
\affiliation{Department of Physics and McGill Space Institute, McGill University, 3600 University Street, Montreal, QC H3A 2T8, Canada}

\author{Anita~Loots}
\affiliation{South African Radio Astronomy Observatory, Cape Town, South Africa}

\author{Yin-Zhe Ma\,\orcidlink{0000-0001-8108-0986}\!\!}
\affiliation{School of Chemistry and Physics, University of KwaZulu-Natal, Westville Campus, Private Bag X54001, Durban, 4000, South Africa}

\author{David~H.E.~MacMahon}
\affiliation{Department of Astronomy, University of California, Berkeley, CA}

\author{Lourence~Malan}
\affiliation{South African Radio Astronomy Observatory, Cape Town, South Africa}

\author{Keith~Malgas}
\affiliation{South African Radio Astronomy Observatory, Cape Town, South Africa}

\author{Cresshim~Malgas}
\affiliation{South African Radio Astronomy Observatory, Cape Town, South Africa}

\author{Matthys~Maree}
\affiliation{South African Radio Astronomy Observatory, Cape Town, South Africa}

\author{Bradley~Marero}
\affiliation{South African Radio Astronomy Observatory, Cape Town, South Africa}

\author{Zachary~E.~Martinot}
\affiliation{Department of Physics and Astronomy, University of Pennsylvania, Philadelphia, PA}

\author{Lisa~McBride}
\affiliation{Department of Physics and McGill Space Institute, McGill University, 3600 University Street, Montreal, QC H3A 2T8, Canada}

\author{Andrei~Mesinger\,\orcidlink{0000-0003-3374-1772}\!\!}
\affiliation{Scuola Normale Superiore, 56126 Pisa, PI, Italy}

\author{Jordan~Mirocha\,\orcidlink{0000-0002-8802-5581}\!\!}
\affiliation{Department of Physics and McGill Space Institute, McGill University, 3600 University Street, Montreal, QC H3A 2T8, Canada}

\author{Mathakane~Molewa}
\affiliation{South African Radio Astronomy Observatory, Cape Town, South Africa}

\author{Miguel~F.~Morales\,\orcidlink{0000-0001-7694-4030}\!\!}
\affiliation{Department of Physics, University of Washington, Seattle, WA}

\author{Tshegofalang~Mosiane}
\affiliation{South African Radio Astronomy Observatory, Cape Town, South Africa}

\author{Julian~B.~Mu\~noz\,\orcidlink{0000-0002-8984-0465}\!\!}
\affiliation{Center for Astrophysics | Harvard \& Smithsonian, Cambridge, MA}

\author{Steven~G.~Murray\,\orcidlink{0000-0003-3059-3823}\!\!}
\affiliation{School of Earth and Space Exploration, Arizona State University, Tempe, AZ}

\author{Vighnesh~Nagpal\,\orcidlink{0000-0001-5909-4433}\!\!}
\affiliation{Department of Astronomy, University of California, Berkeley, CA}

\author{Abraham~R.~Neben\,\orcidlink{0000-0001-7776-7240}\!\!}
\affiliation{Department of Physics and MIT Kavli Institute, Massachusetts Institute of Technology, Cambridge, MA}

\author{Bojan~Nikolic\,\orcidlink{0000-0001-7168-2705}\!\!}
\affiliation{Cavendish Astrophysics, University of Cambridge, Cambridge, UK}

\author{Chuneeta~D.~Nunhokee\,\orcidlink{0000-0002-5445-6586}\!\!}
\affiliation{Department of Astronomy, University of California, Berkeley, CA}

\author{Hans~Nuwegeld}
\affiliation{South African Radio Astronomy Observatory, Cape Town, South Africa}

\author{Aaron~R.~Parsons\,\orcidlink{0000-0002-5400-8097}\!\!}
\affiliation{Department of Astronomy, University of California, Berkeley, CA}

\author{Robert~Pascua\,\orcidlink{0000-0003-0073-5528}\!\!}
\affiliation{Department of Astronomy, University of California, Berkeley, CA}
\affiliation{Department of Physics and McGill Space Institute, McGill University, 3600 University Street, Montreal, QC H3A 2T8, Canada}

\author{Nipanjana~Patra\,\orcidlink{0000-0002-9457-1941}\!\!}
\affiliation{Department of Astronomy, University of California, Berkeley, CA}

\author{Samantha~Pieterse}
\affiliation{South African Radio Astronomy Observatory, Cape Town, South Africa}

\author{Yuxiang~Qin\,\orcidlink{0000-0002-4314-1810}\!\!}
\affiliation{School of Physics, University of Melbourne, Parkville, VIC 3010, Australia}

\author{Nima~Razavi-Ghods\,\orcidlink{0000-0003-2930-5396}\!\!}
\affiliation{Cavendish Astrophysics, University of Cambridge, Cambridge, UK}

\author{James~Robnett}
\affiliation{National Radio Astronomy Observatory, Socorro, NM}

\author{Kathryn~Rosie\,\orcidlink{0000-0003-3611-8804}\!\!}
\affiliation{South African Radio Astronomy Observatory, Cape Town, South Africa}

\author{Mario~G.~Santos\,\orcidlink{0000-0003-3892-3073}\!\!}
\affiliation{Department of Physics and Astronomy, University of Western Cape, Cape Town 7535, South Africa}
\affiliation{South African Radio Astronomy Observatory, Cape Town, South Africa}

\author{Peter~Sims\,\orcidlink{0000-0002-2871-0413}\!\!}
\affiliation{Department of Physics, Brown University, Providence, RI}
\affiliation{Department of Physics and McGill Space Institute, McGill University, 3600 University Street, Montreal, QC H3A 2T8, Canada}

\author{Saurabh~Singh\,\orcidlink{0000-0001-7755-902X}\!\!}
\affiliation{Department of Physics and McGill Space Institute, McGill University, 3600 University Street, Montreal, QC H3A 2T8, Canada}
\affiliation{Raman Research Institute, Sadashivanagar, Bengaluru, India}

\author{Craig~Smith}
\affiliation{South African Radio Astronomy Observatory, Cape Town, South Africa}

\author{Hilton~Swarts} 
\affiliation{South African Radio Astronomy Observatory, Cape Town, South Africa}

\author{Jianrong~Tan\,\orcidlink{0000-0001-6161-7037}\!\!}
\affiliation{Department of Physics and Astronomy, University of Pennsylvania, Philadelphia, PA}

\author{Nithyanandan~Thyagarajan\,\orcidlink{0000-0003-1602-7868}\!\!}
\affiliation{Commonwealth Scientific and Industrial Research Organisation, (CSIRO), Space \& Astronomy, Bentley, WA 6102, Australia}

\author{Michael~J.~Wilensky\,\orcidlink{0000-0001-7716-9312}\!\!}
\affiliation{Jodrell Bank Centre for Astrophysics, University of Manchester, Manchester, M13 9PL, UK}
\affiliation{Department of Physics \& Astronomy, Queen Mary University of London, London, UK}

\author{Peter~K.~G.~Williams\,\orcidlink{0000-0003-3734-3587}\!\!}
\affiliation{Center for Astrophysics | Harvard \& Smithsonian, Cambridge, MA}
\affiliation{American Astronomical Society, Washington, DC}

\author{Pieter~van~Wyngaarden} 
\affiliation{South African Radio Astronomy Observatory, Cape Town, South Africa}

\author{Haoxuan~Zheng\,\orcidlink{0000-0001-8267-3425}\!\!}
\affiliation{Department of Physics and MIT Kavli Institute, Massachusetts Institute of Technology, Cambridge, MA}

\begin{abstract}
We report the most sensitive upper limits to date on the 21\,cm epoch of reionization power spectrum using 94 nights of observing with Phase I of the Hydrogen Epoch of Reionization Array (HERA). 
Using similar analysis techniques as in previously reported limits (\hyperlink{cite.h1c_idr2_limits}{HERA Collaboration 2022a}), we find at 95\% confidence that $\Delta^2(k=\text{\lowzk{}}) \leq \text{\lowzlimit{}}$ at $z=7.9$ and that $\Delta^2(k=\text{\highzk{}}) \leq \text{\highzlimit{}}$ at $z=10.4$, an improvement by a factor of 2.1 and 2.6 respectively. These limits are mostly consistent with thermal noise over a wide range of $k$ after our data quality cuts, despite performing a relatively conservative analysis designed to minimize signal loss.  Our results are validated with both statistical tests on the data and end-to-end pipeline simulations. We also report updated constraints on the astrophysics of reionization and the cosmic dawn. Using multiple independent modeling and inference techniques previously employed by \hyperlink{cite.h1c_idr2_theory}{HERA Collaboration (2022b)}, we find that the intergalactic medium must have been heated above the adiabatic cooling limit at least as early as $z=10.4$, ruling out a broad set of so-called ``cold reionization" scenarios. If this heating is due to high-mass X-ray binaries during the cosmic dawn, as is generally believed, our result's 99\% credible interval excludes the local relationship between soft X-ray luminosity and star formation and thus requires heating driven by evolved low-metallicity stars.
\end{abstract}


\needspace{2cm}\section{Introduction}
\label{sec:intro}   

21\,cm cosmology---the observation of the hyperfine transition of neutral hydrogen at cosmological distances---has long promised to become a sensitive probe of the structure and evolution of the intergalactic medium (IGM) from the Cosmic Dark Ages through to the \edited{cosmic dawn, the epoch of reionization (EoR) \citep{Hogan1979, Madau1997}, and beyond.} By measuring fluctuations in the 21\,cm brightness temperature relative to the Cosmic Microwave Background (CMB) that trace the density, temperature, and ionization state of the IGM, we can precisely constrain our models of cosmology and of the first stars and galaxies \citep{Mao2008, Patil2014, Pober2014, Liu2016b, Greig2016, Ewall-Wice2016a, Kern2017}. For pedagogical reviews see, e.g.\ \citet{Ciardi2005}, \citet{Furlanetto2006c}, \citet{Morales2010}, \citet{Pritchard2012}, \citet{Mesinger2016b}, and \citet{Liu2020}.

A number of low-frequency radio telescopes designed to detect and characterize the cosmic dawn and EoR 21\,cm signal have been built over the last decade and a half. Many are interferometers seeking to statistically detect and ultimately tomographically map 21\,cm fluctuations over a broad range of frequencies and thus redshift. This period has seen increasingly tight limits on the 21\,cm power spectrum from a number of different telescopes, including the Giant Metre Wave Radio Telescope \citep[GMRT;][]{Paciga2013}, the Low Frequency Array \citep[LOFAR;][]{vanHaarlem2013, Patil2017, Gehlot2019, Mertens2020}, the Donald C. Backer Precision Array for Probing the Epoch of Reionization \citep[PAPER;][]{Parsons2010, Cheng2018, Kolopanis2019}, the Murchison Widefield Array \citep[MWA;][]{Tingay2013, Dillon2014, Dillon2015b, Jacobs_2016, Ewall-Wice2016b, Beardsley2016, Barry2019b, Li2019, Trott2020, Yoshiura_2021, Rahimi2021}, and the Owens Valley Long Wavelength Array \citep[LWA;][]{Eastwood2019, Garsden_2021}. 

Additionally, a number of total-power experiments have been conducted to measure the sky-averaged, global 21\,cm signal as it evolves with redshift \citep{Bernardi2016, Singh2017, Monsalve2017}. Recently, the Experiment to Detect the Global EoR Signature \citep[EDGES;][]{Bowman2018} reported the detection of an unexpectedly strong absorption feature in the global signal at $z \approx 17$ which would require either an IGM temperature below the adiabatic cooling limit \citep{Munoz:2015bca,Barkana_2018,Munoz:2018pzp} or a high-redshift radio background in excess of the CMB \citep{Feng_2018, Ewall_Wice_2018,Pospelov:2018kdh, mirocha19}. A number of subsequent analyses have further investigated alternative explanations for this result in terms of instrumental systematics \citep{Bradley2019, Hills2018, Singh2019, Sims2020, Mahesh2021} and the recent non-detection by the Shaped Antenna measurement of the background RAdio Spectrum 3 \citep[SARAS 3;][]{Singh_2022} in an overlapping frequency band is in tension with the EDGES result.

The main challenge facing both interferometric and sky-averaged 21\,cm observations is the roughly five orders of magnitude of dynamic range between the 21\,cm signal and astrophysical foregrounds---largely synchrotron and free-free emission from our Galaxy and other galaxies. While foregrounds are in principle separable from 21\,cm signal using their intrinsic spectral smoothness, that separability is complicated by many real-world factors. Calibration errors due to e.g.\ incomplete sky and instrument models or unaccounted-for non-redundancy can leak foreground power into regions of Fourier space that would otherwise be signal-dominated \citep{Barry2016, Ewall-Wice2017, Orosz2019, Byrne2019, Joseph_2019}. Moreover, interferometers are inherently chromatic instruments with increasing frequency structure with baseline length---the origin of the so-called ``wedge'' feature in 2D power spectra \citep{Datta2010,Vedantham2012,Parsons2012a,Parsons2012b, Liu2014a, Liu2014b}.

The extreme sensitivity and calibration requirements of \edited{high-redshift} 21\,cm cosmology have driven the design of second-generation interferometers including the Hydrogen Epoch of Reionization Array \citep[HERA;][]{DeBoer2017} and the Square Kilometre Array \citep[SKA;][]{Koopmans2015} with larger collecting areas and a diversity of approaches to understanding and controlling instrumental systematics. HERA, when complete, will be a interferometer with 350 fully cross-correlated elements---each a fixed, zenith-pointing 14\,m dish---at the South African Radio Astronomy Observatory site in the Karoo desert. The dishes are designed to minimize the frequency structure of the instrumental response \citep{Thyagarajan2016, Neben2016, Ewall-Wice2016c, Patra2018, Fagnoni2021}. HERA's compact, hexagonally-packed configuration maximizes sensitivity on short baselines, which are intrinsically less chromatic, while enabling relative gain calibration of antennas using redundant baselines \citep{Dillon2016}. 

\begin{figure*}[ht]
    \centering
\includegraphics[width=\textwidth]{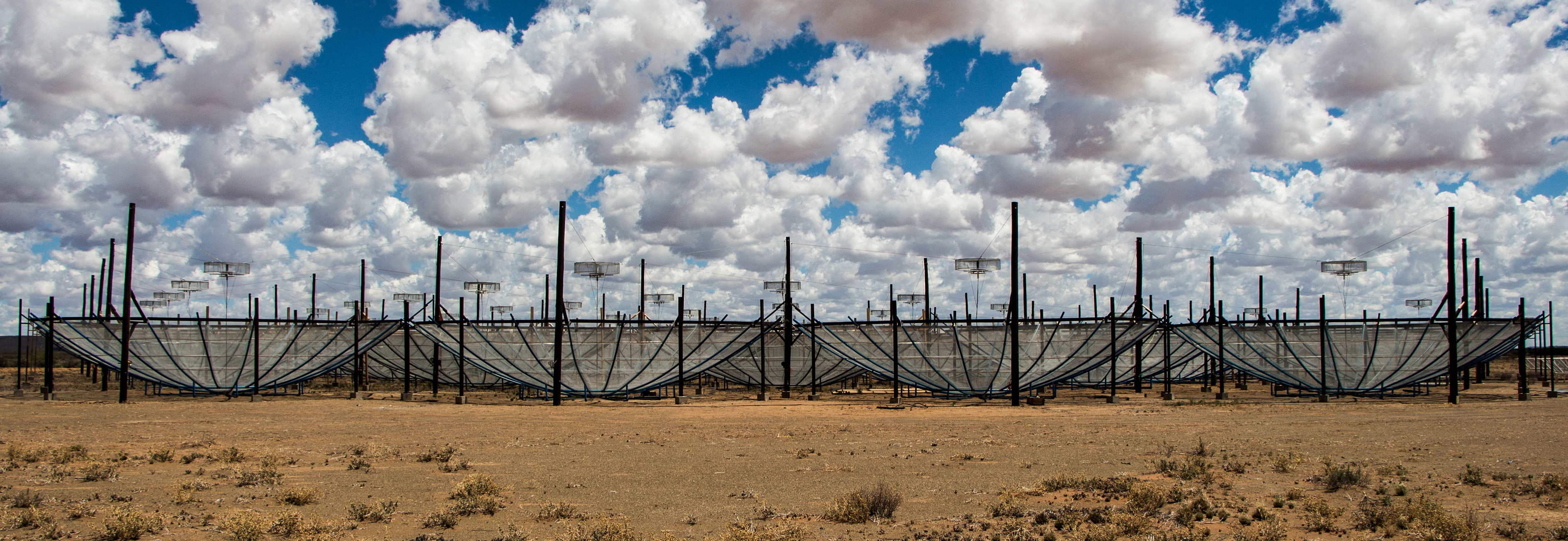}
    \caption{A view of HERA from January 2017. The data in this work were taken during Phase I, when HERA was composed of 14\,m parabolic dishes with sleeved dipole feeds in mesh cages suspended at prime focus. These feeds were later replaced with wide-band Vivaldi feeds, expanding HERA's bandwidth from 100--200\,MHz (Phase I) to 50--250\,MHz (Phase II).}
    \label{fig:hera_pic}
\end{figure*}

During Phase I, which culminated in the 2017--2018 observing season,\footnote{HERA's primary observing seasons are during the Southern summer when both the Sun and the Galactic Center are below the horizon simultaneously at night.} HERA repurposed PAPER's sleeved dipoles, suspended at prime focus (see \autoref{fig:hera_pic}), along with PAPER's correlator and signal chains to observe from 100 to 200\,MHz. 
During that time, the array continued to be built and commissioned. In Phase II, a new signal chain, correlator, and upgraded Vivaldi feeds have extended the bandwidth to 50--250\,MHz \citep{Fagnoni2020}.

Recently, we reported the first upper limits on the 21\,cm brightness temperature power spectrum from HERA in \citet[][hereafter \Ha{}]{h1c_idr2_limits}, using 18 nights of Phase I data and only 39 antennas. \Ha{} built upon a number of supporting papers exploring various aspects of the data analysis. These included redundant-baseline calibration \citep{Dillon2020}, absolute calibration \citep{Kern2020b}, systematics mitigation \citep{Kern2019, Kern2020a}, error estimation \citep{Tan2021}, analysis pipeline architecture \citep{La_Plante_2021}, and end-to-end validation of that pipeline with realistic simulated data \citep{Aguirre_2022}. We focused on the so-called ``foreground-avoidance'' approach to power spectrum estimation \citep{Kerrigan2018,Morales2019}, working predominately in foreground-free regions of Fourier space and applying conservative techniques that minimized potential signal loss or bias. 

Our results, which constrained the ``dimensionless'' brightness temperature power spectrum to $\Delta^2(k)<$~946\,mK$^2$ at $z=7.9$ and $k = 0.19$\,$h$Mpc$^{-1}$ and to $\Delta^2(k)<$~9,166\,mK$^2$ at $z=10.4$ and $k = 0.26$\,$h$Mpc$^{-1}$ (both at the 95\% confidence level), represented the most stringent constraints to date. They allowed us in \citet[][hereafter \Hb{}]{h1c_idr2_theory} to constrain the astrophysics of reionization and the cosmic dawn and show that the IGM was heated above the adiabatic cooling limit by at least $z=7.9$. Evidence from other probes---including the integrated optical depth to reionization \citep{Planck18}, observed galaxy UV luminosity functions, and quasar spectroscopy---indicates that reionization is likely well underway by $z=7.9$ \citep{mason18, Greig_2022}. Our results therefore already rule out some of the most extreme of the so-called ``cold reionization'' models where an adiabatically cooling IGM produces a bright temperature contrast with the CMB, amplifying the 21\,cm power spectrum as it is driven by ionization fluctuations \citep{Mesinger2014}.

In this work, we adapt and apply the analysis techniques of \Ha{} and \Hb{} to a full season of HERA Phase I data. Retaining the philosophy of foreground-avoidance and minimizing (and carefully accounting for) signal loss, we further tighten constraints on the 21\,cm power spectrum at $z=7.9$ and $z=10.4$, and update the astrophysical implications of those limits. While some of our analysis techniques are updated to reflect an improved understanding of our instrument or adapted to better handle the larger volume of data considered, the core analysis techniques remain largely unchanged.\footnote{Following \Ha{}, we also adopt a $\Lambda$CDM cosmology \citep{Planck2016} with $\Omega_\Lambda=0.6844$, $\Omega_b=0.04911$, $\Omega_c=0.26442$, and $H_0=67.27\,\textrm{km}/\textrm{s}/\textrm{Mpc}$.} 

We begin in \autoref{sec:obs} by detailing the observations themselves and the basic cuts performed to ensure data quality. Then in \autoref{sec:analysis} we review the data reduction steps performed to go from raw visibilities all the way to power spectra, highlighting updated analysis techniques and revised analysis choices. These techniques are tested with end-to-end pipeline simulations designed to validate our analysis choices and software in \autoref{sec:validation}, in which we quantify a number of potential small biases and reproduce a few key figures from \citet{Aguirre_2022} in the context of our new limits. In \autoref{sec:upper_limits}, we can then present our final power spectrum estimates, error bars, and upper limits. We build confidence in our results in \autoref{sec:tests} by applying a variety of statistical tests on our power spectra and how they integrate down across baselines and time. In \autoref{sec:theory}, we report the impact of our new limits on the various approaches to astrophysical modeling and inference used in \Hb{}, detailing our updated constraints on the epoch of reionization and the cosmic dawn. We conclude in \autoref{sec:conclusion}, looking forward to potential future analyses of these data and data from the full HERA Phase II system.


\section{Observations and Data Selection} \label{sec:obs}

In this work, we analyze observations with the HERA Phase I system that were performed over the period from September 29, 2017 (JD 2458026) through March 31, 2018 (JD 2458208). In \autoref{tab:hera_specs}, we summarize the key observational parameters of the instrument. For more detail about the precise configuration of the instrument, its signal chain components, and its FX correlator architecture,  we refer the reader to \citet{DeBoer2017} and \Ha{}. In this section, we discuss the process by which a selection of high-quality nights and antennas was performed.

\begin{table}
\centering
\caption{HERA Phase I observing and array specifications.}
\vspace{-4mm}
\label{tab:hera_specs}
\begin{tabular}{p{.29\textwidth}|r}
\toprule
Array Location & $-$30.72$^\circ$S, 21.43$^\circ$E \\ 
Total Antennas Connected & 47--71 \\
Total Antennas Used & 35--41 \\
Shortest Baseline & 14.6\,m \\
Longest Unflagged Baseline & 124.8\,m \\
\midrule
Minimum Frequency & 100\,MHz \\
Maximum Frequency & 200\,MHz \\
Channels & 1024 \\
Channel Width & 97.66\,kHz \\
\midrule
Integration Time & 10.7\,s \\
Nightly Observing Duration & 12~hours \\
Total Nights With Data & 182 \\
Total Nights Used & 94 \\
\bottomrule
\end{tabular}
\end{table}

\needspace{2cm}\subsection{Selection of Nights and Epochs} \label{sec:nights_and_epochs}

Of the 182 nights during this season of simultaneous construction, commissioning, and observing, a significant fraction of nights was discarded for a variety of reasons. Most of these were hardware failures, including network outages, power outages, too many low- and/or high-power antennas, a briefly broadcasting antenna, broken receivers, and broken X-engines. Some were due to site issues, including high winds, a lightning storm, and excess radio frequency interference (RFI). While all nights have significant narrow-band RFI contamination from FM radio, TV broadcasts, and ORBCOMM satellites (see \autoref{sec:rfi}), most nights that were completely flagged for excess RFI showed consistent broadband emission contaminating channels typically free of RFI. These cuts first reduced the 182 nights to 104 nights using contemporaneous observing logs and real-time analysis. After inspecting hundreds of \texttt{jupyter} notebooks\footnote{\url{https://github.com/HERA-Team/H1C_IDR3_Notebooks}} summarizing the nightly results of the  data analysis pipeline after each key stage (see \autoref{sec:analysis}), this was reduced to 94 nights. For more details on the precise selection of nights, see \citet{H1C_IDR3_2_Memo}.

\begin{figure}[tb]
    \centering
    \includegraphics[width=.48\textwidth]{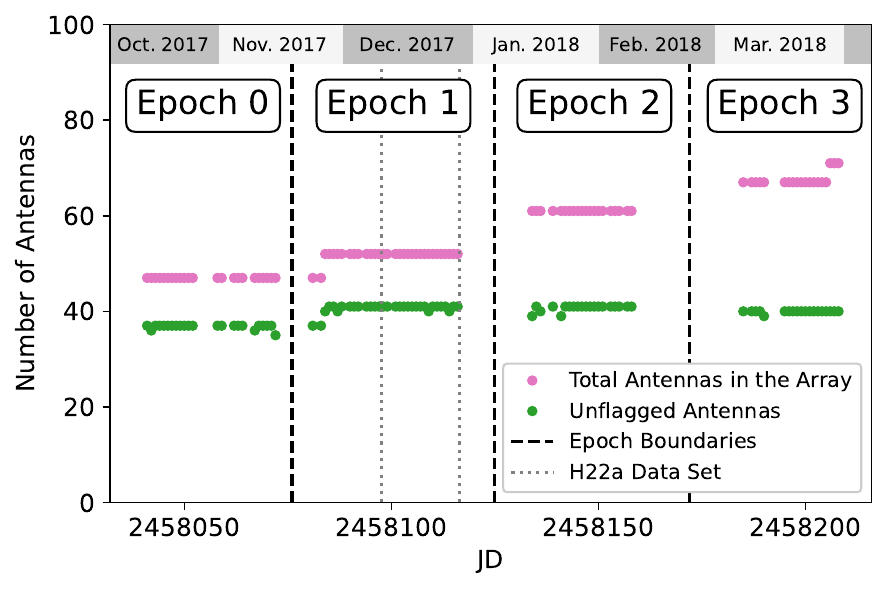}
    \caption{Here we show the observing season, split into epochs, and the number of antennas observing  each night, both in total and after cuts. While the number of total antennas in the array grew from 47 to 71, the number passing all cuts remained roughly constant at $\sim$40. Epochs were defined by natural breaks in the observing season, mostly due to hardware issues. \Ha{} analyzed data from 39 antennas on 18 nights, a subset of Epoch 1.}
    \label{fig:epochs}
    \vspace{5pt}
\end{figure}
In \autoref{fig:epochs}, we show how the nights passing our various data quality checks span the observing season. The breaks in good data (due to a network outage, a correlator malfunction, and a broadcasting antenna) naturally divided the season into four epochs. The data used in \Ha{} were a subset of Epoch 1. While in theory each night could be analyzed independently before binning all of them together at constant local sidereal time (LST), we found it useful to analyze epochs individually, both for systematics mitigation (see \autoref{sec:xtalk_updates}) and for statistical tests on subsets of the data (see \autoref{sec:tests:jk}).

Because we observed during the Southern summer, most 12-hour ``nights'' included some data taken before sunset or after sunrise. These were flagged in our analysis. Further, a number of partial nights were flagged, usually due to excess RFI. The majority of these were due to broadband RFI during the first few hours of the night, possibly attributable to construction activity on site. A few other partial nights were flagged due to suspicious nightly calibration solutions, especially in Epoch 3 when the Galaxy was rising at the end of each night. More detail on the precise subset of nights flagged is given in \citet{H1C_IDR3_2_Memo}.

\begin{figure*}[tbh]
    \centering
    \includegraphics[width=\textwidth]{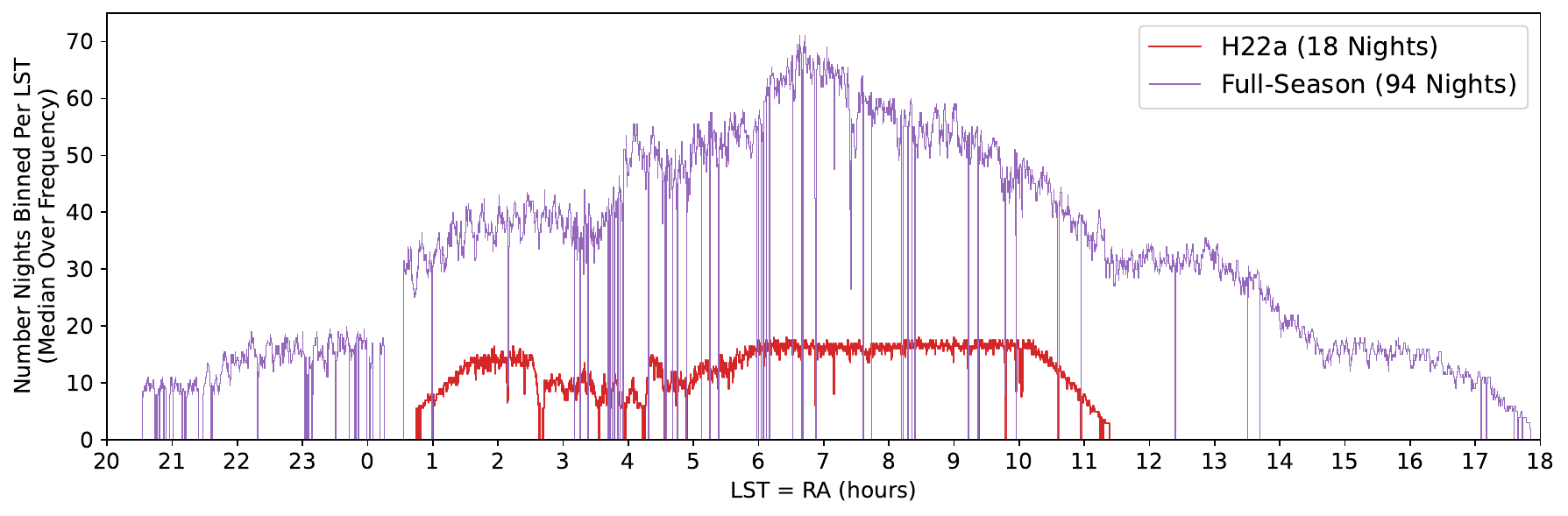}
    \vspace{-10pt}
    \caption{The full season of HERA data that we analyze spans nearly the full 24 hours of LSTs. Since we only observe at night, the time of greatest overlap in LST between nights occurs near the middle of the night in the middle of the season, at roughly 7 hours. At LSTs near that peak, this data set roughly quadruples the depth of the \Ha{} data set. The increased number of flagged times (which cause vertical dips here) is due in large part to the changes in RFI excision (see \autoref{sec:rfi}).}
    \label{fig:lst_coverage}
\end{figure*}

The end result of our data cuts is a set of observations that, when LST-binned together, are significantly deeper than those in \Ha{}, and cover over 21 hours of LST. As we can see in \autoref{fig:lst_coverage}, this data set peaks at $\sim$70 nights' observing around 7~hours of LST, nearly four times deeper than the observations used in \Ha{}.
Roughly speaking, this sets an upper bound on the factor by which our limits might improve due purely to the increase in sensitivity, since the noise on $P(k)$ scales inversely with observing time.

\needspace{2cm}\subsection{Selection of Antennas} \label{sec:antennas}

Antenna selection began with the nightly data quality monitoring system described in \Ha{}. It identified malfunctioning antennas by looking for antennas participating in baselines that were either outliers in total visibility power, or had visibility amplitude spectra significantly different from other baselines measuring the same physical separation on the ground. This procedure ultimately informed the most rigorous identification of malfunctioning antennas described in \citet{Storer_2022}, which was applied to HERA Phase II data.  The results of nightly analysis were synthesized into a set of per-antenna, per-night flags by the HERA commissioning team as part of an internal data release.  

Because our metrics for antenna quality are relative ones, we decided to expand and harmonize this list of flagged antenna-nights under the conservative presumption that antennas that misbehave consistently enough are probably also anomalous at some lower level on other nights.  If, in any given epoch, an antenna was flagged more than 10\% of the days, we flagged it for the whole epoch.  If an antenna was completely flagged for more than 60\% of the epochs that it appeared in (i.e.\ 3 of 4 epochs for antennas that were observing for the whole season), then it was completely removed.

\begin{figure}[tbh]
    \centering
    \includegraphics[width=.48\textwidth]{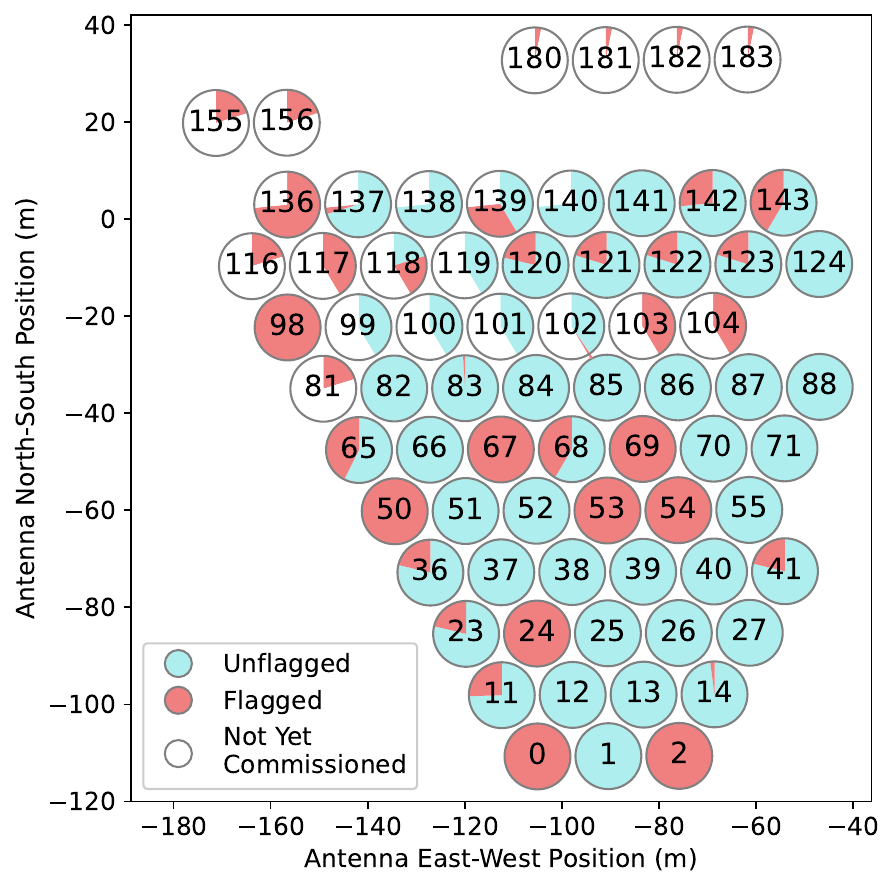}
    \caption{The layout of antennas during the season we analyze. Each antenna is a pie chart showing the fraction of the 94 total nights that each antenna was either flagged, unflagged, or not yet in the data set. The array was under construction as we observed, as can be seen by the northward expansion of the antennas available for observing. All antennas shown here numbered except 155, 156, and 180--183 would eventually become part of the southwest sector of HERA's split-core configuration \citep{Dillon2016,DeBoer2017}.}
    \label{fig:layout}
\end{figure}

Antennas passing this first series of cuts were then used for an initial round of redundant-baseline calibration where per-antenna gains and per-unique-baseline visbilities are solved for simultaneously as part of a large overdetermined system of equations \citep{Liu2010}. Antennas outside the southwest sector of HERA's split-core configuration (155, 156, 180, 181, 182, and 183---see \autoref{fig:layout}---as well as two outriggers not pictured) were excluded as well because they would introduce extra tip-tilt degeneracies \citep{Liu2010, Zheng2014,Zheng_2016, Dillon2018} and thus complicate a subsequent sky-based absolute calibration \citep{Li2018, Kern2020b}. 

As \citet{Dillon2020} describes, redundant-baseline calibration can be cast as a $\chi^2$-minimization problem where $\chi^2$ quantifies how consistent deviations from redundancy are with thermal noise (see \autoref{sec:pipeline_overview} and \autoref{eq:chisq}). If one attributes each baseline's contribution to $\chi^2$ to both of its constituent antennas equally, one can form a $\chi^2$ per antenna statistic that is sensitive to particularly non-redundant antennas. In our first round of redundant-baseline calibration, antennas which made significant excess contributions to $\chi^2$ are flagged in 60-integration (i.e.\ 10.7~minute) chunks, and then calibration is performed again, iteratively, until no outliers remain. These were then likewise harmonized; antennas flagged for non-redundancy more than 20\% of a night or 5\% of an epoch were flagged for the whole night/epoch.

In \autoref{fig:layout}, we show the per-antenna flagging fraction after all of these cuts. The overall impact is that while the array was growing, the number of antennas included in our analysis remained largely static (see also \autoref{fig:epochs}). Likely this set of nightly flags and antenna flags is overly conservative and some good data were thrown out. Due to the extreme dynamic range challenge faced by all of 21\,cm cosmology, we adopted the stance that it was far safer to throw out possibly good data than to risk including bad data. Importantly, all these decisions about data selection were performed without reference to final power spectra and are therefore less likely to introduce experimenter bias. Once the set of good antennas-nights was selected, it was not subsequently changed.


\needspace{2cm}\section{Data Reduction and Systematics Mitigation} \label{sec:analysis}

We now turn to a discussion of our data analysis pipeline, which we designed to reduce nightly visibilities to a final set of power spectra while avoiding systematics contamination. In general, our goal in this work is to apply the methods developed and validated in \Ha{} and its supporting papers to this larger data set. This is a fundamentally iterative approach and likely does not leverage the full constraining power of the data set. Thus, the steps in the analysis pipeline---which we review in \autoref{sec:pipeline_overview}---remain largely unchanged. 

However, a number of changes were incorporated in this work. Some were necessary because this data set is larger and more heterogeneous than the 18 nearly-contiguous nights examined in \Ha{} with the same 39 antennas each night. Others are the result of various tweaks and minor improvements in the HERA team's codebase developed between the \Ha{} analysis and this work. Finally, some are simply minor changes in data analysis parameters and choices motivated by various intermediate data products. In \autoref{sec:analysis_updates} we report changes to our data reduction pipeline and in \autoref{sec:pspec_updates} we similarly detail changes to the estimation of power spectra and their errors and potential biases. 

\needspace{2cm}\subsection{Overview of the Data Analysis Pipeline} \label{sec:pipeline_overview}

\Ha{} gives a detailed description of the analysis steps that take raw visibilities all the way to power spectra. Here we provide a high-level overview of each step in order to give context for the changes and updates in this analysis. We refer the reader to \Ha{} and its supporting papers for more detail. The steps in our data reduction and systematics pipeline are as follows:

\begin{enumerate}
    \item \textbf{Redundant-baseline calibration:} We begin with direction-independent calibration, wherein our observed visbilities, $V_{ij}^\text{obs}$, are modeled as
    \begin{equation}
        V_{ij}^\text{obs} = g_i g_j^* V_{ij}^\text{true} + n_{ij}. \label{eq:gains}
    \end{equation}
    Here $g_i$ is a complex, time- and frequency-dependent gain associated with the $i$th antenna and $n_{ij}$ is the noise on that visibility. Redundant-baseline calibration assumes that all baselines $\mathbf{b}_{ij}$ with the same physical separation and orientation should observe the same true visibility, and thus solves for both gains and unique visibilities as a large $\chi^2$-minimization problem, where
    \begin{equation} \label{eq:chisq}
    \chi^2 = \sum\limits_{i<j}\frac{\left|V_{ij}^\text{obs} - g_ig_j^\ast V_{i-j}^\text{sol}\right|^2}{\sigma_{ij}^2}.
    \end{equation}
    $V_{i-j}^\text{sol}$ here is the visibility solution for all redundant baselines with the same physical separation as $V_{ij}$.
    We calibrate by minimizing $\chi^2$ for every time and frequency independently, using only internal degrees of freedom and without reference to any model of the instrument or the sky. For an exploration of redundant-baseline calibration in the context of HERA, see \citet{Dillon2020}.
    
    \item \textbf{Absolute calibration:} The internal consistency of redundant baselines alone cannot solve for three important degrees of freedom, namely the overall gain amplitude and two phase tip/tilt terms. These degeneracies must be solved, per time and frequency, by reference to externally calibrated (or simulated) visibilities \citep{Kern2020b}. In \Ha{} and in this work, we performed absolute calibration using a set of visibilities synthesized from three nights with LSTs spanning this data set (JDs 2458042, 2458116, and 2458207). These were calibrated on three separate fields using the MWA GLEAM catalog \citep{Hurley-Walker2017} and CASA \citep{CASA}, as described in Section~3.3 of \Ha{}.
    
    \item \textbf{RFI flagging:} RFI is identified and flagged using an iterative outlier detection algorithm described in \Ha{}. Essentially, several sets of waterfalls (visibilities, gains solutions, etc.) are converted into time- and frequency-dependent measures of ``outlierness'' expressed as a $z$-score or modified $z$-score. This is done by looking at a 17$\times$17 pixel region
    centered on each pixel of the waterfall and measuring how consistent each it is with its neighbors in time and frequency. After averaging together $z$-scores of baselines or antennas to improve the signal-to-noise ratio (SNR), 5$\sigma$ outliers and 2$\sigma$ outliers neighboring 5$\sigma$ outliers are flagged. This is done independently for each set of waterfalls and the flags are then combined. We start RFI flagging with median filters and modified $z$-scores to reduce the effect of really bright RFI, then use those flags as a prior on a second round using mean filters and standard $z$-scores. Finally, we examine these statistics over the whole day, looking for whole channels or whole integrations that are 7$\sigma$ outliers and their 3$\sigma$ outlier neighbors. The same flags are applied to all baselines. 
    
    \item \textbf{Gain smoothing:} After flagging, all gains solutions are smoothed to mitigate the effect of noise and calibration errors, taking as a prior that the gains should be stable and relatively smooth in frequency. This smoothing is performed with a CLEAN-like deconvolution algorithm \citep{Hogbom1974, Parsons2009}, filtering gains in 2D Fourier space on a 6 hour timescale and a 10\,MHz frequency scale (or equivalently, 100\,ns delay scale). These are the scales on which we have evidence for intrinsic gain variation in time \citep{Dillon2020} and frequency \citep{Kern2020b}. For more implementation details, see \Ha{}.
    
    \item \textbf{LST-binning:} Having calibrated and flagged each night's data, we then coherently average nights together on a fixed LST grid. This 21.4\,s cadence grid---double the integration time of raw visibilities---is created by assigning each observation to the nearest gridded time and then rephasing that visibility to account for sky rotation due to the difference in LST. An additional round of flagging is performed on a per-LST, per-frequency, and per-baseline basis, looking for $5\sigma$ outliers\footnote{\Ha{} mistakenly stated that $4\sigma$ outliers were thrown out as part of the ``sigma-clipping" procedure. The cut was actually $5\sigma$ in both this work and in \Ha{}.} in modified $z$-score among the list of rephased visibilities to be averaged together. This cut is designed to identify low-level residual RFI and calibration failures; it is highly unlikely to flag outliers due to noise.
    
    \item \textbf{Hand-flagging:} After LST-binning, a final set of additional flags are added by manually examining high-pass delay-filtered residuals. This filtering was performed on using an iterative delay CLEAN to remove power below the 2000\,ns scale in order to highlight spectrally compact features. Clear outlier channels and/or times that are consistent across baselines are flagged upon visual inspection. The same additional flags are applied to all baselines.
    
    \item \textbf{Inpainting flagged channels:} When using Fast Fourier Transforms (FFTs) to form power spectra, flagged channels introduce discontinuities that leak foreground power to high delays. To avoid this, we use the same delay-based CLEAN algorithm to low-pass filter the data on 2000\,ns scales and inpaint the flagged channels with the filtered result. Entirely flagged times are not inpainted. For more details and a demonstration of this procedure, see \Ha{}.
    
    \item \textbf{Cable reflection calibration:} When a signal bounces off of both ends of a cable before being transmitted, the result is a copy of the signal at a fixed delay associated with twice the light-travel time along the cable. \citet{Kern2020a} showed how the 20\,m and 150\,m cables in the signal chain produce these reflections, which can be represented as complex gain terms. While these gains are in principle calibratable with redundant-baseline calibration, our gain smoothing procedure completely removes them. Thus, following the procedure outlined in \citet{Kern2019}, we iteratively model and calibrate out reflection and sub-reflection terms using autocorrelations, which have much higher SNR than cross-correlations. Since cable reflections are stable over many nights, this is done after LST-binning. 
    
    \item \textbf{Crosstalk subtraction:} \citet{Kern2020a} also demonstrated the pernicious impact of crosstalk systematics across a large range of delay modes in HERA Phase I data. It was hypothesized that this was due to over-the-air crosstalk that led to autocorrelations leaking to cross-correlations at high delays. In \autoref{appendix:crosstalk}, we show how this effect's delay and amplitude structure can be explained by an emitter in the signal chain. Because autocorrelations are non-fringing and thus quite stable in time, the effect can be mitigated by modeling each baseline's excess power near zero fringe-rate. \citet{Kern2019} does this using singular value decomposition (SVD) to find the delay and time modes affected and uses Gaussian process regression to limit the range of fringe-rates modeled and subtracted so as to avoid EoR signal loss. That fringe-rate range is symmetric about zero and limited by the east-west projected baseline length $|b_\text{E-W}|$ such that
    \begin{equation}
         \left|f_\text{max}\right| = 0.024\text{\,mHz}\left(\frac{|b_\text{E-W}|}{1\text{\,m}}\right)  - 0.28\text{\,mHz}. \label{eq:xtalk_frate}
    \end{equation}
    The signal loss due to crosstalk subtraction was calculated in \citet{Aguirre_2022} and corrected for in \Ha{}; we repeat the calculation in \autoref{sec:validation:noisefree} and find similar results. Crosstalk subtraction is not applied to baselines with projected east-west distances less than 14\,m, since the zero fringe-rate mode overlaps with the fringe-rates associated with the main lobe of the primary beam \citep{Parsons2016}. 
    
    \item \textbf{Coherent time averaging:} Following \Ha{}, we next coherently average visibilities from the 21.4\,s cadence after LST-binning down to a 214\,s cadence, using the same rephasing procedure described above to account for sky-rotation. This timescale was chosen to keep signal loss at the $\sim$1\% level \citep{Aguirre_2022}; we repeat that calculation for this data set in \autoref{sec:validation:noisefree} and find consistent results.
    
    \item \textbf{Forming pseudo-Stokes I visibilities:} Before forming power spectra, we construct pseudo-Stokes visibilities. As \Ha{} showed, this limits the leakage of Faraday-rotated foregrounds into high-delays, though $Q\rightarrow I$ leakage from primary beam asymmetry is still possible \citep{Moore2013, Asad2016, Kohn2016, Nunhokee2017, Asad2018}. For the pseudo-$I$ channel, this step consists of simply averaging together the $EE$- and $NN$-polarized visibilities for the same baseline, where $E$ and $N$ denote the east- and north-aligned linearly-polarized feeds respectively \citep{Kohn2019}. 
    
    \item \textbf{Power spectrum estimation:} We compute power spectra using the delay approximation, in which we substitute a Fourier transform along the frequency axis of a visibility (i.e.\ a delay transform) for a line-of-sight Fourier transform \citep{Parsons2012b}. This strategy avoids mapmaking entirely \citep{Dillon_2013, Dillon2015a, Xu_2022}. We thus approximate $\tau$ and $u$ (the magnitude of the baseline in units of wavelengths) as mapping linearly to line-of-sight Fourier modes, $k_\|$, and transverse Fourier modes, $k_\perp$, respectively. The power spectra are estimated by taking the real part of
    \begin{equation} \label{eq:dspec}
        \hat{P}(\kperp, \kpara) = \frac{X^2Y}{\Omega_{pp}B}\widetilde{V}_\text{1}(u, \tau)\widetilde{V}_\text{2}^\ast(u, \tau),
    \end{equation}
    where $\widetilde{V}$ is a Fourier transformed visibility in frequency. $\Omega_{pp}$ is the full-sky integral of the squared primary beam response---we use the beam simulated in \citet{Fagnoni2021}---and $B$ is the bandwidth. $X$ and $Y$ are scalars mapping angles and frequency to cosmological distances, defined via $k_\| = 2 \pi \tau / X$, $k_\perp = 2 \pi |\mathbf{b}| / (Y \lambda)$, with \edited{$X = D(z)$ and $Y = c(1 + z)^2 [H(z)]^{-1} / (1420\text{\,MHz})$,} where $H(z)$ is the Hubble parameter, and $D(z)$ is the transverse comoving distance \citep{Hogg1999}.\footnote{\edited{We note that the definitions of $X$ and $Y$ were erroneously swapped in the text following Equation~14 \Ha{}. However, the power spectrum calculations themselves were performed with the correct definitions of $X$ and $Y$.}}
    $\hat{P}(\kperp, \kpara)$ here is in units of ${\rm mK^2}\ h^{-3}\ {\rm Mpc^3}$, though it conventional to report the ``dimensionless'' power spectrum (which has units of ${\rm mK^2}$):
    \begin{equation}
        \Delta^2(k) \equiv P(k) \frac{k^3}{2\pi^2}. \label{eq:deltasq}
    \end{equation}
    \Ha{} shows how this can be recast into the language of quadratic estimators \citep{Tegmark1997,Liu2011,Dillon_2013,Trott_2016}. In that formalism, we use a diagonal normalization matrix $\mathbf{M}$ (i.e\ no decorrelation of bandpower uncertainties). In lieu of any inverse covariance weighting, we use a Blackman-Harris taper to prevent foreground leakage into the EoR window. When computing \autoref{eq:dspec}, we cross-multiply Fourier transformed visibilities from alternating 214\,s blocks of time (i.e.\ $\widetilde{V}_1$ and $\widetilde{V}_2$), using all pairs of baselines in a redundant baseline group. In the (quite accurate) approximation that visibilities interleaved in this way have uncorrelated noise, this produces an estimate of the power spectrum free of noise bias. 
    
    \item \textbf{Error estimation:} As in \Ha{}, we employ the noise estimation formalism of \citet{Tan2021}. The noise power spectrum is given by 
    \begin{equation}\label{eq:PN}
        P_{N} = \frac{X^2Y\Omega_{\rm eff}T^2_{\rm sys}}{t_{\rm int}N_{\rm coherent}\sqrt{2N_{\rm incoherent}}},
    \end{equation}
    where $T_{\rm sys}$ is the system temperature, $t_{\rm int}$ is the integration time, and $N_{\rm coherent}$ and $N_{\rm incoherent}$ are the numbers of integrations averaged together coherently or incoherently---i.e.\ averaged as visibilities with phase information or averaged as power spectra. $\Omega_{\rm eff}$ is the effective beam area, defined as $\Omega_{\rm eff} \equiv \Omega_p^2 / \Omega_{pp}$ in Appendix B of \citet{Parsons2014}.
    We use the inverse square of the noise power spectrum to perform inverse variance-weighted averaging of the power spectra. For reporting final errorbars on power spectra, we use an unbiased estimator of the noise and signal-noise cross-terms developed by \citet{Tan2021},
    \begin{align}
        \hat{P}_{\rm SN} = P_N & \left (  \sqrt{\sqrt{2}(\hat{P}/P_N) + 1} \right . \nonumber \\
          & \left . ~ \left[\sqrt{1 / \sqrt{\pi} + 1} - 1\right] \right ). \label{eq:PSN}
    \end{align}    
    
    \item \textbf{Incoherent power spectrum averaging:} Here and in \Ha{}, power spectra are averaged incoherently over several axes to produce the final limits. First, all baseline-pairs within a redundant baseline group are averaged, ignoring auto-baseline pairs (power spectra formed from the same pair of antennas at interleaved times).  This preserves most of the sensitivity of coherently averaging visibilities within a redundant group before forming power spectra, while excluding the pairs most likely to exhibit correlated systematics. Then power spectra are averaged incoherently in several disjoint LST ranges, which we call ``fields'' since they correspond to different parts of the sky at zenith. Power spectra are estimated independently for the LST ranges in the separate fields and we perform no further averaging in \Ha{} or this analysis when reporting power spectrum upper limits. 
    Finally, we perform a binning to spherical $k = \sqrt{k_\|^2 + k_\perp^2}$, excising baselines based on their length and certain sets of delay modes based on their proximity to the ``horizon wedge," which is set by the light travel time along the baseline
    \begin{equation}
        \tau_\text{wedge} = |\mathbf{b}| / c. \label{eq:wedge}
    \end{equation}
    $P_{N}$ and $\hat{P}_{\rm SN}$ are propagated through each of these averaging steps.
\end{enumerate}

For more details on the implementation of these techniques, we refer the interested reader to \Ha{} and its supporting papers.

\needspace{2cm}\subsection{Updates to the Data Reduction and Systematics Mitigation Pipeline Since H22a} \label{sec:analysis_updates}

With the full context of the analysis pipeline established above, we now detail the ways it has changed since \Ha{}. While most are relatively minor tweaks \citep{H1C_IDR3_2_Memo}, we detail them here for completeness and reproducibility.  

\needspace{2cm}\subsubsection{Updates to Redundant-Baseline Calibration}

Two minor changes were made to redundant-baseline calibration. The first is the addition of a step in \texttt{firstcal}---the solver for per-antenna delays and phase offsets \citep{Dillon2020}---to also solve per-antenna polarity flips. A polarity flip, which could result from rotating the feed by 180$^\circ$, simply flips the sign on the measured voltage from the antenna or equivalently multiplies the gain by $-1$. Solving for polarities allows \texttt{firstcal} to converge faster and more reliably, but does not appreciably change the result. 

The second change was an increase of the maximum number of iterations allowed in \texttt{omnical}, which uses damped fixed-point iteration to minimize $\chi^2$ in \autoref{eq:chisq} \citep{Dillon2020}. This was increased from 500 to 10,000. This likely makes little difference in practice, since \texttt{omnical} usually only converges that slowly for a given time and frequency in the presence of bright RFI contamination. Since allowing more steps did not substantially increase runtime (individual times and frequencies can converge independently), we felt it was safer give the algorithm as long as necessary to minimize $\chi^2$, even if doing so had little impact after gain smoothing.

\needspace{2cm}\subsubsection{Updates to Absolute Calibration} \label{sec:abscal}

Two important changes were made to absolute calibration as compared to \Ha{}. The first is a change to how flags are propagated from the sky-calibrated reference visibilities. Previously, antennas flagged or otherwise not included in the reference set of visibilities were also flagged on a nightly basis after absolute calibration. In this work, these antennas are simply given zero weight when solving for the degeneracies of redundant-baseline calibration---which are then applied to all gain solutions. Per-antenna flags, once set (see \autoref{sec:antennas}), did not change during the nightly calibration.

Second, we added a new step in absolute calibration to fix the bias discovered in the course of validating the \Ha{} pipeline. In \citet{Aguirre_2022}, we found that absolute calibration produced gains that were biased high and that the bias got larger with decreasing visibility SNR. This is particularly worrisome because gains affect both power spectrum and error estimates quartically, and high gains lead to artificially low power spectrum estimates. In \citet{Aguirre_2022} we calculated the size of this effect and in \Ha{} we increased our measurements and error bars to compensate for this $\sim$10\% bias on our power spectra.

A detailed mathematical account of the origin of the bias appears in Appendix B of \citet{Byrne_2021}. However, it can be understood intuitively as follows: when solving for the overall gain degree of freedom in absolute calibration, noise turns individual visibilities in the complex plane into samples of a circularly symmetric distribution whose center is displaced from the origin (the ``true'' visibility). When measuring magnitudes, that probability distribution is Ricean and always has a larger mean than the magnitude of the ``true" visibility. Simply put, adding symmetric noise is more likely to increase the amplitude of a complex number than decrease it. However, by calibrating the overall multiplicative amplitude as a complex number and then only taking the absolute value at the very end, one can avoid this bias---as we show in \autoref{sec:validation:endtoend} and \autoref{fig:validation_gain_errors}.

\needspace{2cm}\subsubsection{Updates to RFI Excision} \label{sec:rfi}

\begin{figure*}[t!]
    \centering
    \includegraphics[width=\textwidth]{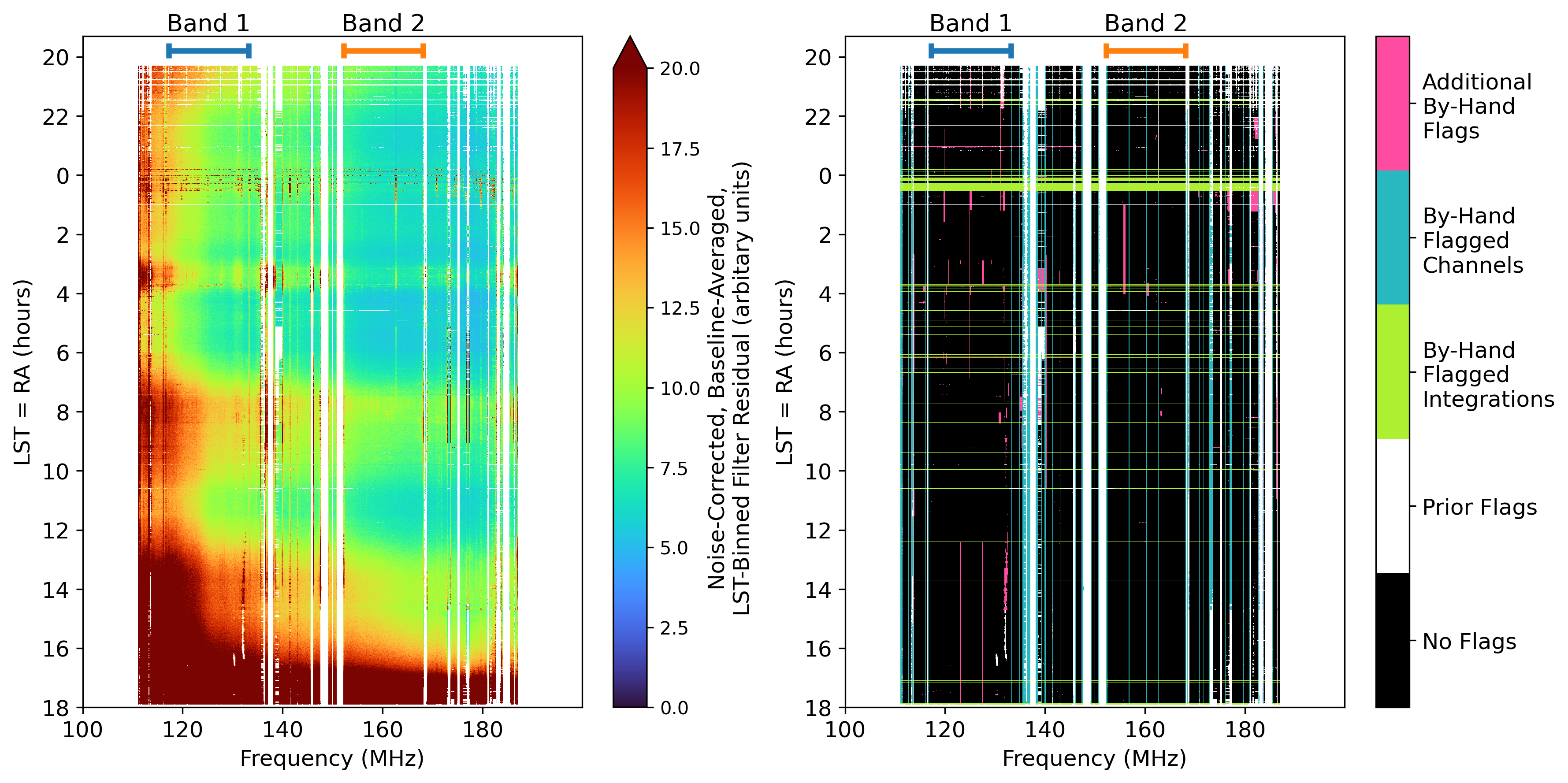}
    \caption{Here we illustrate the process for RFI excision after binning together all 94 days. On the left, we show 2000\,ns high-pass filter residuals of all epochs combined. It shows some clear temporal and spectra structure which necessitates further flagging. These additional flags, shown on the right, are performed first by looking for outlier channels or integrations, then by hand-flagging contiguous regions of excess structure. We attribute these outliers to low-level RFI, as well as the interplay between the night-to-night variation in the RFI flagging mask and night-to-night changes in calibration errors. Most of the additional flagged pixels in this waterfall were already flagged on a significant fraction of the nights.}
    \label{fig:flag_origins}
\end{figure*}

While the fundamental algorithm for RFI excision remains unchanged, we made two updates to how it was performed on a nightly basis. The first is related to how the analysis dealt with data file boundaries. Previously, every data file was analyzed in parallel for RFI. Since the outlier identification algorithm relies on neighboring times and frequencies, this became less reliable near file boundaries where there are roughly half as many neighbors to compare to, and likely led to some of the $\sim$10~minute periodicity we saw in the flags in \Ha{}. In this analysis, we parallelized the pipeline in overlapping time chunks, so that every point was compared to exactly the same number of neighbors---except at the beginning and end of the night and at the edges of the band.

Second, we modified the set of data products used in searching for outliers. In both analyses we used raw visibilities (albeit only in the mean filter round); gains from both redundant-baseline calibration and absolute calibration; and $\chi^2$ from both calibration steps. Based on experiments we performed on which data products were providing unique information and not leading to overflagging, we removed a global cut on outliers in $\chi^2$---which likely led to the over-flagging around Fornax A in \Ha{}---and added uncalibrated autocorrelations for their high SNR and computational tractability compared to the full set of visibilities. The result is still a very expansive set of flags and likely contains a significant number of false positive identifications of RFI, especially around ORBCOMM at 137\,MHz and the clock line at 150\,MHz (see e.g.\ \autoref{fig:flag_origins} and \autoref{fig:bands}). 
Given the extreme dynamic range requirements of 21\,cm cosmology, it is far safer to over-flag than under-flag \citep{Kerrigan_2019}.

In that spirit, we also revisited how a final set of by-hand flags were identified. In \Ha{}, this flagging was performed by examining a handful of key baselines after high-pass filtering to remove structure below the 2000\,ns scale. In this work, the four epochs were first combined together without any inpainting, reflection calibration, or crosstalk subtraction. After performing the same per-baseline high-pass filter on every baseline, their amplitudes were averaged together, inverse variance weighted by noise and then corrected for the noise bias.

\begin{figure*}[t!]
    \centering
    \includegraphics[width=\textwidth]{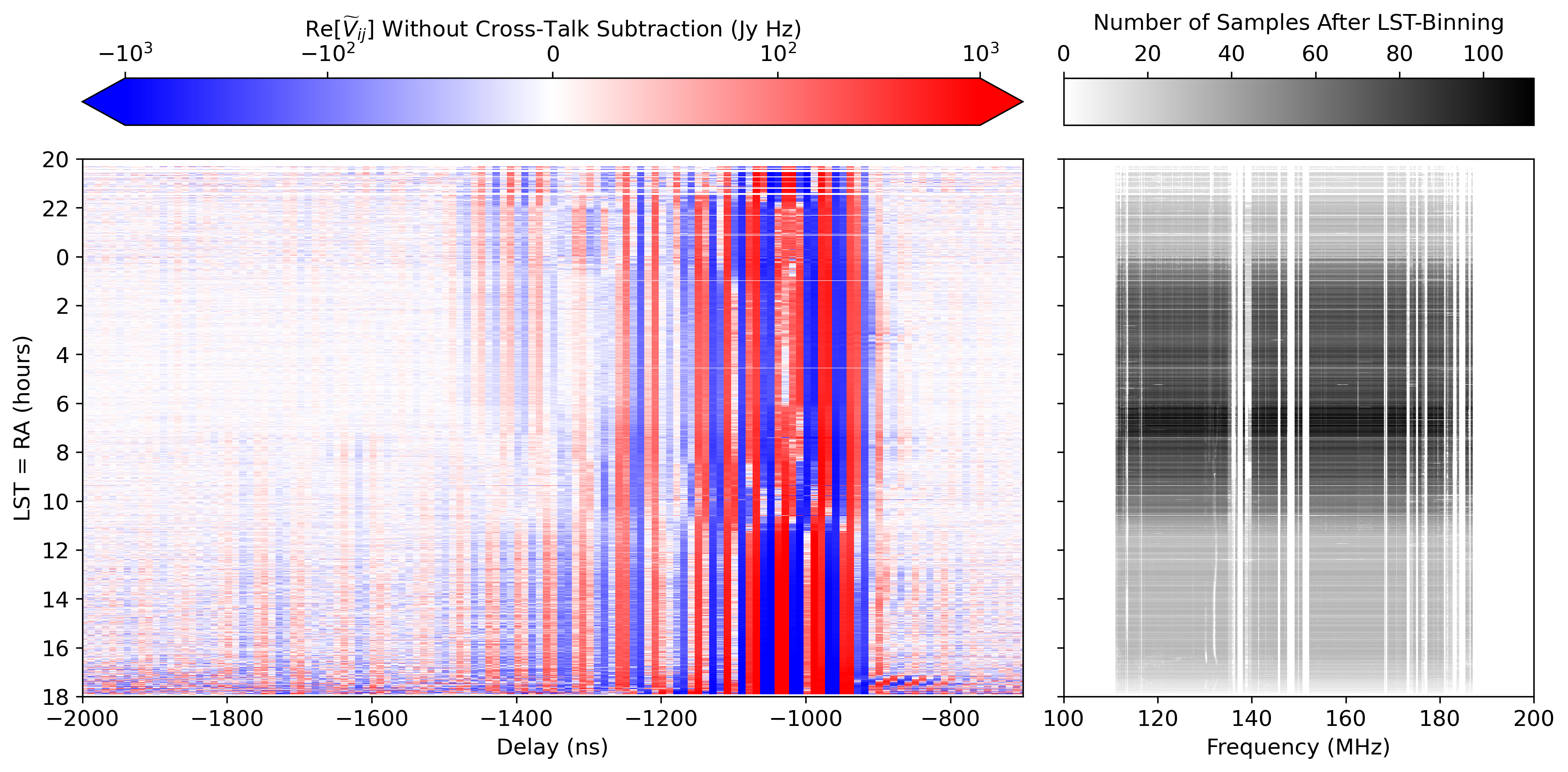}
    \caption{In \citet{Kern2020a}, the observation that the phase of crosstalk systematics remained stable in delay in space was key to removing them down to nearly the noise level. This technique proved foundational to our results in \Ha{}. However, when we combined all four epochs together, we began to see discontinuities in the phase structure of the crosstalk. In the left panel, this effect is particularly clearly illustrated in Band 1 of the north-polarized baseline between antennas 11 and 66. We see a number of delay modes where the sign of the crosstalk feature abruptly shifts from positive to negative (or vice versa) as a function of LST. These discontinuities appear to be associated with epoch boundaries, which give rise to discontinuities in the number of samples binned together (right panel). We hypothesize that these discontinuities arise due to the effects of the ongoing construction of HERA, either on the source of the crosstalk emission or on how it is transmitted through the array (see \autoref{appendix:crosstalk} for more detail).}
    \label{fig:xtalk_discon}
\end{figure*}

The result, shown in the left-hand panel of \autoref{fig:flag_origins}, highlights residual frequency structure. 
Much of this has low $N_\text{samples}$ and/or borders on previously identified RFI, indicating its origin as low-level, inconsistently flagged RFI. As the right-hand panel shows, outlier channels and integrations were first identified by averaging along each axis. Next, individual areas of concern were flagged by hand, by converting the waterfall to a bitmap image and individually marking times and frequencies in Adobe Photoshop. An effort was made to flag coherent rectangular regions near previously identified flags to avoid cherry-picking, though fundamentally this step involved a series of subjective judgment calls. Once the final flagging waterfall was developed, it was not revisited after estimating power spectra in order to avoid experimenter bias.

\needspace{2cm}\subsubsection{Updates to Reflection Systematics and Crosstalk Mitigation} \label{sec:xtalk_updates}

\begin{figure*}
    \centering
    \includegraphics[width=\textwidth]{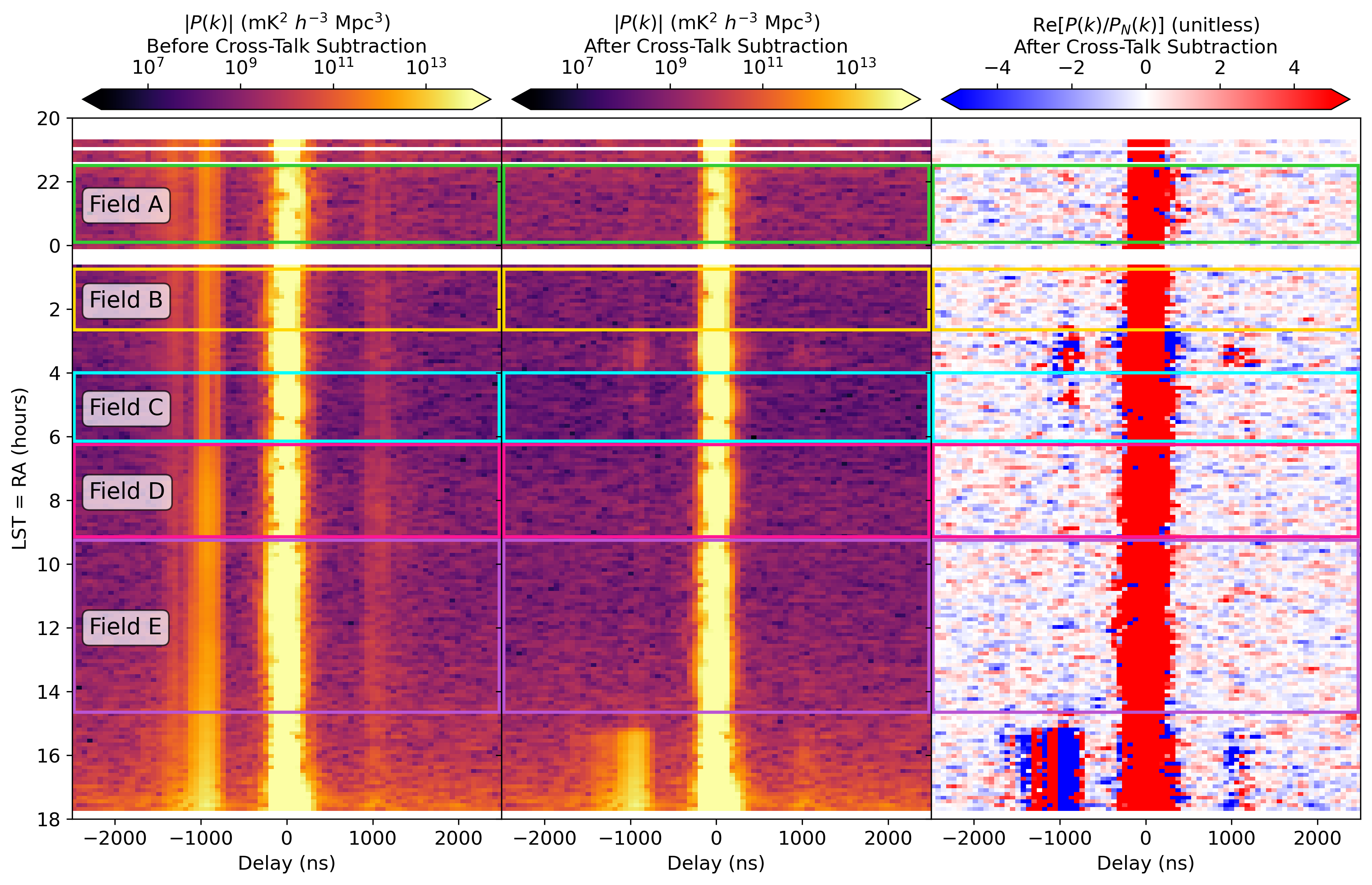}
    \caption{Here we show the effect of our crosstalk subtraction algorithm on the power spectrum of a  baseline-pair (in this case, 38--66 crossed with 52--82 in Band 1) with a particularly strong crosstalk signal at negative delay. The technique of \citet{Kern2019}, applied on a per-epoch basis, removes the crosstalk features seen here at roughly $\pm$1000\,ns down to a level nearly consistent with noise even after averaging all four epochs together. Since the crosstalk is proportional to the amplitude of the autocorrelations (see \autoref{appendix:crosstalk}), which rise very steeply when the Galactic center transits the beam at 17.8~hours LST, and since our removal algorithm depends on the temporal stability of the crosstalk, we give LSTs between $\sim$15.3 and 21~hours zero weight. This results in cleaner residuals in the more sensitive fields, at the price of greater residuals at high LST. This also motivated our definition of the five fields in which the power spectrum was independently estimated; see \autoref{sec:bands_and_fields} and \autoref{fig:fields} for more details on the definition of the fields.}
    \label{fig:xtalk_demonstration}
\end{figure*}

In \Ha{}, all the steps in \autoref{sec:pipeline_overview} after LST-binning were performed on the full-sensitivity, 18-night data set. At first, we attempted performing the same analysis with all 94 nights binned together but found that the level of residual crosstalk had substantially higher SNR than anything seen in \Ha{}. Some baselines were particularly bad, exhibiting a delay- and time-averaged SNR greater than 10 in the affected delay range. Upon examining the pre-subtraction waterfalls of the baselines where crosstalk subtraction performed the worst, we found clear evidence for temporal structure in the delays contaminated by crosstalk. In \autoref{fig:xtalk_discon}, we show one such baseline. 
Plotting the real part of its waterfall in delay space shows clear temporal structure at certain delays, including some where it flips from positive to negative and vice versa. This is potentially disastrous for the crosstalk subtraction technique of \citet{Kern2019}, which relies on stability in delay space. 

Furthermore, there appears to be a correlation between discontinuities in $N_\text{samples}$ (right-hand panel of \autoref{fig:xtalk_discon}) and discontinuities in the delay structure of $\tilde{V}_{ij}$. Since the former are largely attributable to epoch boundaries, which affect how often each LST is observed, we hypothesized that the changing and growing array was affecting the precise structure of the crosstalk. This ultimately pointed us toward a new understanding of the physical origin of the effect, namely that all antennas' signals were being broadcast from one point on the west side of the array, likely the refrigerated enclosures which contained the analog receivers. We discuss this model and the evidence for it in detail in \autoref{appendix:crosstalk}. 

The upshot of this result is twofold. First, it confirms that the model of \citet{Kern2019}, of autocorrelations leaking into cross-correlations, is correct. Second, it implies that as long as the array is stable, the effect should be stable in LST-binned data as well. We therefore decided to perform inpainting, cable reflection calibration, and crosstalk subtraction on a per-epoch basis before binning together the four epochs. This proved a substantially better approach---as \autoref{fig:xtalk_demonstration} shows---and got us much closer to consistency with noise after crosstalk subtraction.

A few minor tweaks were also made to cable reflection calibration and crosstalk subtraction. The total number of terms used to fit the cable reflections was increased from 28 terms between 75 and 1500\,ns to 35 terms between 75 and 2500\,ns, to both better model the 20\,m cables and to be able to model a few extra long cables whose reflection timescale were larger than 1500\,ns but were not in the \Ha{} data set. The SVD used in crosstalk subtraction was previously limited to 30 delay modes; we increased it to 50 based on experiments where it made the crosstalk residuals a bit more consistent with noise. Finally, we revised how weights were applied before computing the SVD. First, we weighted each time by the frequency-averaged number of samples. \Ha{} used an unweighted SVD, but as \autoref{fig:lst_coverage} shows, the approximation of weights that are constant in time breaks down when considering such a large and discontinuous data set. Second, we also set the weight in the SVD to zero from 15.3--21~hours to prevent bright galactic emission at the edge of the data set from introducing temporal structure that caused the crosstalk subtraction procedure to perform worse for the most sensitively measured LSTs.

\subsection{Updates to Power Spectrum Estimation Since H22a} \label{sec:pspec_updates}

Just as with the data reduction pipeline, we sought to apply the same power spectrum analysis procedures and choices as were used in \Ha{}. However, differences in LSTs observed, flagging, and systematics removal motivated slightly different approaches to how the data should be reduced and cut. These decisions were made without reference to the final power spectra in order to minimize experimenter bias.

\subsubsection{Picking Bands and Fields} \label{sec:bands_and_fields}

\begin{figure}[b!]
    \centering
    \includegraphics[width=.48\textwidth]{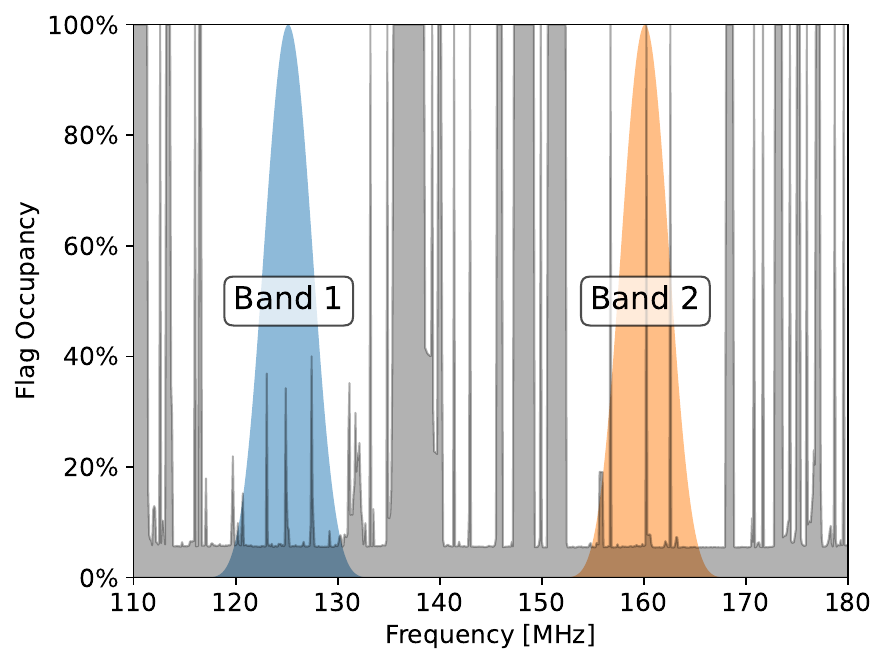}
    \caption{The two frequency bands used in this analysis were chosen to avoid sections of the band with heavy flagging. Here we illustrate the bands as Blackman-Harris windows, indicating the relative weight of the different channels in our final power spectra. These two bands are from  117.19--133.11\,MHz and 152.25--167.97\,MHz, corresponding to $z=10.4$ and $z=7.9$ respectively. These bands differ very slightly from the bands with the same names used in \Ha{}.}
    \label{fig:bands}
\end{figure}

The two frequency bands analyzed in \Ha{} were Blackman-Harris tapered ranges from 117.09--132.62\,MHz and 150.29--167.77\,MHz. These were motivated by the two contiguous regions of low flag occupancy (see Figure~12 in \Ha{}). 
We reproduce that same plot in \autoref{fig:bands} and following the same logic---albeit with somewhat different flags---pick Band 1 and Band 2 to range from 117.19--133.11\,MHz and 152.25--167.97\,MHz, respectively. The bands still center on approximately the same redshifts: $z=10.4$ and $z=7.9$.

\begin{figure*}[t!]
    \centering
    \includegraphics[width=\textwidth]{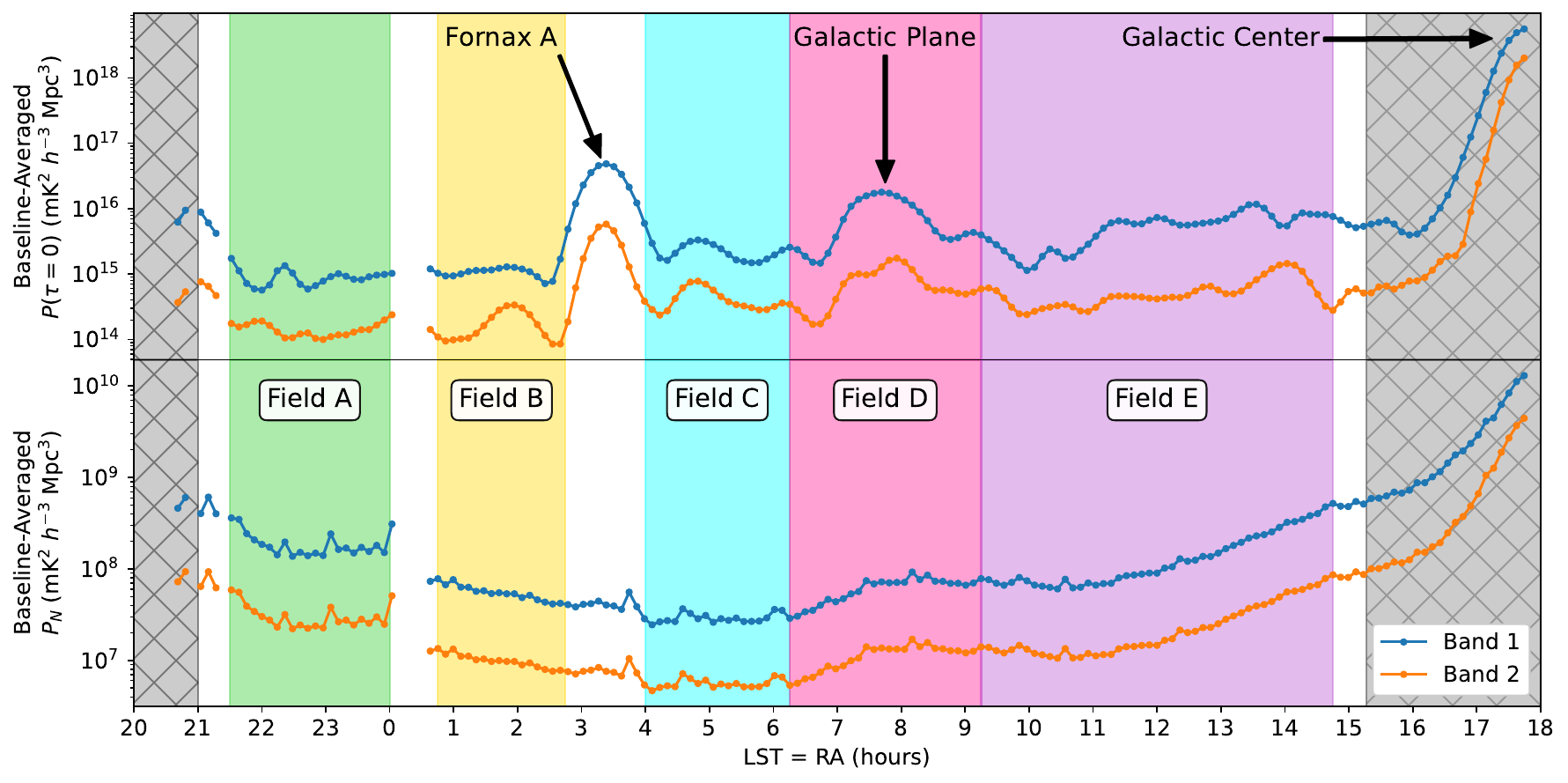}
    \caption{Two metrics---the baseline-averaged power spectrum at $\tau=0$ in the top panel, and noise power spectrum amplitude in the bottom panel---were used to divide the data set into fields in which to independently estimate the power spectrum. Following \Ha{}, which divided the data set into three fields, we divide the data set into five and label them A, B, C, D, and E since they do not directly correspond to the fields used in \Ha{}. As we describe in \autoref{sec:bands_and_fields}, we picked the fields to avoid gaps in the data, Fornax A (at 3.4~hours), and the galactic center (at 17.8~hours). These are all features that introduce sharp temporal changes that make crosstalk subtraction more difficult. We also wanted to avoid the times where the SVD in the crosstalk subtraction was given zero weight (gray hatched region) to avoid affecting the more sensitive fields.}
    \label{fig:fields}
\end{figure*}

Because we observed a larger range of LSTs than in \Ha{}, we need to define new fields (i.e.\ new LST ranges) in which to independently estimate power spectra. In order to motivate the choice of fields without reference the final power spectra, we looked at two other statistics, which we plot in \autoref{fig:fields}.
These two statistics are the inverse-variance-weighted, baseline-averaged delay-zero power, $P(\tau = 0)$---a proxy for foreground power---and $P_N$, which is flat in delay and tells us about both foreground power and total observation time. We used these to define a total of five fields: A (21.5--0.0~hours), B (0.75--2.75~hours), C (4.0--6.25~hours), D (6.25--9.25~hours), and E (9.25--14.75~hours).\footnote{While fields B, C, and D correspond most closely to fields 1, 2, and 3 in \Ha{}, they are different enough that we chose to change the nomenclature to prevent conflation of the two. Band 1 and 2 are close enough to those in \Ha{} to be treated as equivalent for most purposes.} We restricted our field boundaries to quarter-hour increments to avoid cherry-picking integrations.

The rationale for defining the fields is as follows. We wanted fields B, C, and D, to correspond reasonably well to fields 1, 2, and 3 in \Ha{}, so to cover the new LST ranges, we added fields A and E. Field A was set by the flagging gaps at either end, intentionally avoiding the last integration before the flagging gap between 0 and 1~hours due to the potential for signal loss from crosstalk subtraction \citep{Aguirre_2022}. Likewise, field B was defined to exclude the first integration after the gap and keep Fornax A in the main lobe no brighter than its brightest point in the first sidelobe around 2 hours. Field C was defined to start at a roughly symmetrical place to where Field B ended with respect to Fornax A and to include the range of maximum sensitivity from roughly 4--6\,hrs. Thus, the upper field boundary was set by the sidelobe of the Galactic plane at 6.25 hours. The boundary between fields D and E was set to include the roughly symmetrical sidelobe at $\sim$9 hours within Field D, keeping the Galactic plane contained to a single field. Field E ends a bit before where the cross-talk subtraction gets zero weight in the SVD. Once these field definitions were established, they were not allowed to change.

\needspace{2cm}\subsubsection{Excluding Baseline Pairs with Substantial Residual Crosstalk} \label{sec:xtalk_residual}

Despite subtracting crosstalk on a per-epoch basis, we still found strong evidence for residual crosstalk on certain baselines. In the baseline examined in \autoref{fig:xtalk_demonstration}, we can see clear residual power as an excess SNR in the delay range of $800 < |\tau| < 1500$\,ns (right-hand panel). This is most prominent near the Galactic center, which got zero weight in the SVD, and near Fornax A, but there appears to be a slight excess at other LSTs as well. 
To quantify this, we averaged $|$SNR$|$ over that delay range and over the three most sensitive fields, B, C, and D. This average was performed separately for positive and negative delays, since we now know that those two signals have independent origin (see \autoref{appendix:crosstalk}). 

\begin{figure}
    \centering
    \includegraphics[width=.48\textwidth]{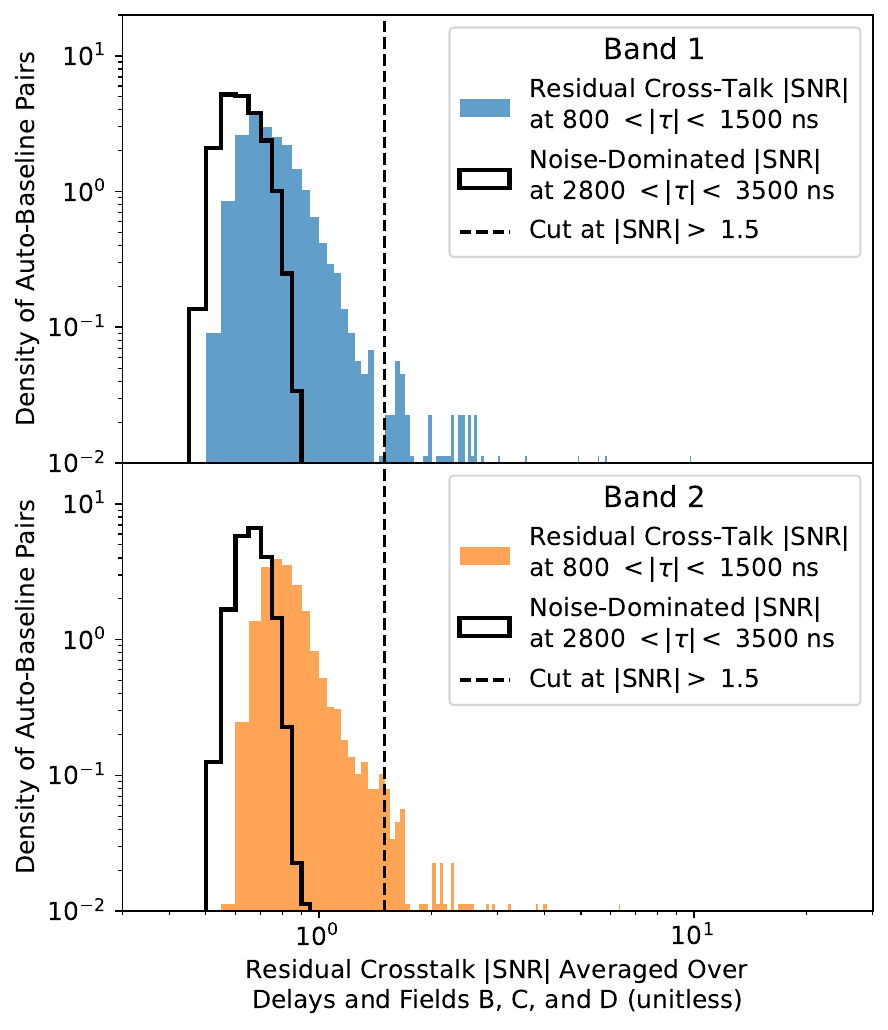}
    \caption{While our crosstalk subtraction algorithm removes the vast majority of the systematic, the technique is not perfect. We quantify the level of residual crosstalk by taking the average magnitude of the SNR in our most sensitive fields---B, C and D---and in the delay range most affected by crosstalk, $800 < |\tau| < 1500$\,ns. Comparing this quantity to the same quantity computed in the noise-dominated delay-range of $2800 < |\tau| < 3500$\,ns, we find significant excess. Using auto-baseline-pairs, which are not included in the final power spectrum, we eliminate baselines with $|$SNR$| > 1.5$. Note that our most sensitive upper limits come from lower delays, so this is not a potentially biasing cut on SNR in the final quantity of astrophysical interest. Because the cross-talk at positive and negative delay is sourced independently (see \autoref{appendix:crosstalk}), we perform this cut separately for positive and negative delays. Because, by convention baselines are east-west-orientated and not west-east-oriented and because the crosstalk emitter is close to the west side of the array, more negative delays are flagged than positive delays. In all, this cut flags 3.5\% and 3.8\% of baselines at negative delays for bands 1 and 2, respectively, and 1.0\% and 0.9\% of baselines at positive delays.}
    \label{fig:xtalk_histogram}
\end{figure}

We computed this averaged $|$SNR$|$ for every auto-baseline pair (i.e.\ power spectra formed from the same baseline at interleaved times, rather than power spectra formed from different but redundant baselines) and plotted a histogram of it for each band in \autoref{fig:xtalk_histogram}, treating positive and negative delays as independent samples. 
Compared to an equivalently sized delay range from $2800 < |\tau| < 3500$\,ns (solid black), we see evidence for a mild excess on most baselines. This is perhaps not too surprising; the crosstalk subtraction algorithm of \citet{Kern2019} attempts to model and subtract the crosstalk down to the noise---in our case, the noise in a single epoch. To the extent that crosstalk remains correlated from epoch to epoch, integrating down should reveal more crosstalk. That said, there is a tail of outliers in \autoref{fig:xtalk_histogram} that motivated us to perform a cut at $|$SNR$| > 1.5$. The cut was performed separately for positive and negative delays, so some baseline pairs are ``half-flagged.'' The vast majority of auto-baseline pairs ($>$95\%) were kept. More baseline-pairs were cut at negative delays than positive delays because the antenna ordering means that negative delays were more often associated with antennas nearer the crosstalk source (see \autoref{appendix:crosstalk}).

We also computed the $|$SNR$|$ for cross-baseline pairs as well and found that they were highly correlated with the $|$SNR$|$ of the two corresponding auto-baseline pairs. However, we decided to more conservatively use only the auto-baseline pairs---which are not included in the final power spectra---for our cut. Any cross-baseline pair with one baseline participating in a flagged auto-baseline pair (and delay sign) was flagged. This cut is the most surgical and perhaps more worrisome analysis change from \Ha{}, in the terms of removing individual power spectra before averaging. However, by only looking at high $|\tau|$---well above the corresponding $k$ values where we set our tightest upper limits---and by using only the auto-baseline-pairs, we insulate ourselves from the risk of cherry-picking and signal loss.

\begin{figure}
    \centering
    \includegraphics[width=.48\textwidth]{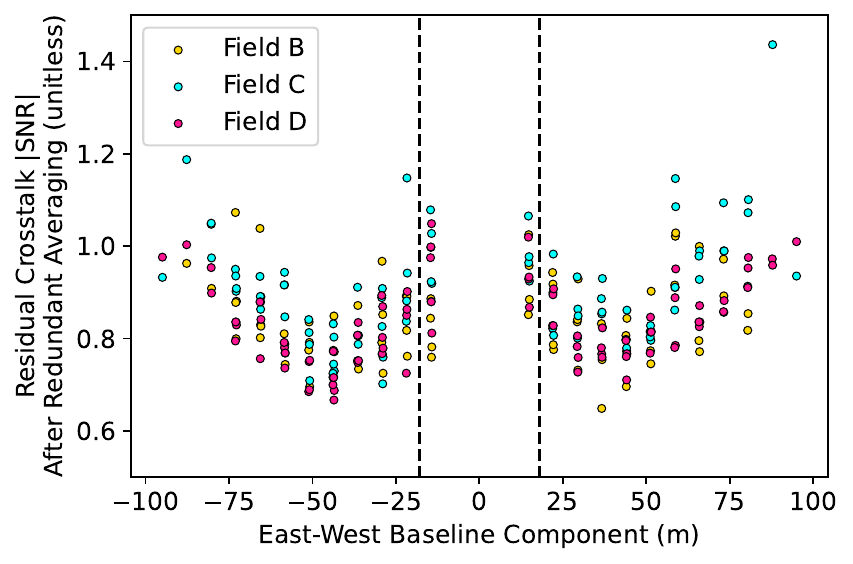}
    \caption{Even after flagging baselines for residual crosstalk, averaging the power spectra of cross-baseline pairs incoherently within redundant groups shows residual crosstalk. This is not surprising, since  \autoref{fig:xtalk_histogram} showed ubiquitous excess in this delay range. What averaging reveals more clearly is an important trend in crosstalk $|$SNR$|$ with projected east-west baseline.  Baselines with small east-west components are the slowest fringing, which means that the crosstalk subtraction algorithm of \citet{Kern2019} only attempts to remove crosstalk in a relatively small range of fringe rates for fear of removing cosmological signal. That is why baselines with east-west components less than $14$\,m were removed before power spectrum estimation in \Ha{} and do not appear here either. Based on this metric, we decided to extend that cut to baselines with east-west components less than 15\,m (those inside the dashed lines), which includes all single-unit separated baselines in the hexagonal grid (see \autoref{fig:layout}). While some other long baselines also show strong residuals, this noted jump among the most sensitive baselines was concerning enough to merit a cut. 
    }
    \label{fig:xtalk_vs_baseline}
\end{figure}

One other key change from \Ha{} is the shortest east-west projected baseline length allowed to be included in our final spherical power spectra. Even after averaging cross-baseline pairs within redundant groups, we still see a substantial uptick in $|$SNR$|$ in the crosstalk delay range for baselines with 14.6\,m projected east-west baselines, the shortest baselines used in \Ha{} (see Figure~\ref{fig:xtalk_vs_baseline}). 
This makes sense physically---these baselines have their main lobe closest to zero fringe-rate, where the crosstalk subtraction algorithm only operates on a very narrow range of fringe-rates (\autoref{eq:xtalk_frate}) for fear of signal loss. While the crosstalk is centered at 0\,mHz, it has some width in fringe-rate space. We expect, therefore, that these baselines should be the first to show residual cross-talk as we integrate down. It was also likely true in \Ha{}; the lower noise level in these data simply makes the systematics clearer.

We decided to conservatively increase that cut to 15\,m, throwing out several redundant baseline groups, including the most sensitive single baseline group used in \Ha{}, the single-unit 14.6\,m east-west baseline. To keep this baseline without simply accepting excess systematics, we would have had to find a way to more aggressively filter crosstalk, which would then have necessitated a more thorough and precise quantification of baseline-dependent signal loss using end-to-end simulations. Since our aim in this work is to apply the analysis of \Ha{} as directly as possible, we defer such an investigation to future work.

\needspace{2cm}\subsubsection{Changes to $k$ Cuts and Bins} \label{sec:kcuts}

The final key analysis change between \Ha{} and this work is an increase in the area of power spectrum modes that were excised from within the EoR window (but still near the wedge). In \Ha{}, the modes excluded from the spherical power spectra were all those within 200\,ns of the horizon wedge (\autoref{eq:wedge}). This ``wedge buffer" has a long history in the field, going back to \citet{Parsons2012b}, which suggested that a combination of foreground and beam chromaticity and the application of tapering functions in the delay power spectrum can extend power $\sim$0.15\,$h$Mpc$^{-1}$ beyond the horizon wedge. The choice of 200\,ns in \Ha{} (equivalent to 0.11\,$h$Mpc$^{-1}$ at $z=7.9$ and 0.10\,$h$Mpc$^{-1}$ at $z=10.4$) was motivated by Figures 14 and 15 of that paper, which show power spectrum SNRs after cylindrically binning to $k_\|$-$k_\perp$ space. The buffer was picked to mostly exclude the region of $k$-space with SNR consistently larger than 1, while balancing that exclusion of foreground-dominated modes at low $k_\perp$ against the admission of noise-dominated modes at high $k_\perp$.

Reassessing the same question in light of our equivalent plots of SNR in cylindrical $k$-space (\autoref{fig:cyl_pk_1} and \autoref{fig:cyl_pk_2}), we increase the wedge buffer to 300\,ns, to achieve roughly the same balance of excluding and admitting modes. We picked 300\,ns as a round number to avoid cherry-picking. This produces a wedge buffer at $k=0.15$\,$h$Mpc$^{-1}$ in Band 1 and $k=0.17$\,$h$Mpc$^{-1}$ in Band 2, which is more in line with the value suggested by \citet{Parsons2012b}, and used in early HERA forecasts \citep{Pober2014}. That said, increasing the wedge buffer is another sensitivity hit, so finding other ways to mitigate foreground emission near the wedge (e.g.\ \citealt{Ewall-Wice2021}) in future work might increase the constraining power of this data set.

One other minor change to our spherical power spectra is our precise binning in $k$. In most of the EoR window, $k$ is dominated by $k_\|$, which maps to $\tau$. \Ha{} picked $k$ bin centers and widths with the intention that two $\tau$ modes would fall into each bin. However, using a fixed $\Delta k$ of 0.064\,$h$Mpc$^{-1}$ for both bands, with the first bin centered at $k=0$, did not quite achieve this. Some fixed-$\tau$ modes got split between $k$ bins and some $k$ bins had more power spectra averaged together than others. To achieve a better alignment with $k$ bin centers nearer the average $k$ value of modes in the bin, we used $\Delta k = 0.0619$\,$h$Mpc$^{-1}$ in Band 1 and $\Delta k = 0.0709$\,$h$Mpc$^{-1}$ in Band 2m with the first bin centered at $3 \Delta k / 4$. For more details on this change, see \citet{H1C_IDR3_2_Pspec_Memo}.

\needspace{2cm}\subsubsection{More Precisely Calculating Power Spectrum Window Functions} \label{sec:window_updates}

The expectation value of the estimated power spectrum for a given baseline and delay, $\hat{P}(\mathbf{u},\tau)$, is actually a weighted sum of the neighboring true bandpowers $P(\mathbf{k})$. These weights are usually referred to as the window functions $W$, defined through
\begin{equation}\label{eq:def_wf}
\hat{P}(u,\tau) \propto  \int \mathrm{d}^3 \mathbf{k} \,P(\mathbf{k})\,W(\mathbf{k};\mathbf{u},\tau).
\end{equation}
In \Ha{}, the horizontal error bars on the spherical power spectrum were evaluated with the same assumptions made to analyze the data, i.e.\ the delay approximation, in which the delays and line-of-sight Fourier modes are treated interchangeably. This leads to underestimating the tails of the window functions, and in particular the foreground power leaking into the EoR window at low $k_\parallel$ \citep{Liu2014a}. In this work, we estimate the exact window functions by lifting this approximation in the derivative of the covariance
with respect to each bandpower, and hence obtain an accurate description of the mapping between instrumental space $(\mathbf{u},\tau)$ and cosmological space $(k_\perp, k_\parallel)$. In doing so, we can account for the delay approximation when comparing theory to data because we now know exactly which $k_\perp$ and $k_\parallel$ modes contribute to a given bandpower and in what proportion.
Note that, in order to account for the frequency dependence of the HERA primary beam, 
we have used the simulations introduced in \citet{Fagnoni2021}. For details on the derivation of the window functions, as well as a complete illustration of their importance in the analysis of low-frequency radio data in general, and of the HERA data in particular, we refer the interested reader to \citet{Gorce2022}. We discuss the impact of the improved window function calculation in \autoref{sec:sphere_limits}.

\needspace{2cm}\subsubsection{Quantifying Decoherence Due to Non-Redundancy} \label{sec:nonredundancy}

Because we average together power spectra of pairs of different baselines within a redundant group, we must quantify the effect of non-redundancy on the power spectrum. We know HERA's putatively redundant baselines are not quite redundant \citep{Dillon2020, Carilli_2020}, so we should expect some level of decoherence when cross-multiplying baselines that see a slightly different beam-weighted sky. Following \Ha{}, we compare incoherent power spectrum averages---which are decoherence-free by construction because they only use auto-baseline pairs---to forming power spectra from visibilities coherently averaged within a redundant baseline group. This yields a metric for decoherence given by 
\begin{equation}
    \Delta \kappa (t, \tau) \equiv \frac{P_\text{coherent}(t, \tau) - P_\text{incoherent}(t, \tau)}{\langle P_\text{incoherent}(t,\tau)\rangle}, \label{eq:non-redundancy-decoherence}
\end{equation}
where the angle brackets indicate a rolling time average over 1~hour timescales to ameliorate the effects of nulls in power for certain baselines (as was done in \Ha{}).

\begin{figure}
    \centering
    \includegraphics[width=.48\textwidth]{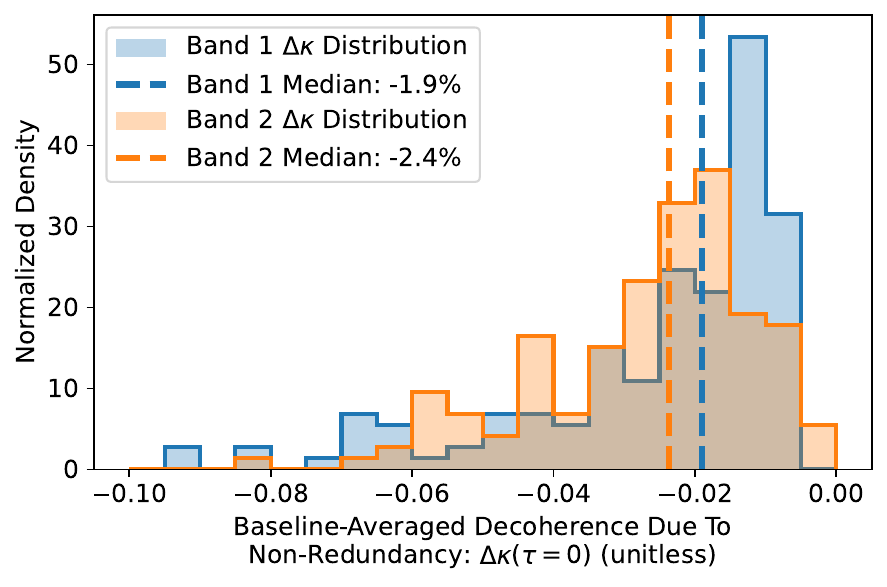}
    \caption{Following \Ha{}, we show here our estimate for signal loss due to non-redundancy while forming and averaging cross-baseline power spectra within redundant baseline groups. Our metric $\Delta \kappa$, defined in \autoref{eq:non-redundancy-decoherence}, looks at the difference between the power in an incoherent average and a coherent average over baselines. We compute those powers at $\tau = 0$, the delay of sources at zenith, as a proxy for cosmological signal loss in the main beam. As in \Ha{}, that difference is normalized by the incoherently-averaged power smoothed on 1~hour timescales to minimize the effect of nulls. This quantity is then weighted by the inverse noise variance, and we show here the histogram over unique baselines and over our five fields.  Taking the median of this histogram, we estimate a 1.9\% correction for Band 1 and a 2.4\% correction for Band 2, as we discuss in \autoref{sec:correcting_bias}.}
    \label{fig:non-redundacy}
\end{figure}

In particular, we examine $\Delta \kappa(t, \tau=0)$, which we take as a foreground-dominated (and thus high SNR) metric of decoherence of sky signal in the primary beam. In \autoref{fig:non-redundacy}, we show the histogram of decoherence levels at zero delay, using different LSTs and unique baselines as samples of decoherence.
\Ha{} performed this analysis over a short range of LSTs around the galactic plane crossing (7.2--8.3~hours) and found a $\sim$1\% signal loss due to decoherence. We measure $\Delta \kappa$ by taking a median over a wider range of LSTs---all five fields---in part to account for LST- or JD-dependent gain errors and the possibility that the array became more or less redundant over the season. Assuming that the signal loss due to non-redundancy should relatively be stable in LST, and with the knowledge that nulls in power on certain baselines can create spurious temporal outliers in $\Delta \kappa$ that always create extra apparent decoherence, even with the 1-hour smoothing, we take the median of this histogram as our signal loss estimate. This yields a slightly larger estimate of the loss: 1.9\% for Band 1 and 2.4\% for Band 2. This will be accounted for in our final power spectrum upper limits, as we will discuss presently.

\needspace{2cm}\subsubsection{Correcting for Potential Biases and Signal Loss} \label{sec:correcting_bias}

\Ha{} performed a careful accounting for four potential sources of bias in the final power spectrum. Three of these are forms of signal loss---ways in which true sky power from the 21\,cm signal can be removed due to the analysis choices we made. The fourth, due to absolute calibration, produced an overall bias in \Ha{} that affected both our measured power spectrum and also our noise and error estimates, which are ultimately derived from autocorrelations which were also biased \citep{Aguirre_2022}. Each of these corrections was applied separately per band.

In \autoref{tab:signal_loss}, we report all of the per-band bias corrections used in this work.
\begin{table}
    \centering
    \caption{Summary of fractional signal power lost over the course of the analysis. The effect of non-redundancy is estimated from the data directly (\autoref{sec:nonredundancy}); the rest are derived from simulations of sky signal, noise, and instrumental systematics (\autoref{sec:validation}).}
    \vspace{-2mm}
    \label{tab:signal_loss}
    \begin{tabular}{p{.26\textwidth}|cc}
    \toprule
    \textbf{Potentially Lossy or Biased Analysis Step} & 
    \vtop{\hbox{\strut \textbf{Band 1}}\hbox{\strut ($z=10.4$)}} & 
    \vtop{\hbox{\strut \textbf{Band 2}}\hbox{\strut ($z=7.9$)}} \\
    \midrule
    Absolute calibration & $\lesssim 0$\% & $\lesssim 0$\% \\
    Crosstalk subtraction & 2.4\% & 3.3\% \\
    Coherent time-averaging in LST & 1.2\% & 1.5\% \\
    Redundant-baseline averaging & 1.9\% & 2.4\% \\
    \midrule
    \textbf{Total signal loss correction} & \textbf{5.5\%} & \textbf{7.2\%} \\
    \bottomrule
    \end{tabular}
\end{table}
The corrections for crosstalk subtraction, coherent time averaging, and redundant-baseline averaging are all forms of signal loss that do not affect our autocorrelations and thus our estimate of the thermal noise $P_N$ (though they can have a small effect on $\hat{P}_{SN}$). All these effects are taken into account when reporting power spectra, errors, and upper limits in \autoref{sec:upper_limits}. 

We discussed our evaluation of the signal loss due to non-redundancy in \autoref{sec:nonredundancy}. The other three sources of bias are quantified using the realistic sky and instrument simulations we use to validate our analysis pipeline, as we discuss in \autoref{sec:validation}. The first two, signal loss due to coherent time averaging and signal loss due to crosstalk subtraction, are evaluated in simulations performed without noise in order to precisely measure the effect on a known EoR-like signal in \autoref{sec:validation:noisefree}. As discussed in \autoref{sec:abscal}, the absolute calibration bias that was due to the effect of low-SNR visibilities should have been eliminated in this work. Indeed, we will show in \autoref{sec:validation:endtoend} that the effect has been reduced in magnitude to less than 1\% in the gains, and has reversed sign---our gains now appear to be biased very slightly low. If this is correct, then that would lead to an overestimate of our power spectra, error bars, and upper limits. Since we are far more concerned with the possibility of signal loss leading us to report an upper limit lower than the data justify, we choose conservatively not to adjust for this effect in the limits we report in \autoref{sec:upper_limits}.


\needspace{2cm}\section{Analysis Pipeline Validation with Simulations} \label{sec:validation}

Before we present our final upper limits in \autoref{sec:upper_limits}, we first report the results of our extensive simulation-based validation of the analysis pipeline. The techniques used for simulating our instrument and the analyses performed on the output of those simulations are very similar to those in \citet{Aguirre_2022}, which was written to support \Ha{}. We present here a brief summary of how we applied those techniques to this work, highlighting the relevant updates, and then show some key results that both validate the overall pipeline and help us to quantify specific signal loss biases that we correct for (see \autoref{sec:correcting_bias}).

\needspace{2cm}\subsection{Visibility and Systematics Simulations}

Our primary method for validating the analysis pipeline presented in \autoref{sec:analysis} is via an end-to-end simulation, wherein we generate realistic visibility simulations of foregrounds, noise, and a boosted EoR-like signal that should be easily detectable given our sensitivity. This allows us to holistically evaluate the performance of the analysis and identify any unknown sources of bias. \edited{One could also approach the same problem by injecting a signal of known amplitude (larger than the real EoR) into the visibilities and analyzing that data in the same way. However, this technique requires high confidence in one's calibration solutions (injected visibilities must be ``uncalibrated'' before injection) and it is difficult to disentangle residual systematics form power spectrum biases.}

Our sky simulations consist of three unpolarized components: diffuse Galactic emission, a point source catalog, and an EoR analogue created as a Gaussian random field drawn from a known power spectrum. We present each of these here and discuss how they differ from the corresponding components in \citet{Aguirre_2022}.  For the diffuse Galactic emission, we use the Global Sky Model \citep[GSM;][]{Oliveira2008}, computed at every frequency we measure using \texttt{pygdsm}\footnote{\url{https://github.com/telegraphic/pygdsm}}. The result is a HEALPix map \citep{Gorski2005} with resolution $N_\text{side} = 512$. We smooth the map with a $1^{\circ}$ Gaussian kernel and degrade it to $N_\text{side} = 128$ to speed up our simulations, since that provides sufficient resolution to cover the angular scales HERA is sensitive to. Each HEALPix pixel is treated as a single point source at its center using the pixel area for normalization.

Our model of point source emission differs significantly from \citet{Aguirre_2022} in our handling of spatial gaps in the MWA GLEAM catalog \citep{Hurley-Walker2017} due to the Galactic plane. Since \Ha{} used observations spanning a much smaller range of LSTs than those used in this work, it was decided to slightly further restrict the range of LSTs simulated to 1.5--7~hours in order to avoid those gaps. In this work, we wanted to be able to simulate the full 24~hours of LST, so we developed a GLEAM-analogue with simulated point sources. We created 14,073,688 random synthetic sources with uniformly random positions across the sky. Their fluxes were drawn to match the GLEAM source count distribution in \citet{franzen_2019} from 0.001--87\,Jy. Their spectral indices were drawn from a Gaussian distribution with mean $-0.79$ and standard deviation 0.05, again following \citet{franzen_2019} (see also, e.g., \citealt{Offringa_2016, Carroll_2016}). The 200,000 brightest sources were simulated at their random positions, the remainder (those below 0.1975\,Jy) were treated as confusion noise. For the sake of computational expediency, these were added to the GSM at the nearest HEALPix grid point. Above 87\,Jy, we added a several real sources in their true positions. This includes GLEAM J215706-694117, GLEAM J043704+294009, GLEAM J122906+020251, GLEAM J172031-005845, and the ``A-Team" sources reported in Table 2 of \citet{Hurley-Walker2017}. We also include Fornax A, which we model as a single 750\,Jy source with a spectral index of $-0.825$ at a right ascension of 3$^\text{h}$22$^\text{m}$42$^\text{s}$ and a declination of $-37^\circ$12$'2''$ \citep{MCKinley2015}.

Finally, our boosted EoR analogue is created with the same simulator\footnote{\url{https://github.com/zacharymartinot/redshifted_gaussian_fields}} used in \citet{Aguirre_2022}. We used a higher amplitude mock EoR\footnote{We had originally intended to use a similar signal amplitude as in \citet{Aguirre_2022} so as to show distinct regions of the power spectrum dominated by foregrounds, signal, or noise. The signal used in our end-to-end simulations was accidentally created substantially higher than this, so that almost no $k$ modes are noise-dominated (see \autoref{fig:validation_final_pspec}). This has the advantage of measuring biases more precisely, but it has the disadvantage of making the blinded comparison of end-to-end simulations with and without a mock boosted EoR---a test we had intended to perform---trivially easy. \edited{While we do not expect any nonlinearities in the response of our analysis to this elevated signal level, if there were it would likely result an in overestimate of the signal loss and thus overly conservative upper limits.}} and a slightly different power spectrum slope: $P(k) \propto k^{-2.7}$. That EoR simulation was binned to the same HEALPix grid as the diffuse foregrounds, where again each pixel center is treated as a point source.

To actually simulate visibilities, we use \texttt{vis\_cpu} \citep{viscpu_memo},\footnote{\url{https://github.com/HERA-Team/vis\_cpu}} a fast visibility simulator validated against \texttt{pyuvsim} \citep{Lanman2019a},\footnote{\url{https://github.com/RadioAstronomySoftwareGroup/pyuvsim}} a reference simulator designed for accuracy \citep{Pascua2022}. The simulator used in \citet{Aguirre_2022}, \texttt{RIMEz},\footnote{\url{https://github.com/UPennEoR/RIMEz}} calculates visibilities in spherical harmonic space. \texttt{vis\_cpu} takes a much simpler approach. It calculates a per-antenna visibility factor for each point-like sky component---essentially the square root of the source flux with a phase factor that depends on frequency and antenna position---and multiplies them by a Jones matrix. These are then cross-multiplied to form visibilities and summed over sources. We use the primary beam calculated in \citet{Fagnoni2021} and interpolated in azimuth and zenith angle. We simulate the full 24~hours of LST at a five second cadence for each unique baseline and frequency observed by HERA.

Our end-to-end validation began with producing two sets of 94 nights of simulated data, one with the mock boosted EoR and one without. Both included foregrounds, noise, and systematics. For each night, we interpolated the original set of simulated visibilities using a cubic spline onto each night's 10.7 second cadence, since the LST grid of the data varied from night to night. Unlike in \citet{Aguirre_2022}, where we used only a subset of the antennas, in this work we inflated the redundant baselines to produce data files with all baselines not flagged in the real data. The final simulated data set completely matched the real data in terms of baselines, times, and frequencies. No non-redundancy due to antenna-to-antenna variation of beams or positions offsets \citep{Orosz2019,Dillon2020,Choudhuri2021} was simulated. 

We then ``uncalibrate'' the data by applying per-antenna complex bandpasses and cable reflection terms and then add noise to the visibilities, both steps performed exactly the same way as in \citet{Aguirre_2022}. Finally, our simulations have per-baseline crosstalk added to them. Again, this is done with nearly the same procedure as in the prior validation work. For each baseline, we model the crosstalk as a series of copies of the autocorrelation of each antenna in the baseline, each multiplied by a complex delay term and an amplitude that decreases exponentially with delay by a factor of 100 from the peak delay out to 2000\,ns.  In this work, the amplitudes and delays of the crosstalk peak, and how they depend on position in the array, are motivated by the new physical model of the crosstalk (see \autoref{appendix:crosstalk}) in which we attribute crosstalk to an emitter on the west side of the array.  (In our model we use $\alpha = -2.29$ and $\tau_{\rm offset} = 0$\,ns; see \autoref{eq:xtalk_model} for definitions.) The crosstalk structure is allowed to change per epoch, but not per night. The amplitudes are on the low end of what is observed in real data in order to avoid cross-talk effects on certain baselines becoming crosstalk-dominated, since the visibility simulation itself is also somewhat underpowered on long baselines relative to the real sky. We do not explicitly model any of the multipath effects hypothesized to be responsible for the breadth of the crosstalk spectrum, relying instead on that series of peaks as in \citet{Aguirre_2022}. All of the techniques for visibility corruption, along with an interface to \texttt{vis\_cpu}, are packaged into \texttt{hera\_sim}.\footnote{\url{https://github.com/HERA-Team/hera_sim}}

\subsection{Validation Results from End-to-End Simulations} \label{sec:validation:endtoend}

With our procedure for turning visibility simulations into full nights of ``uncalibrated," systematics-corrupted data matching the real data, we then apply our analysis pipeline almost exactly as described in \autoref{sec:analysis}. We perform redundant-baseline calibration and absolute calibration, then flag, then smooth our gain solutions. We next bin together individual epochs and perform inpainting, cable reflection calibration, and crosstalk subtraction. After binning together the four epochs, we form pseudo-Stokes visibilities, time-average, and estimate power spectra. We run the entire end-to-end pipeline twice---once without the mock EoR and once with it---including the various power spectrum cuts described in \autoref{sec:pspec_updates}. In order to faithfully validate the analysis pipeline, the same software packages---\texttt{pyuvdata}\footnote{\url{https://github.com/RadioAstronomySoftwareGroup/pyuvdata}} \citep{Hazelton2017}, \texttt{hera\_cal},\footnote{\url{https://github.com/HERA-Team/hera_cal/}} \texttt{hera\_qm},\footnote{\url{https://github.com/HERA-Team/hera_qm}} and \texttt{hera\_pspec}\footnote{\url{https://github.com/HERA-Team/hera_pspec}}---with the same \texttt{git} hashes that were used to run the end-to-end validation.\footnote{With one exception: the first round of LST-binning was performed with a newer version of \texttt{hera\_cal} because the version run on data assumes that flagged baselines are left in the data files and just flagged. Our end-to-end validation run skipped simulating these antennas, which required an update to the original code.}

The one major step that was performed differently was RFI flagging. Following \citet{Aguirre_2022}, we do not inject RFI and then attempt to detect it. Rather, we simply cross-apply the real data's flags to the simulated data at the same step in the pipeline. Since the times, frequencies, and baselines match perfectly, this step is very straightforward. Likewise, we take the final flagging mask generated from a manual inspection of the data (see \autoref{sec:rfi}) and apply it to each epoch of simulated data at precisely the same point in the real pipeline that those flags are applied to real data.

\begin{figure}
    \centering
    \includegraphics[width=.48\textwidth]{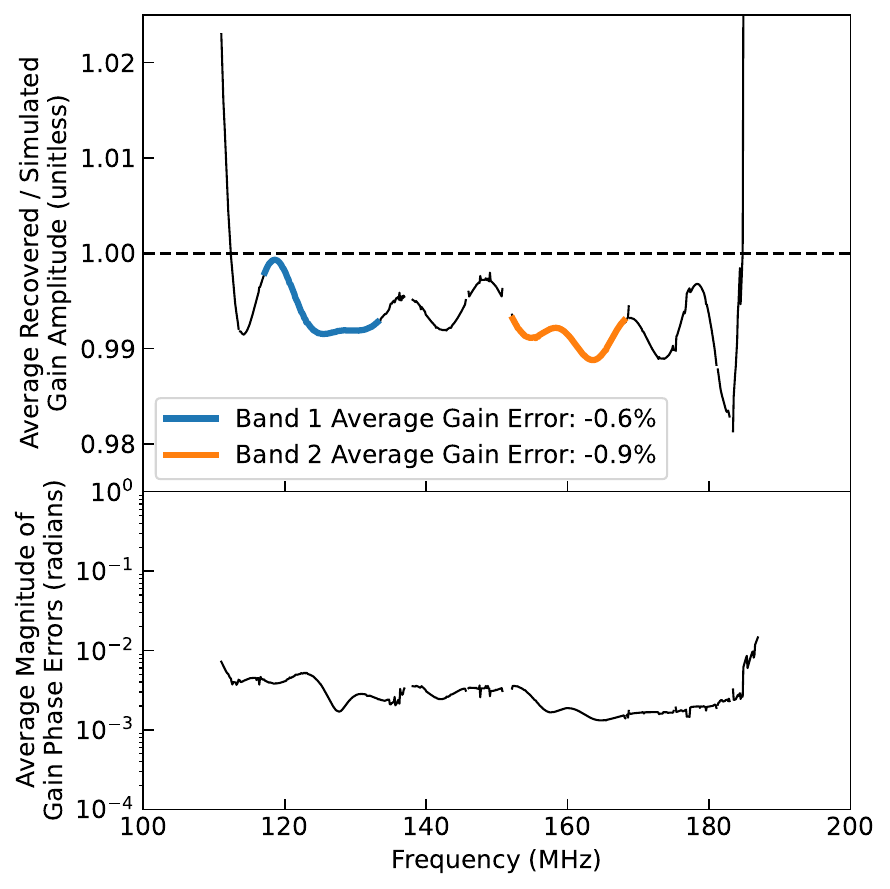}
    \caption{In our end-to-end tests with simulated data, we can compare the known input gains to the derived gains after redundant-baseline calibration, absolute calibration, and calibration smoothing. Here we show the average gain errors after averaging over antennas, times, and nights for Epoch 1, the longest epoch. In the bottom panel, we show the error in our gain phases; these are comparable to those found in the validation of our \Ha{} limits in \citet{Aguirre_2022}. In the top panel we show our gain amplitude errors. These are substantially smaller than those found in \citet{Aguirre_2022} due to our new algorithm for absolute amplitude calibration which is not biased when visibility SNR is low (see \autoref{sec:abscal} for details). In the bands of interest, our gains are correct to within 1\%. While we do have some evidence that they are biased slightly low, which would lead to a power spectrum that is slightly too high, we choose conservatively to not to adjust the power spectrum to compensate (see \autoref{sec:correcting_bias}).}
    \label{fig:validation_gain_errors}
\end{figure}

Before we discuss the final power spectra, we can now evaluate whether the fix offered in \autoref{sec:abscal} eliminates the bias in absolute calibration discovered in \citet{Aguirre_2022} and corrected for in \Ha{}. In \autoref{fig:validation_gain_errors}, we compare the calibration solutions produced by the pipeline for Epoch 1 (the longest epoch) with the known simulated gains, averaging over antennas, times, and nights.
By comparison to Figure 11 of \citet{Aguirre_2022}, we can see that the absolute calibration bias is largely eliminated and that the phases are still well-recovered across the band. The gain errors are largest at the band edges, due to edge effects of the low-pass filter used in gain smoothing.

Interestingly, we actually now see a slight negative bias for most frequencies (0.6\% for Band 1, 0.9\% for Band 2), which could lead to overestimating power spectra and error bars by a few percent, since gain errors impact power spectra quartically. One possible origin of the effect is a rare failure mode of the absolute calibration bias fix described in \autoref{sec:abscal}. When fitting for a single overall amplitude, solutions are biased but quite stable in time. When fitting for a complex number and then taking the absolute value of that, we have seen rare instances where the data and reference visibilities are so far apart that the gains are driven to 0 in order to minimize $\chi^2$. In real data, these sorts of collapses are easily identified as discontinuities and flagged as RFI. This justifies our conservative choice in \autoref{sec:correcting_bias} to not correct for any remaining absolute calibration bias since we do not expect this effect to be as large on our final upper limits. However, in our end-to-end pipeline validation, we flag precisely where we did in the real data, ignoring any potential new discontinuities. These artificially low gains can thus be the consequence of a rare calibration failure getting spread out and diluted by gain smoothing. 

\begin{figure*}
    \centering
    \includegraphics[width=\textwidth]{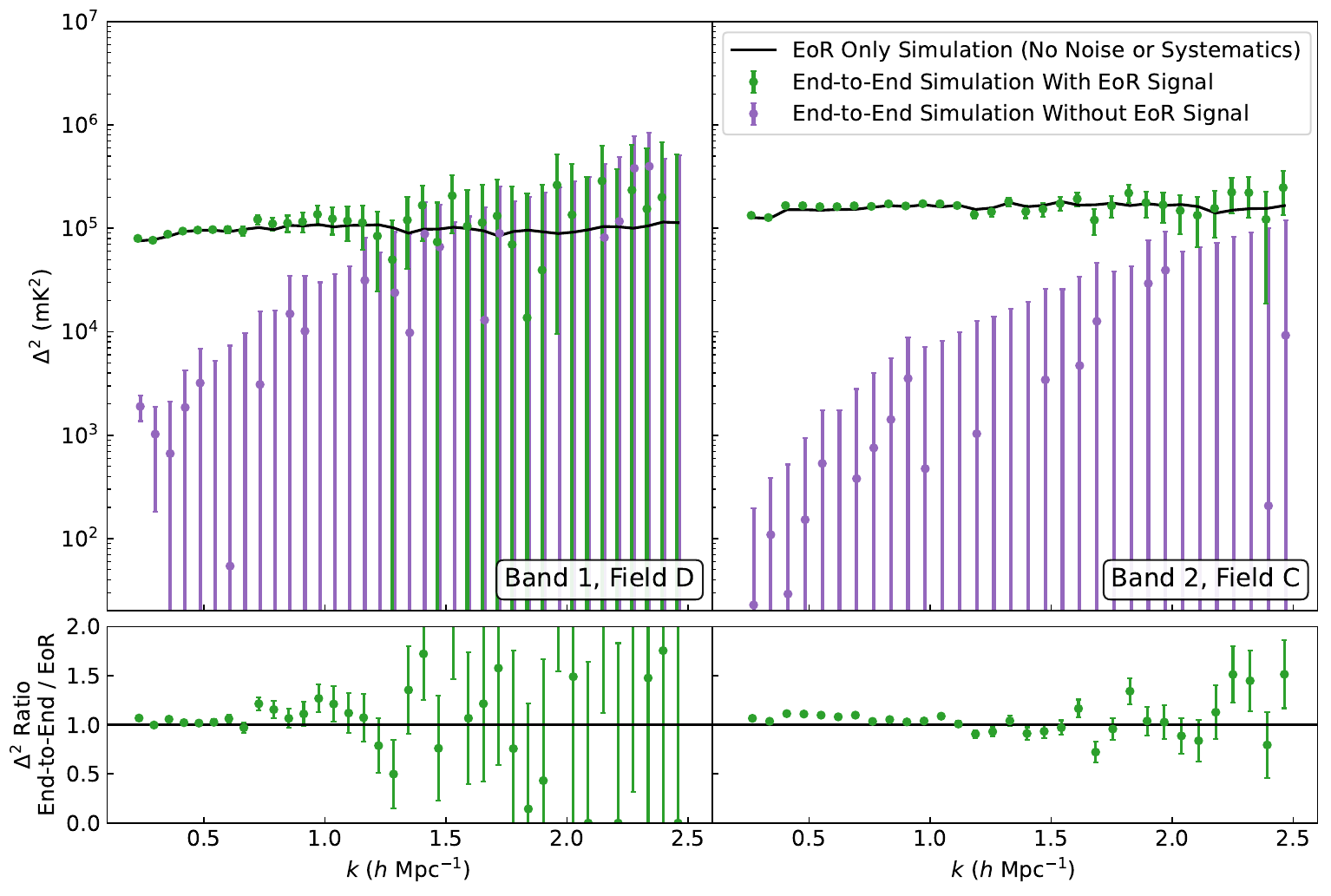}
    \caption{Here we show the final result of our end-to-end test of our data reduction pipeline on simulated data. We simulated visibilities both with (green) and without (purple) a boosted EoR analogue Gaussian random field, along with foregrounds, noise, and instrumental systematics, for all 94 nights. (For visual clarity, the points along the $k$ axis were slightly off set between green and purple.) This was then processed with almost all the same code as was used to process the real data in order to thoroughly test our analysis chain. As the bottom panels showing the ratio of our end-to-end-test (the green points) to the particular boosted mock EoR realization (black solid line), our results match the EoR realization quite well. There is no evidence for additional signal loss not accounted for in \autoref{sec:correcting_bias}. While this calculation was performed for all bands and fields, we show here just the two fields where our lowest limits are derived at each band (see \autoref{sec:upper_limits}).}
    \label{fig:validation_final_pspec}
\end{figure*}

We turn now to the final result of our end-to-end test: spherically-averaged power spectra for both bands and all five fields, with and without our mock boosted EoR. In \autoref{fig:validation_final_pspec}, we show one field for each band (corresponding to our lowest limits, see \autoref{sec:upper_limits}).
The power spectra include signal loss corrections for crosstalk subtraction and coherent time averaging, as described in \autoref{sec:correcting_bias}, but no correction for non-redundancy as none is included in the simulation. For comparison, we also show the results of a simulation run with only the mock EoR with no foregrounds, noise, systematics, or flags (black line). It is interpolated directly onto the grid of the final LST-binned data set, time-averaged, converted to pseudo-Stokes I, and formed into power spectra. 

In general, we find the results of our end-to-end test in good agreement with the mock EoR-only power spectrum, as the bottom panels of \autoref{fig:validation_final_pspec} show. \edited{Though the error bars on the power spectrum with EoR signal may be underestimates (since sample variance is not accounted for in our real pipeline), there} is no evidence for an additional, unaccounted-for contribution to signal loss that might lead us to report artificially low upper limits. If anything, we see some evidence that our results are biased slightly high. This is likely due to a number of factors; the bias high due to rare absolute calibration failures, a possible overestimation of the impact of signal loss, and the effect of flagging and inpainting. When the same pipeline is run without the EoR, the results are consistent with noise at almost all $k$---a result of the somewhat aggressive cuts we discuss in \autoref{sec:xtalk_updates} and \autoref{sec:kcuts}---though we unsurprisingly do see some residual foregrounds at very low $k$, especially in Band 1. This is consistent with what we see in the real data in most fields (see \autoref{sec:sphere_limits}).

\needspace{2cm}\subsection{Noise-Free Tests for Quantifying Potential Signal Loss} \label{sec:validation:noisefree}

\begin{figure}
    \centering
    \includegraphics[width=.48\textwidth]{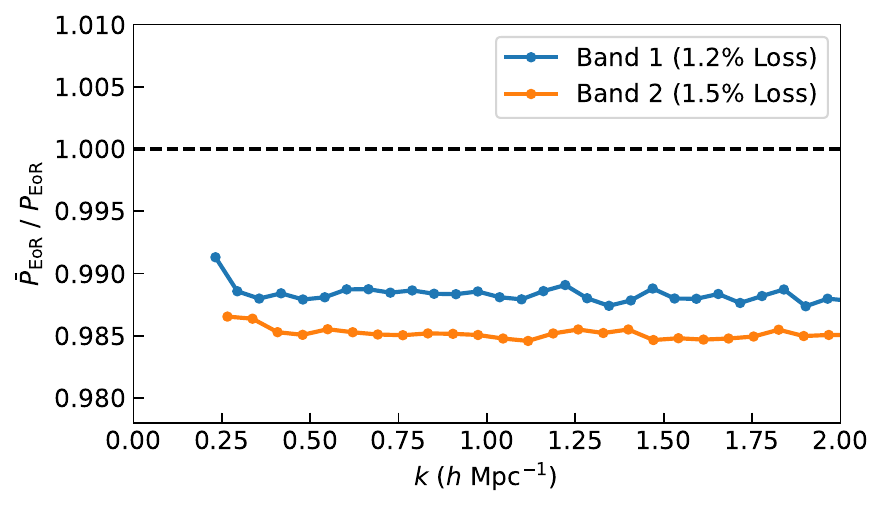}
    \caption{Here we repeat the test of decoherence from time-averaging shown Figure 17 of \citet{Aguirre_2022}. We compare power spectra computed from EoR-only visibility simulations at a 21.4\,s cadence---which is what we obtain after LST-binning---to the same data averaged to a 214\,s cadence before forming power spectra. After averaging over all fields and baselines, we find a result in agreement with \citet{Aguirre_2022}. In \Ha{}, the two bands are averaged together and the effect is taken to be a 1\% correction; here we keep the two bands separate to arrive at our signal loss correction factors (see \autoref{sec:correcting_bias} and \autoref{tab:signal_loss}).}
    \label{fig:validation_tavg_signal_loss}
\end{figure}

Separately from our end-to-end simulations, we also repeat two tests of signal loss performed in \citet{Aguirre_2022} that were used in \Ha{} to correct the final power spectrum measurements (see \autoref{sec:correcting_bias}). The first is the effect of coherently time-averaging LST-binned visibilities from 21.4\,s to 214\,s before cross-multiplying interleaved visibilities to form power spectra. This is done by interpolating mock EoR-only visibilities onto the 21.4\,s grid of the final LST-binned data set, forming pseudo-Stokes I, and then either forming power spectra directly or forming power spectra after coherently averaging to a 214\,s cadence with rephasing as described in \autoref{sec:pipeline_overview}. If we take the result with 21.4\,s integrations to be loss-free, which it should be to a good approximation, then we find a 1.2\% signal loss for Band 1 and a 1.5\% signal loss for Band 2 (see \autoref{fig:validation_tavg_signal_loss}).

It is not surprising that the signal loss should be slightly higher for Band 2, since the primary beam is smaller at higher frequencies and the rephasing of visibilities before averaging cannot account for the changing beam-weighted sky. \edited{Likewise, if we were to open up additional degrees of freedom in our signal loss correction, we would likely find that this decoherence depends on baseline length and orientation---baselines that fringe more quickly should see more loss here. Since we are using the same weighting of baselines here as in analysis of the real data, the overall loss estimate should be correct to first order. Therefore, because the overall signal loss is quite small, it is not necessary to further complicated the signal loss correction by making it baseline dependent.} Regardless, these values are consistent with the 1\% loss figure used in \Ha{} and calculated in \citet{Aguirre_2022}, which did not separate out the two bands. 

\begin{figure}
    \centering
    \includegraphics[width=.5\textwidth]{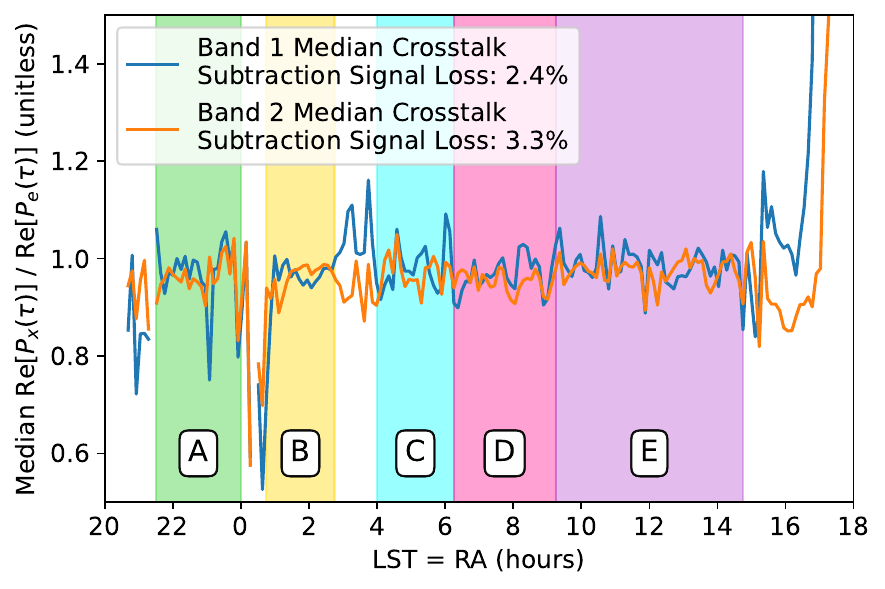}
    \caption{Our final test of potentially biasing signal loss examines the effect of our crosstalk subtraction algorithm on the EoR signal. Here we compare simulations with just our boosted EoR model---the numerator in \autoref{fig:validation_tavg_signal_loss}---to noise-free, per-epoch simulations with EoR, foregrounds, and crosstalk systematics. Crosstalk subtraction is performed on a per-epoch basis, after which epochs are combined, formed into power spectra, and then averaged as usual to get spherical power spectra as a function of time. Taking the median over delays less than 4000\,ns and LSTs in our five fields, we estimate 2.4\% loss in Band 1 and 3.3\% loss in Band 2, which we report in \autoref{tab:signal_loss}.}
    \label{fig:validation_xtalk_signal loss}
\end{figure}

The second specialized test is to examine the impact of crosstalk subtraction (described above in \autoref{sec:pipeline_overview} and \autoref{sec:xtalk_updates}). The crosstalk subtraction algorithm devised in \citet{Kern2019}, demonstrated in \citet{Kern2020a}, and employed for \Ha{} removes power near fringe-rate zero. The maximum extent of this removal in fringe-rate space (\autoref{eq:xtalk_frate}) is designed to keep signal loss at the $\sim$1\% level. To measure this effect, we interpolate to get one data set per epoch with foregrounds, mock EoR, and crosstalk injected, but no noise or calibration errors. To this data set we apply our final flagging mask, inpaint, subtract cross-talk, LST-bin the four epochs together, form pseudo-Stokes visibilities, coherently time-average, and then form power spectra. We spherically average those power spectra over baselines using the same noise-based weights as were applied to the data. 
In \autoref{fig:validation_xtalk_signal loss}, we compare those per-time-step power spectra to mock EoR-only power spectra with the same averaging performed. We take a median over delay up to 4000\,ns, the highest delay in the crosstalk subtraction SVD, to produce a single bias estimate per LST and per band. As \Ha{} argues, we expect the crosstalk subtraction bias to scale-independent.

The final result is a bit difficult to interpret. Clearly the result is LST-dependent; just as in \citet{Aguirre_2022}, we see higher levels of signal loss near gaps in the data---this was one of the reasons we avoided the range from 0--0.75~hours of LST. We also see some evidence under-subtraction around Fornax A (in Band 1) and the Galactic center, though the latter is expected due the weighting in the SVD. To avoid the effect of outliers in both directions, we estimate our per-band signal loss by taking the median over LSTs in the five fields. This yields 2.4\% signal loss in Band 1 and 3.3\% loss in Band 2, which is basically consistent with the 1\% and 3\% used in \Ha{}.


\section{Upper Limits on the 21\,cm Power Spectrum} \label{sec:upper_limits}

Having surveyed our technique for reducing 94 nights of visibilities to our final power spectra, and having validated that technique and quantified the signal loss biases we must correct for, we are finally in a position to present our upper limits on the 21\,cm EoR power spectrum. 

\needspace{2cm}\subsection{Cylindrically-Averaged Power Spectra} \label{sec:cylindrical}

We begin with cylindrically-averaged power spectra. These are first averaged over all cross-baseline pairs within each redundant baseline group (excluding those with high residual crosstalk, see \autoref{sec:xtalk_updates}). Next they are averaged incoherently in time over each field. Finally, they are incoherently averaged in $|\mathbf{u}|$, combining together baselines with the same length. All averages use $P_N$ (\autoref{eq:PN}) to perform inverse variance weighting; $P_N$ and $\hat{P}_{SN}$ (\autoref{eq:PSN}) errors are propagated during each step. This produces power spectra in $\tau$ and $u$, which is equivalent in the delay approximation to $k_\|$ and $k_\perp$ \citep{Parsons2012b}. These cylindrical power spectra are the most sensitive data products that still keep $k_\|$ and $k_\perp$ separate, which is useful because different scales along those two axes are measured in a fundamentally different way instrumentally and correspond to different ways of measuring distance cosmologically. While we are not yet sensitive enough to constrain the cosmological signal's dependence on line-of-sight velocity effects, cylindrical power spectra are still useful for evaluating how foregrounds and other systematics appear upon deep integration.  

\begin{figure*}
    \centering
    \includegraphics[width=\textwidth]{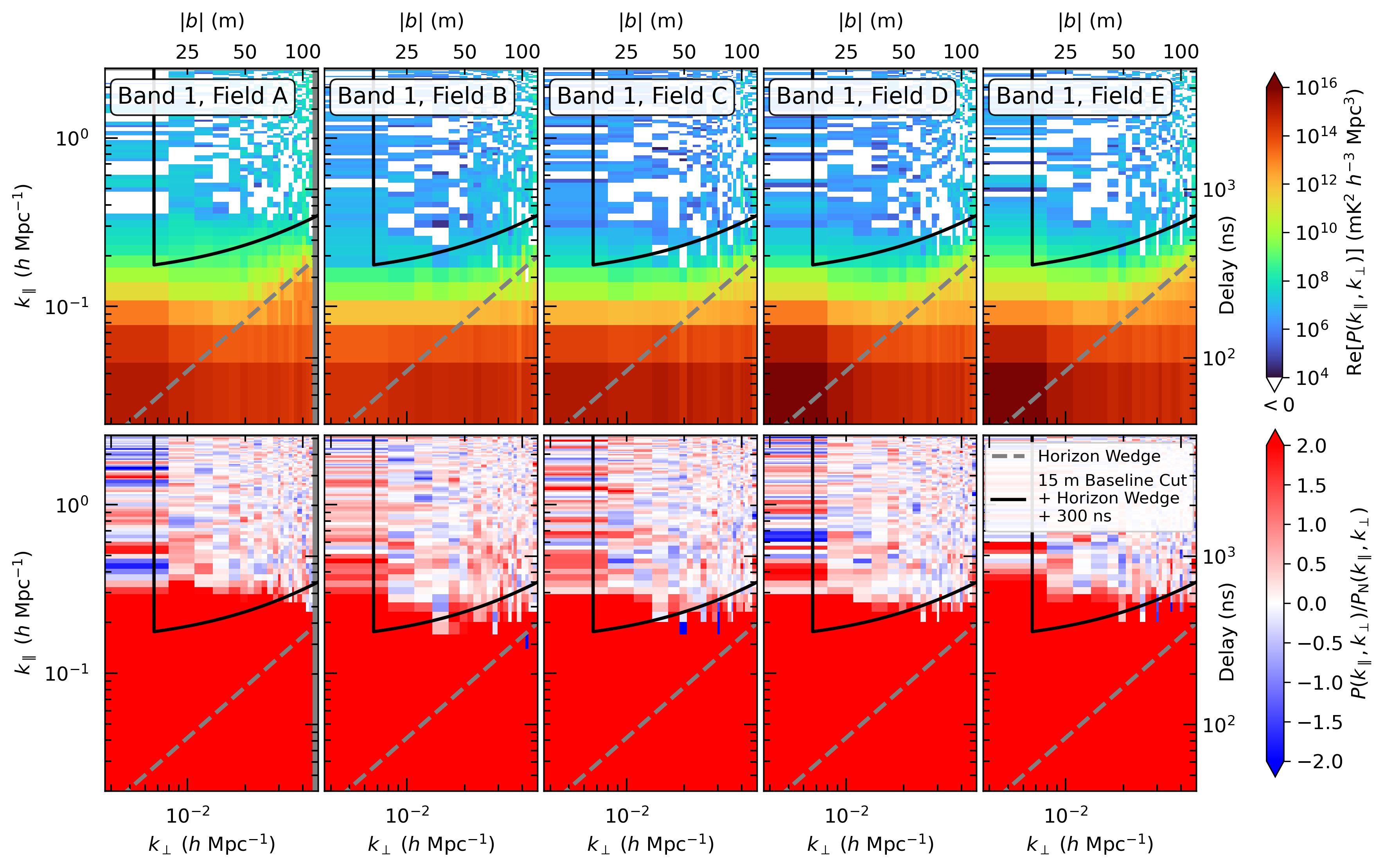}
    \caption{Here we show our cylindrically-binned power spectra for all five fields for Band 1 ($z = 10.4$). In the top row, we show the real part of $P(k)$; in the bottom row, we show the same power spectra normalized by the power spectrum of expected noise. Since noise-dominated regions of this space are equally likely to be positive or negative, we can see that most of the EoR window is noise-dominated. These power spectra are formed by averaging together all cross-baseline pairs within a redundant baseline group and then by incoherently averaging in time in each field. Both averages are performed with inverse noise variance weighting. We convert from baseline and delay to $k_\perp$ and $k_\|$ respectively using the delay spectrum approximation \citep{Parsons2012b}.  Note that bins that fall below the horizon wedge (\autoref{eq:wedge}) plus a 300\,ns buffer, along with baselines with projected east-west distances less than 15\,m, are included here to help illustrate systematics but are excluded from all spherically binned power spectra.}
    \label{fig:cyl_pk_1}
\end{figure*}

\begin{figure*}
    \centering
    \includegraphics[width=\textwidth]{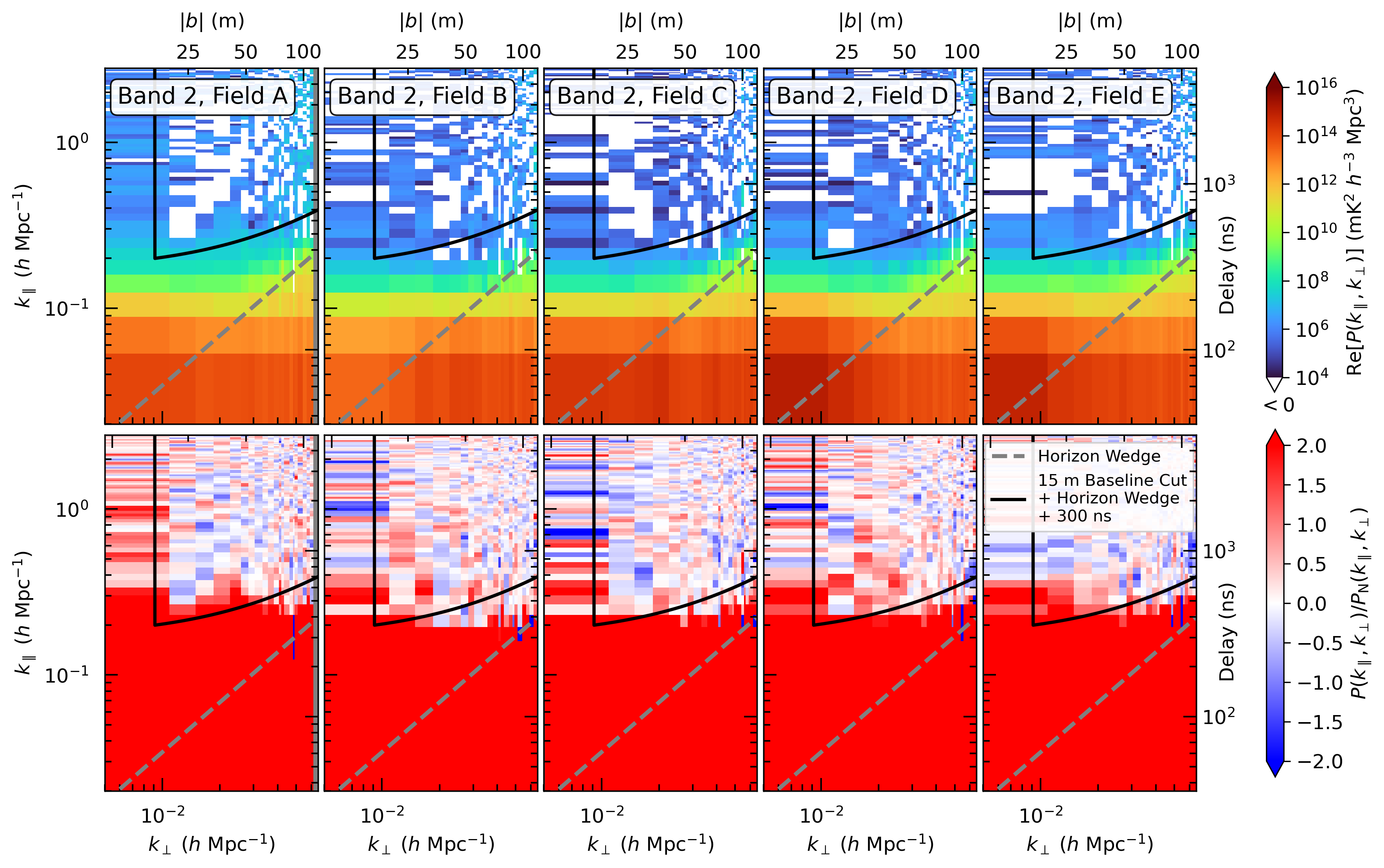}
    \caption{The same plot of cylindrically-binned $P(k)$ and SNRs as in \autoref{fig:cyl_pk_1}, but for Band 2 ($z = 7.9$).}
    \label{fig:cyl_pk_2}
\end{figure*}

We show our cylindrically-averaged power spectra for each of the five fields in \autoref{fig:cyl_pk_1} for Band 1 and \autoref{fig:cyl_pk_2} for Band 2. 
In the top row of each figure, we show the real part of the power spectrum.\footnote{The power spectrum is complex because it involves a cross-multiplication of two independent times. We expect the imaginary part to be noise-dominated in the EoR window and dominated in the wedge by signal-noise cross terms \citep{Tan2021}. This is exactly what we see, both in cylindrical and spherical power spectra, but we do not show them here for the sake of brevity. Their consistency with noise, however, is shown in \autoref{tab:pvalues}.} We also show the power spectrum SNR, $P / P_N$ in the bottom row. In regions that are noise-dominated, we expect roughly half of the power spectrum bins to be negative (white) and we expect the SNR to have mean 0 and standard deviation 1. 

What we see is fairly consistent with that expectation across most of the EoR window. However, the foreground-dominated region clearly extends well beyond the horizon wedge (gray dashed line). This justifies the need for a buffer beyond the horizon wedge and our choice specifically to expand it from 200\,ns as in \Ha{} to 300\,ns (black solid line). As we discussed in \autoref{sec:kcuts}, our choice of the buffer was set by examining these SNR plots and trying to balance the amount of foreground leakage into the window at low $k_\perp$ with the amount of noise-dominated modes excised at high $k_\perp$.

Note that a number of data points that we excise from our spherically-averaged power spectra are still shown in \autoref{fig:cyl_pk_1} and \autoref{fig:cyl_pk_2}. We have not yet removed the modes under the horizon wedge and buffer, nor have we removed baselines with projected east-west distance less than 15\,m (though baselines with a projected east-west distance less than 14\,m have already been removed because crosstalk subtraction could not be performed on them without substantial signal loss, see \autoref{sec:pipeline_overview}).

\subsection{Spherically-Averaged Power Spectra and the Deepest Upper Limits} \label{sec:sphere_limits}

\begin{figure*}
    \centering
    \includegraphics[width=\textwidth]{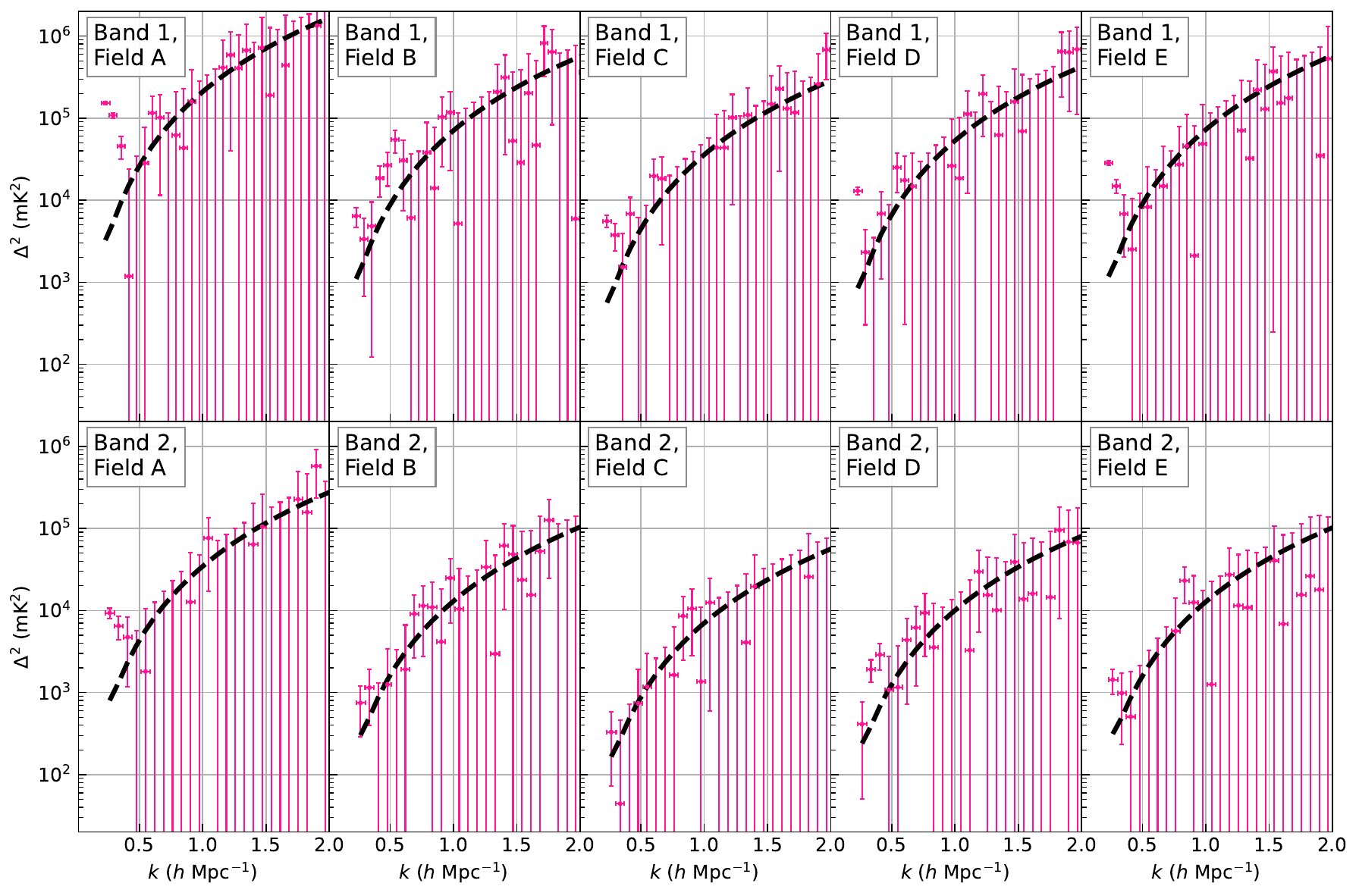}
    \caption{Our final limits on the 21\,cm brightness temperature power spectra $\Delta^2(k)$, for all five fields and both Band 1 ($z = 10.4$) and Band 2 ($z = 7.9$). The points are the real parts of the bandpowers. The vertical error bars are 2$\sigma$ and the black dashed lines denote 1$\sigma$ errors. The horizontal error bars denote the range from the 16th to the 84th percentile of the window functions. Upper limits derived from negative measurements are conservatively set at $2\sigma$; all other limits are set at $\Delta^2(k) + 2\sigma$. Any points whose error bars pass through the y-axis are consistent with noise at 2$\sigma$. Our lowest 2$\sigma$ upper limits are \highzlimit{} at \highzk{} in Band 1, Field D and \lowzlimit{} at \lowzk{} in Band 2, Field C. The precise values of the power spectrum, error bars, and upper limits at low $k$ are reproduced in \autoref{tab:band1_limits} and \autoref{tab:band2_limits}. Following \Ha{}, we leave the incoherent averaging of different fields together for future work.}
    \label{fig:all_limits}
\end{figure*}

We now turn to our spherically-averaged power spectrum measurements and upper limits. These are produced from averaging together all baselines incoherently after excising modes below the horizon wedge (\autoref{eq:wedge}) plus a 300\,ns buffer and all baselines with a projected east-west length less than 15\,m (which includes the single-unit east-west spacing; see \autoref{fig:layout}). The result, for all bands and fields, is shown as a ``dimensionless" power spectrum (\autoref{eq:deltasq}) in \autoref{fig:all_limits}.\footnote{Following \Ha{}, we do not attempt to further combine fields, though we will use the results from all fields jointly in our astrophysical constraints in \autoref{sec:theory}.}

We estimate vertical errors by propagating $\hat{P}_{SN}(k)$ (\autoref{eq:PSN}) and show the measurements $\pm2\sigma$ after converting to $\Delta^2$ using \autoref{eq:deltasq}. When measurements are negative, we adopt the strong prior that the true power spectrum must be zero or positive and plot upper limits at precisely $2\sigma$. The $1\sigma$ error level derived from $\hat{P}_{SN}(k)$ is shown as a black dashed line. Our absolute lowest $2\sigma$ upper limits on the power spectrum are $\Delta^2(k=\text{\highzk{}}) \leq \text{\highzlimit{}}$ at $z=10.4$ (Band 1, Field D) and $\Delta^2(k=\text{\lowzk{}}) \leq \text{\lowzlimit{}}$ at $z=7.9$ (Band 2, Field C). These limits are 2.6 and 2.1 times deeper, respectively, than those presented in \Ha{}. We report the measured power spectra, $1\sigma$ error bars, and and $2\sigma$ upper limits results at \edited{most values of $k$---including low $k$,} which is where most of our astrophysical constraining power comes from---in \autoref{tab:band1_limits} for Band 1 and \autoref{tab:band2_limits} for Band 2.\footnote{\edited{Additionally, upper limits from this work and many other 21\,cm experiments are available at \url{https://github.com/HERA-Team/eor_limits}.}}

\begin{deluxetable*}{c | c c c c | c c c c | c c c c }[t!]
    \tabletypesize{\footnotesize}
    \tablecaption{Band 1 ($z=10.4$) power spectra, errors, and upper limits from \autoref{fig:all_limits}. The upper limit is taken to be $2\sigma$ above the power spectrum measurement or above 0, whichever is greater. The lowest upper limit is in bold. \label{tab:band1_limits}}
    
    \tablehead{\colhead{} & \colhead{$k$} & \colhead{$\Delta^2$} & \colhead{$1\sigma$} & \colhead {$\Delta^2_{\rm UL}$} & \colhead{$k$} & \colhead{$\Delta^2$} & \colhead{$1\sigma$} & \colhead {$\Delta^2_{\rm UL}$} & \colhead{$k$} & \colhead{$\Delta^2$} & \colhead{$1\sigma$} & \colhead {$\Delta^2_{\rm UL}$} \\
    \colhead {}  & \colhead {$(h\ {\rm Mpc}^{-1})$} & \colhead {$({\rm mK})^2$} & \colhead {$({\rm mK})^2$} & \colhead {$({\rm mK})^2$} & \colhead {$(h\ {\rm Mpc}^{-1})$} & \colhead {$({\rm mK})^2$} & \colhead {$({\rm mK})^2$} & \colhead {$({\rm mK})^2$} & \colhead {$(h\ {\rm Mpc}^{-1})$} & \colhead {$({\rm mK})^2$} & \colhead {$({\rm mK})^2$} & \colhead {$({\rm mK})^2$} }
    
    \startdata
	\multirow{8}{*}{\rotatebox[origin=c]{90}{\textbf{Field A}}}
	& 0.23 & 153,691 & 2,933 & 159,556 & 0.73 & -32,771 & 58,377 & 116,753 & 1.22 & 588,465 & 274,340 & 1,137,146 \\
	& 0.29 & 107,667 & 4,271 & 116,208 & 0.79 & 61,900 & 74,547 & 210,994 & 1.28 & 406,694 & 318,088 & 1,042,869 \\
	& 0.36 & 45,688 & 7,105 & 59,898 & 0.85 & 43,307 & 93,159 & 229,625 & 1.35 & 669,812 & 366,722 & 1,403,255 \\
	& 0.42 & 1,182 & 11,336 & 23,854 & 0.91 & 158,855 & 115,453 & 389,760 & 1.41 & -22,002 & 418,809 & 837,617 \\
	& 0.48 & -6,756 & 17,231 & 34,462 & 0.97 & -2,569 & 140,221 & 280,442 & 1.47 & 721,387 & 476,106 & 1,673,599 \\
	& 0.54 & 28,245 & 24,666 & 77,577 & 1.04 & -25,200 & 167,921 & 335,842 & 1.53 & 190,383 & 537,247 & 1,264,877 \\
	& 0.60 & 115,543 & 33,997 & 183,537 & 1.10 & -206,058 & 199,136 & 398,273 & 1.59 & -63,312 & 604,314 & 1,208,628 \\
	& 0.66 & 101,535 & 45,017 & 191,569 & 1.16 & 412,511 & 235,335 & 883,180 & 1.65 & 440,694 & 676,991 & 1,794,676 \\
	\midrule
	\multirow{8}{*}{\rotatebox[origin=c]{90}{\textbf{Field B}}}
	& 0.23 & 6,414 & 878 & 8,169 & 0.73 & -16,887 & 19,741 & 39,482 & 1.22 & -3,270 & 90,946 & 181,893 \\
	& 0.29 & 3,352 & 1,343 & 6,038 & 0.79 & 38,162 & 25,246 & 88,654 & 1.28 & -23,054 & 105,349 & 210,698 \\
	& 0.36 & 4,815 & 2,346 & 9,506 & 0.85 & 14,007 & 31,608 & 77,223 & 1.35 & 208,370 & 121,329 & 451,027 \\
	& 0.42 & 18,593 & 3,852 & 26,297 & 0.91 & 103,161 & 38,899 & 180,959 & 1.41 & 312,314 & 138,131 & 588,576 \\
	& 0.48 & 26,651 & 5,896 & 38,443 & 0.97 & 117,071 & 46,962 & 210,995 & 1.47 & 52,833 & 157,295 & 367,424 \\
	& 0.54 & 54,508 & 8,475 & 71,457 & 1.04 & 5,188 & 55,544 & 116,277 & 1.53 & 28,767 & 178,728 & 386,224 \\
	& 0.60 & 30,557 & 11,578 & 53,714 & 1.10 & -21,580 & 66,145 & 132,291 & 1.59 & 202,144 & 200,619 & 603,381 \\
	& 0.66 & 6,089 & 15,308 & 36,705 & 1.16 & -38,769 & 77,809 & 155,618 & 1.65 & 46,875 & 225,195 & 497,265 \\
	\midrule
	\multirow{8}{*}{\rotatebox[origin=c]{90}{\textbf{Field C}}}
	& 0.23 & 5,565 & 472 & 6,509 & 0.73 & -5,911 & 10,003 & 20,006 & 1.22 & 101,849 & 46,514 & 194,876 \\
	& 0.29 & 3,763 & 687 & 5,137 & 0.79 & -1,040 & 12,731 & 25,461 & 1.28 & -42,117 & 53,624 & 107,247 \\
	& 0.36 & 1,525 & 1,185 & 3,895 & 0.85 & -555 & 15,963 & 31,926 & 1.35 & 109,485 & 62,018 & 233,520 \\
	& 0.42 & 6,806 & 1,993 & 10,793 & 0.91 & -35,281 & 19,656 & 39,312 & 1.41 & -59,321 & 70,760 & 141,521 \\
	& 0.48 & -6,007 & 3,067 & 6,134 & 0.97 & -51,204 & 23,825 & 47,651 & 1.47 & -42,848 & 80,462 & 160,924 \\
	& 0.54 & -7,550 & 4,320 & 8,641 & 1.04 & -27,920 & 28,398 & 56,795 & 1.53 & 148,253 & 90,978 & 330,208 \\
	& 0.60 & 19,718 & 5,898 & 31,514 & 1.10 & 43,689 & 33,761 & 111,212 & 1.59 & 227,669 & 102,654 & 432,978 \\
	& 0.66 & 18,353 & 7,755 & 33,864 & 1.16 & 43,764 & 39,661 & 123,085 & 1.65 & 130,546 & 114,700 & 359,945 \\
	\midrule
	\multirow{8}{*}{\rotatebox[origin=c]{90}{\textbf{Field D}}}
	& 0.23 & 13,006 & 677 & 14,360 & 0.73 & -11,090 & 14,727 & 29,454 & 1.22 & 197,836 & 68,914 & 335,664 \\
	& 0.29 & 2,325 & 1,011 & 4,347 & 0.79 & -5,471 & 18,791 & 37,582 & 1.28 & -45,122 & 79,627 & 159,254 \\
	& 0.36 & -207 & 1,748 & \textbf{3,496} & 0.85 & -29,815 & 23,439 & 46,878 & 1.35 & 62,242 & 91,821 & 245,885 \\
	& 0.42 & 6,866 & 2,888 & 12,643 & 0.91 & -21,743 & 29,005 & 58,010 & 1.41 & -95,510 & 105,107 & 210,215 \\
	& 0.48 & -1,255 & 4,423 & 8,846 & 0.97 & 26,219 & 35,219 & 96,657 & 1.47 & 158,372 & 119,620 & 397,612 \\
	& 0.54 & 25,094 & 6,301 & 37,696 & 1.04 & 18,461 & 42,035 & 102,530 & 1.53 & 69,220 & 134,761 & 338,742 \\
	& 0.60 & 17,496 & 8,595 & 34,687 & 1.10 & 112,195 & 50,065 & 212,326 & 1.59 & -3,969 & 151,666 & 303,332 \\
	& 0.66 & 14,751 & 11,370 & 37,490 & 1.16 & -46,370 & 58,914 & 117,827 & 1.65 & -256,395 & 169,976 & 339,953 \\
	\midrule
	\multirow{8}{*}{\rotatebox[origin=c]{90}{\textbf{Field E}}}
	& 0.23 & 28,531 & 957 & 30,444 & 0.73 & -29,956 & 20,066 & 40,133 & 1.22 & -4,695 & 94,165 & 188,330 \\
	& 0.29 & 14,846 & 1,394 & 17,634 & 0.79 & 27,200 & 25,575 & 78,350 & 1.28 & 70,547 & 109,087 & 288,722 \\
	& 0.36 & 6,793 & 2,382 & 11,557 & 0.85 & 45,734 & 32,039 & 109,812 & 1.35 & 32,410 & 125,598 & 283,606 \\
	& 0.42 & 2,510 & 3,946 & 10,401 & 0.91 & 2,109 & 39,559 & 81,228 & 1.41 & 220,926 & 143,720 & 508,367 \\
	& 0.48 & -2,082 & 6,016 & 12,031 & 0.97 & 48,558 & 48,118 & 144,795 & 1.47 & 129,091 & 163,068 & 455,227 \\
	& 0.54 & 8,265 & 8,530 & 25,324 & 1.04 & -99,003 & 57,725 & 115,450 & 1.53 & 370,465 & 185,109 & 740,683 \\
	& 0.60 & -21,997 & 11,634 & 23,267 & 1.10 & -58,513 & 68,608 & 137,217 & 1.59 & 153,201 & 207,953 & 569,107 \\
	& 0.66 & 14,858 & 15,432 & 45,722 & 1.16 & -78,339 & 80,512 & 161,025 & 1.65 & 175,314 & 232,387 & 640,089 \\
    \bottomrule
    \enddata
\end{deluxetable*}

\begin{deluxetable*}{c | c c c c | c c c c | c c c c }[t!]
    \tabletypesize{\footnotesize}
    \tablecaption{Band 2 ($z=7.9$) power spectra, errors, and upper limits from \autoref{fig:all_limits}. The upper limit is taken to be $2\sigma$ above the power spectrum measurement or above 0, whichever is greater.  The lowest upper limit is in bold. \label{tab:band2_limits}}
    
    \tablehead{\colhead{} & \colhead{$k$} & \colhead{$\Delta^2$} & \colhead{$1\sigma$} & \colhead {$\Delta^2_{\rm UL}$} & \colhead{$k$} & \colhead{$\Delta^2$} & \colhead{$1\sigma$} & \colhead {$\Delta^2_{\rm UL}$} & \colhead{$k$} & \colhead{$\Delta^2$} & \colhead{$1\sigma$} & \colhead {$\Delta^2_{\rm UL}$} \\
    \colhead {}  & \colhead {$(h\ {\rm Mpc}^{-1})$} & \colhead {$({\rm mK})^2$} & \colhead {$({\rm mK})^2$} & \colhead {$({\rm mK})^2$} & \colhead {$(h\ {\rm Mpc}^{-1})$} & \colhead {$({\rm mK})^2$} & \colhead {$({\rm mK})^2$} & \colhead {$({\rm mK})^2$} & \colhead {$(h\ {\rm Mpc}^{-1})$} & \colhead {$({\rm mK})^2$} & \colhead {$({\rm mK})^2$} & \colhead {$({\rm mK})^2$} }
    
    \startdata
	\multirow{8}{*}{\rotatebox[origin=c]{90}{\textbf{Field A}}}
	& 0.27 & 9,331 & 649 & 10,628 & 0.83 & -14,403 & 14,954 & 29,908 & 1.40 & 63,705 & 68,760 & 201,225 \\
	& 0.34 & 6,479 & 1,027 & 8,532 & 0.90 & 12,644 & 18,997 & 50,637 & 1.47 & 104,906 & 79,266 & 263,438 \\
	& 0.41 & 4,731 & 1,774 & 8,279 & 0.97 & -7,283 & 23,849 & 47,699 & 1.54 & -115,042 & 91,161 & 182,322 \\
	& 0.48 & -2,629 & 2,857 & 5,713 & 1.05 & 76,346 & 29,519 & 135,384 & 1.61 & -93,039 & 104,349 & 208,699 \\
	& 0.55 & 1,803 & 4,376 & 10,556 & 1.12 & -4,423 & 35,449 & 70,899 & 1.68 & -80,538 & 118,562 & 237,124 \\
	& 0.62 & -12,812 & 6,296 & 12,592 & 1.19 & -95,045 & 41,893 & 83,785 & 1.75 & 227,088 & 134,070 & 495,227 \\
	& 0.69 & -2,143 & 8,621 & 17,243 & 1.26 & -19,638 & 49,987 & 99,974 & 1.83 & 157,690 & 151,649 & 460,988 \\
	& 0.76 & -4,635 & 11,486 & 22,973 & 1.33 & -44,250 & 58,807 & 117,614 & 1.90 & 576,790 & 169,926 & 916,643 \\
	\midrule
	\multirow{8}{*}{\rotatebox[origin=c]{90}{\textbf{Field B}}}
	& 0.27 & 747 & 229 & 1,204 & 0.83 & 10,939 & 5,601 & 22,140 & 1.40 & 61,788 & 25,729 & 113,245 \\
	& 0.34 & 1,149 & 377 & 1,902 & 0.90 & 4,164 & 7,130 & 18,424 & 1.47 & 48,349 & 29,834 & 108,017 \\
	& 0.41 & -770 & 657 & 1,314 & 0.97 & 24,889 & 8,943 & 42,774 & 1.54 & 23,599 & 34,335 & 92,268 \\
	& 0.48 & 1,254 & 1,080 & 3,415 & 1.05 & 10,417 & 11,027 & 32,470 & 1.61 & 15,477 & 39,195 & 93,868 \\
	& 0.55 & -1,770 & 1,655 & 3,310 & 1.12 & -33,006 & 13,209 & 26,418 & 1.68 & 52,495 & 44,690 & 141,876 \\
	& 0.62 & 1,916 & 2,369 & 6,654 & 1.19 & -11,140 & 15,659 & 31,319 & 1.75 & 125,943 & 50,570 & 227,083 \\
	& 0.69 & 9,099 & 3,231 & 15,562 & 1.26 & 33,775 & 18,651 & 71,077 & 1.83 & -41,091 & 56,640 & 113,280 \\
	& 0.76 & 11,389 & 4,315 & 20,019 & 1.33 & 2,963 & 22,054 & 47,071 & 1.90 & -14,827 & 63,497 & 126,993 \\
	\midrule
	\multirow{8}{*}{\rotatebox[origin=c]{90}{\textbf{Field C}}}
	& 0.27 & 330 & 129 & 587 & 0.83 & 8,578 & 3,048 & 14,674 & 1.40 & 19,670 & 13,891 & 47,452 \\
	& 0.34 & 44 & 206 & \textbf{457} & 0.90 & 10,570 & 3,868 & 18,306 & 1.47 & -14,754 & 16,094 & 32,188 \\
	& 0.41 & -258 & 360 & 720 & 0.97 & 1,360 & 4,824 & 11,009 & 1.54 & -7,811 & 18,511 & 37,022 \\
	& 0.48 & 736 & 595 & 1,925 & 1.05 & 12,516 & 5,958 & 24,431 & 1.61 & -40,971 & 21,178 & 42,355 \\
	& 0.55 & 1,174 & 921 & 3,016 & 1.12 & -13,271 & 7,158 & 14,315 & 1.68 & -41,603 & 24,006 & 48,012 \\
	& 0.62 & -1,277 & 1,305 & 2,609 & 1.19 & -18,287 & 8,473 & 16,946 & 1.75 & -21,962 & 27,134 & 54,269 \\
	& 0.69 & -379 & 1,778 & 3,556 & 1.26 & -23,016 & 10,058 & 20,116 & 1.83 & 25,762 & 30,636 & 87,034 \\
	& 0.76 & 1,636 & 2,341 & 6,317 & 1.33 & 4,083 & 11,855 & 27,794 & 1.90 & -47,835 & 34,297 & 68,593 \\
	\midrule
	\multirow{8}{*}{\rotatebox[origin=c]{90}{\textbf{Field D}}}
	& 0.27 & 413 & 181 & 775 & 0.83 & 3,556 & 4,295 & 12,146 & 1.40 & -5,572 & 19,750 & 39,500 \\
	& 0.34 & 1,912 & 294 & 2,500 & 0.90 & -609 & 5,482 & 10,963 & 1.47 & 38,934 & 22,954 & 84,841 \\
	& 0.41 & 2,898 & 511 & 3,921 & 0.97 & -12,554 & 6,839 & 13,678 & 1.54 & 13,747 & 26,422 & 66,590 \\
	& 0.48 & 1,079 & 836 & 2,751 & 1.05 & -3,677 & 8,465 & 16,931 & 1.61 & 15,991 & 30,140 & 76,271 \\
	& 0.55 & 1,161 & 1,282 & 3,725 & 1.12 & 3,285 & 10,238 & 23,761 & 1.68 & -28,833 & 34,159 & 68,319 \\
	& 0.62 & 4,365 & 1,824 & 8,013 & 1.19 & 29,719 & 12,123 & 53,965 & 1.75 & 14,543 & 38,550 & 91,642 \\
	& 0.69 & 6,178 & 2,487 & 11,151 & 1.26 & 15,449 & 14,403 & 44,255 & 1.83 & 95,139 & 43,587 & 182,313 \\
	& 0.76 & 9,354 & 3,307 & 15,967 & 1.33 & 10,098 & 16,922 & 43,942 & 1.90 & 68,406 & 48,901 & 166,208 \\
	\midrule
	\multirow{8}{*}{\rotatebox[origin=c]{90}{\textbf{Field E}}}
	& 0.27 & 1,437 & 240 & 1,918 & 0.83 & 23,107 & 5,493 & 34,092 & 1.40 & -616 & 25,170 & 50,340 \\
	& 0.34 & 984 & 375 & 1,733 & 0.90 & 12,550 & 6,995 & 26,541 & 1.47 & -40,653 & 29,194 & 58,387 \\
	& 0.41 & 507 & 649 & 1,806 & 0.97 & -6,756 & 8,755 & 17,510 & 1.54 & 40,604 & 33,594 & 107,793 \\
	& 0.48 & -1,879 & 1,070 & 2,141 & 1.05 & 1,256 & 10,779 & 22,815 & 1.61 & 6,847 & 38,395 & 83,636 \\
	& 0.55 & -3,158 & 1,632 & 3,264 & 1.12 & -7,231 & 13,009 & 26,017 & 1.68 & -2,468 & 43,729 & 87,457 \\
	& 0.62 & -5,496 & 2,316 & 4,631 & 1.19 & 27,219 & 15,431 & 58,081 & 1.75 & 15,566 & 49,408 & 114,382 \\
	& 0.69 & -542 & 3,164 & 6,328 & 1.26 & 11,508 & 18,321 & 48,149 & 1.83 & 26,283 & 55,545 & 137,374 \\
	& 0.76 & 5,617 & 4,209 & 14,036 & 1.33 & 10,852 & 21,545 & 53,943 & 1.90 & 18,011 & 62,323 & 142,657 \\
    \bottomrule
    \enddata
\end{deluxetable*}

In general, most of our measurements are consistent with the expected noise level. At low $k$, especially in Band 1, we see evidence for residual foregrounds beyond our horizon plus buffer cut. This makes sense; foregrounds are brighter at low frequency. For a few fields and bands, there is some evidence for residual crosstalk, which should appear between roughly 0.4--1.0\,$h$Mpc$^{-1}$. 
\edited{Close inspection of \autoref{tab:band1_limits} and \autoref{tab:band2_limits} also reveals a handful of large negative power spectra at the roughly $-2\sigma$ level, especially in Band 2, Field E. There are more such points than might be expected from random noise, even after accounting for that fact that, as \Ha{} showed, errors are correlated between neighboring $k$ bins at the $\sim$25\% level.}

\edited{This is potentially concerning, since ignoring auto-baseline-pairs as we do here and as was done in \Ha{} has the potential to introduce negative power spectrum biases \citep{Morales:2022}, which would then lead to artificially low upper limits. Tracing back these points to their corresponding cylindrical band powers in \autoref{fig:cyl_pk_1} and \autoref{fig:cyl_pk_2} reveals only a handful of negative power spectra in excess of what is expected from thermal noise, the largest of which are all near $|\tau|$=900\,ns. Because these excesses appear at higher $k$ than the best limits and because we conservatively chose to use $2\sigma$ as the upper limit wherever $\Delta^2$ is negative, it is unlikely that our deepest power spectra or the likelihoods used in \autoref{sec:theory} are significantly affected.} That the systematics in this range remain mostly marginal speaks to the efficacy of our crosstalk subtraction and baseline selection techniques (see \autoref{sec:xtalk_updates}).

\begin{figure}
    \centering
    \includegraphics[width=.5\textwidth]{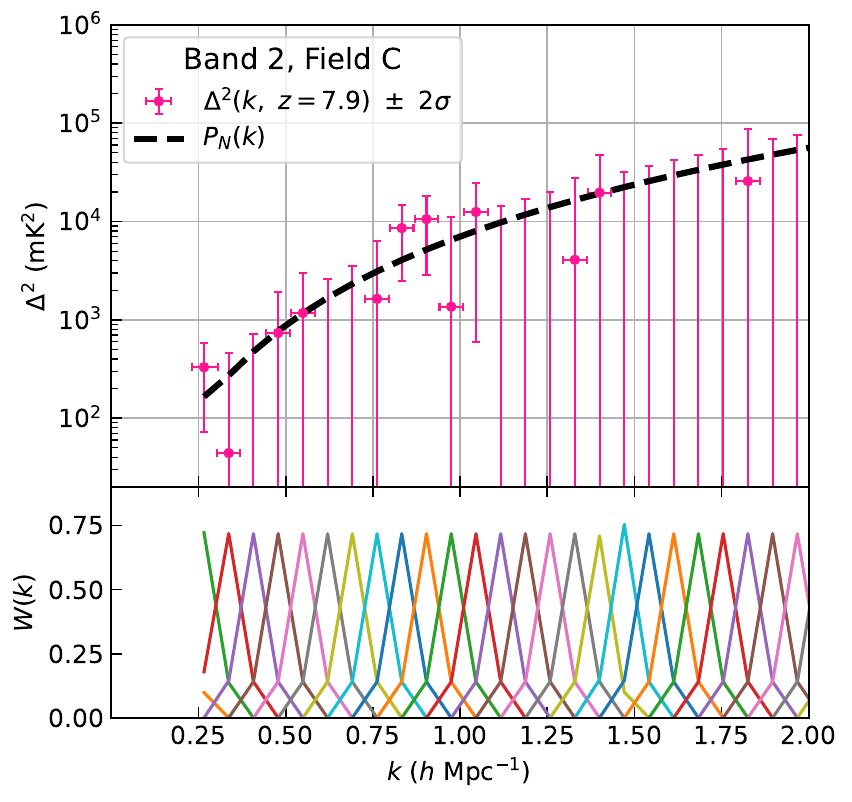}
    \caption{Here we show in greater detail the most sensitive 2$\sigma$ upper limit on the 21\,cm power spectrum we report, namely \lowzlimit{} at \lowzk{} Band 2 ($z=7.9$), Field C. The information in the top panel is identical to that in \autoref{fig:all_limits}. We also include the window functions $W(k)$ in the bottom panel, which tell us how each measurement is expected to be a linear combination of the underlying bandpowers. All band window functions peak at the measured $k$ and are relatively narrow, hence the relatively narrow horizontal errors bars which are interpolated from the window functions to span the 16th to the 84th percentile.}
    \label{fig:best_limit}
\end{figure}

We use horizontal error bars in \autoref{fig:all_limits} to represent the range of $k$ values whose contribution sums to 68\% of each measurement. To examine both the window functions and our deepest upper limit more closely, we reproduce our Band 2, Field C results from \autoref{fig:all_limits} in the upper panel of \autoref{fig:best_limit}, along with the window functions themselves in the lower panel.
The window function matrix $\mathbf{W}$, defined in \autoref{eq:def_wf} and propagated to the spherical power spectra, tells us the extent to which different bandpower measurements are expected to be the linear combination of the true bandpowers across $k$ modes. Each row---shown as a different colored line in the bottom row of \autoref{fig:best_limit}---sums to 1 by definition. The fact that each row is peaked along the diagonal of the matrix means dominant contribution to each measurement is in fact the $k$ mode at which the measurement is reported.

As discussed in \autoref{sec:window_updates}, these window functions more precisely take into account the delay approximation than the more simpler calculation used in \Ha{}. While they may look similar to the window functions presented in \Ha{} because of the linear scale used in the figure, there are as many as three orders of magnitude of difference in the amplitude of the tails, with the delay approximation leading to a large underestimate \citep{Gorce2022}. This is particularly important if the true 21\,cm power spectrum has features in $k$ space or deviates substantially from a roughly flat power law in $\Delta^2(k)$, though that is not the case in most ``vanilla" models of reionization.


\section{Statistical Tests of the Power Spectrum Upper Limits} \label{sec:tests}

In this section we report on a series of statistical tests designed to build confidence in our upper limits. Our goal is to test the self-consistency of our results by showing that they either integrate down like noise or are inconsistent with noise in ways that are well understood. Likewise, we split our data set in various ways to look for signs of possible residual systematic effects. By repeating key tests from \Ha{}, we can help ensure that no new failure modes have cropped up.

\needspace{2cm}\subsection{Noise Integration Tests at High Delay} \label{sec:tests:allan}

As in \Ha{}, we performed a series of noise integration tests to determine whether bandpowers outside the wedge region are consistent with being noise-dominated. By cumulatively averaging the samples that went into the fully-averaged power spectra shown in Figure~\ref{fig:all_limits}, we are able to test whether the samples average together as would be expected for bandpowers formed from uncorrelated (white) random noise, or whether an additional correlated or non-random signal, such as a source of systematic contamination, may be present.

Our first test is to simply integrate all bandpowers above a fixed $k$. Following \Ha{}, we randomly draw 300,000 pure-noise realizations of the bandpowers using the error covariance matrix propagated through our various averaging steps and used to set the error bars on our power spectra. We then compute the fraction of noise realizations that are larger than the data, which is by definition the $p$-value of the measurement if the null hypothesis is that measured values are drawn from the noise covariance and have mean zero.  In \autoref{tab:pvalues}, we show the results of this test for both the real and imaginary parts of both bands' and all fields' power spectra.
\begin{table}[ht]
\caption{Statistical $p$-values, testing the consistency of the real and imaginary components of the spherical $\Delta^2(k)$ with noise over different overlapping ranges of $k$. At high $k$, most measurements are consistent with noise. At low $k$, we do see decisive evidence for residual systematics (values when $p < 0.01$, in bold)---mostly foregrounds and perhaps also crosstalk, especially in Band 1. We do not quote numbers below $p < 0.001$, since we cannot precisely calculate such low $p$-values with our 300,000 random draws.}
\vspace{-2mm}
\label{tab:pvalues}
\centering
\begin{tabular}{l c c c}
\toprule
\multicolumn{1}{l}{} &
\multicolumn{3}{c}{$p$-value} \\
\cmidrule(lr){2-4}
\multicolumn{1}{c}{Data Selection} & 
\multicolumn{1}{c}{$k\ge0.2$} &
\multicolumn{1}{c}{$k\ge0.5$}  &
\multicolumn{1}{c}{$k\ge1.0$}  \\
\toprule
Re[$\Delta^2$], Band 1, Field A & \textbf{$<$0.001} & 0.378 & 0.464 \\
Im[$\Delta^2$], Band 1, Field A & \textbf{$<$0.001} & 0.052 & 0.089 \\
\midrule
Re[$\Delta^2$], Band 1, Field B & \textbf{$<$0.001} & 0.015 & 0.730 \\
Im[$\Delta^2$], Band 1, Field B & 0.773 & 0.633 & 0.844 \\
\midrule
Re[$\Delta^2$], Band 1, Field C & \textbf{$<$0.001} & 0.016 & 0.050 \\
Im[$\Delta^2$], Band 1, Field C & 0.182 & 0.790 & 0.859 \\
\midrule
Re[$\Delta^2$], Band 1, Field D & \textbf{$<$0.001} & 0.015 & 0.024 \\
Im[$\Delta^2$], Band 1, Field D & 0.810 & 0.925 & 0.686 \\
\midrule
Re[$\Delta^2$], Band 1, Field E & \textbf{$<$0.001} & 0.745 & 0.939 \\
Im[$\Delta^2$], Band 1, Field E & 0.343 & 0.686 & 0.506 \\
\toprule
Re[$\Delta^2$], Band 2, Field A & \textbf{$<$0.001} & 0.182 & 0.124 \\
Im[$\Delta^2$], Band 2, Field A & 0.262 & 0.320 & 0.144 \\
\midrule
Re[$\Delta^2$], Band 2, Field B & 0.077 & 0.385 & 0.654 \\
Im[$\Delta^2$], Band 2, Field B & 0.499 & 0.398 & 0.356 \\
\midrule
Re[$\Delta^2$], Band 2, Field C & 0.198 & 0.214 & 0.314 \\
Im[$\Delta^2$], Band 2, Field C & 0.041 & 0.120 & 0.095 \\
\midrule
Re[$\Delta^2$], Band 2, Field D & \textbf{0.007} & 0.767 & 0.937 \\
Im[$\Delta^2$], Band 2, Field D & 0.100 & 0.116 & 0.366 \\
\midrule
Re[$\Delta^2$], Band 2, Field E & \textbf{0.006} & 0.276 & 0.724 \\
Im[$\Delta^2$], Band 2, Field E & 0.383 & 0.714 & 0.819 \\
\bottomrule
\end{tabular}
\end{table}

In general, we find our imaginary power spectra consistent with noise across $k$, as expected. The one exception is in Field A for Band 1, where we are likely seeing noise-foreground cross terms dominating the imaginary power spectrum. Our real power spectra are another matter. We see strong evidence for inconsistency with noise at low $k$, which is expected given the foreground leakage just outside the wedge buffer in both cylindrical power spectra (\autoref{fig:cyl_pk_1} and \autoref{fig:cyl_pk_2}) and at low $k$ in spherical power spectra (\autoref{fig:all_limits}). In Band 1, we also see more marginal evidence for inconsistency with noise even after $k < 0.5\,h{\rm Mpc}^{-1}$ is excluded. This is likely attributable to residual crosstalk, which appears in our final power spectra at the few-$\sigma$ level.

\begin{figure}
    \centering
    \includegraphics[width=.48\textwidth]{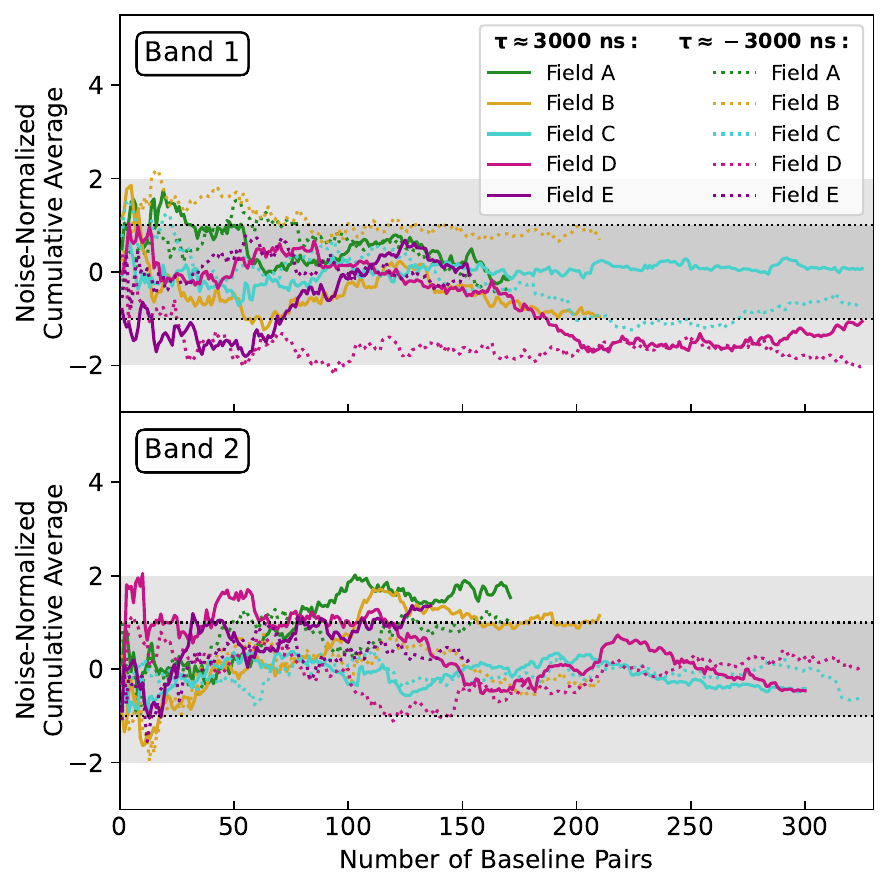}
    \caption{Cumulative average of the (crosstalk-subtracted) delay power spectrum for a single redundant baseline group (29\,m east-west) as a function of the number of baseline pairs, for our two frequency bands and five fields. Only the results for single bandpowers at $\tau \approx \pm 3000$~ns are shown here, corresponding to $k \approx 1.48$~$h\, {\rm Mpc}^{-1}$ (Band 1) and $k \approx 1.68$~$h\, {\rm Mpc}^{-1}$ (Band 2). This is well above the crosstalk contaminated delay range. Solid lines denote positive delays, dotted denote negative delays, and an incoherent time average within each field has been performed for each baseline pair before the cumulative average. Different amounts of flagging apply to each band, field, and bandpower, hence the different lengths of the lines. The cumulative average is normalized by the expected noise variance, calculated from the mean noise power over each field/band; see the text for more details. The gray bands show regions of 1$\sigma$ and 2$\sigma$ corresponding to where the cumulatively-averaged power spectra of white noise with the same (inhomogeneous) noise variance as the data, $P_N$, would be expected to fall 68\% and 95\% of the time. We see no strong evidence for any violation of the null hypothesis that these baseline groups average down like noise at this delay.}
    \label{fig:noise_integ}
\end{figure}

While the $p$-values are a useful test of the null hypothesis on the final power, a lot can be obscured by collapsing an entire power spectrum to just a few numbers. To further probe the  consistency of our results with noise, especially at high delay, we perform two additional tests. The first assesses how baselines integrate down within a redundant group. In \autoref{fig:noise_integ} we examine how a single delay mode at $\tau \approx \pm 3000$\,ns integrates down as more baselines are added to the cumulative average. 
We use our most sensitive redundant group---the 29\,m east-west baseline. We normalize the cumulative average by the expected variance of that cumulative average derived from $P_N$. We show 1$\sigma$ and 2$\sigma$ regions where we expect 68\% and 95\% of pure noise realizations to fall in gray. We have validated this calculation with a set of 100 white noise-only simulations that have been passed through the same flagging, weighting, and power spectrum estimation steps as the real data. The simulations are constructed to have noise variance that depends on time, frequency, and baseline in the same way as the data, according to our empirical estimates of the noise power spectrum $P_N$.

The overall conclusion from Figure~\ref{fig:noise_integ} is that there are no strong deviations from the expected noise-like behavior for any band or field in our data set at $|\tau| \approx 3000$~ns, according to this statistic. Band 1, Field D has the most substantial deviations, but these remain largely within the 95\% region. Band 2, Field A also approaches the edge of the 95\% region. Bands 1 and 2 of Field C have averages that are more consistently close to zero than for any of the others, though this is not necessarily a statistically significant anomaly.

\begin{figure*}
    \centering
    \includegraphics[width=\textwidth]{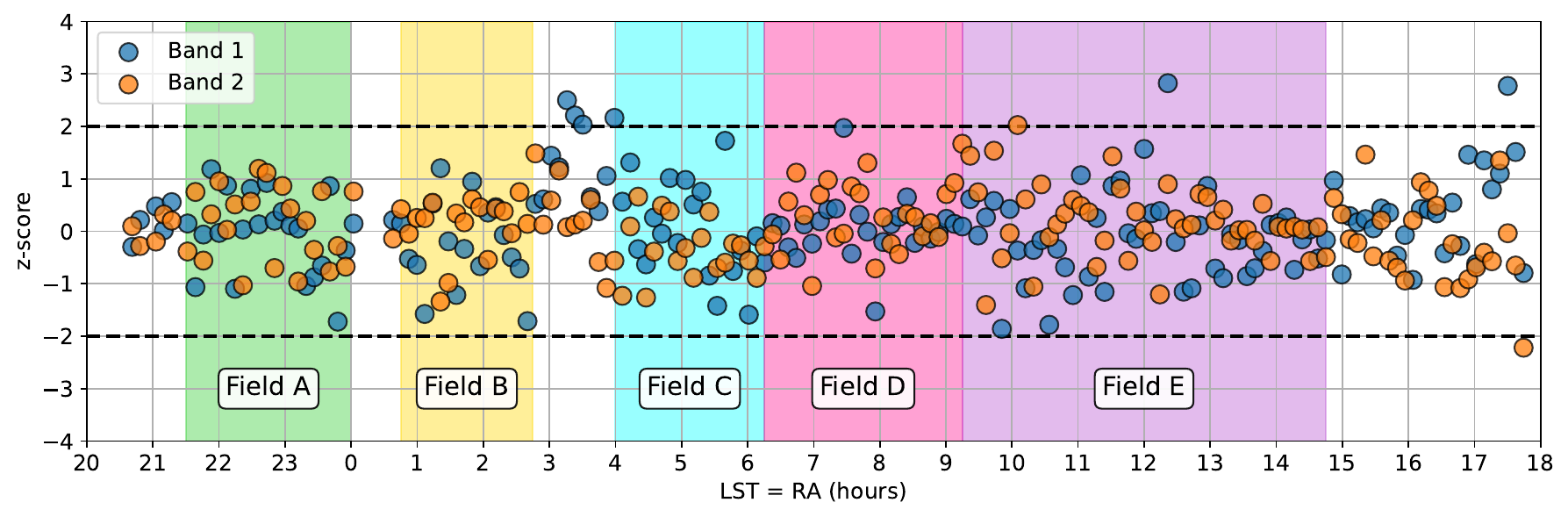}
    \vspace{-10pt}
    \caption{We examine our data for LST-dependent systematics by computing the $z$-score of the bandpowers averaged over a delay range of 2000--4000\,ns for the 29\,m east-west baseline group. The expected standard deviation at each LST is estimated by recomputing the bandpower average for 100 noise realizations matched to the thermal noise in the data. Dashed horizontal lines mark a $z$-score of 2, showing very few significant outlier in the data across the LST range.}
    \label{fig:LST_z_score}
\end{figure*}

Another similar test is shown in \autoref{fig:LST_z_score} where we examine the same baseline group, but average over baselines in the group and over a wider range of delays ($\pm$2000--4000\,ns) to look for evidence of LST-dependent systematics. 
We compare each average to the distribution of 100 noise realizations which are drawn from $P_N$ and reflect the real data's sampling and cuts. In the fields of interest, we see little evidence for significant outliers, indicating that there is no strongly LST-dependent high-delay systematic affecting our results for this baseline group. Outside our fields, we do see some large outliers from Fornax A and the Galactic center which might be concerning in the future if they appear elsewhere in HERA data as we integrate deeper.

\needspace{2cm}\subsection{Testing Systematics Mitigation}

Though we have taken several steps to mitigate the effect of crosstalk---including subtracting it from our visibilities and cutting baselines that exhibit substantial residuals (see \autoref{sec:xtalk_updates})---it is useful to understand how those choices may have affected the distribution of bandpowers. In \Ha{} we performed a one-sample Kolmogorov-Smirnov (KS) test comparing the cumulative distribution function (CDF) of power spectra over 0.4 hours of LST to a validated analytic noise model. 

\begin{figure}
    \centering
    \includegraphics[width=.48\textwidth]{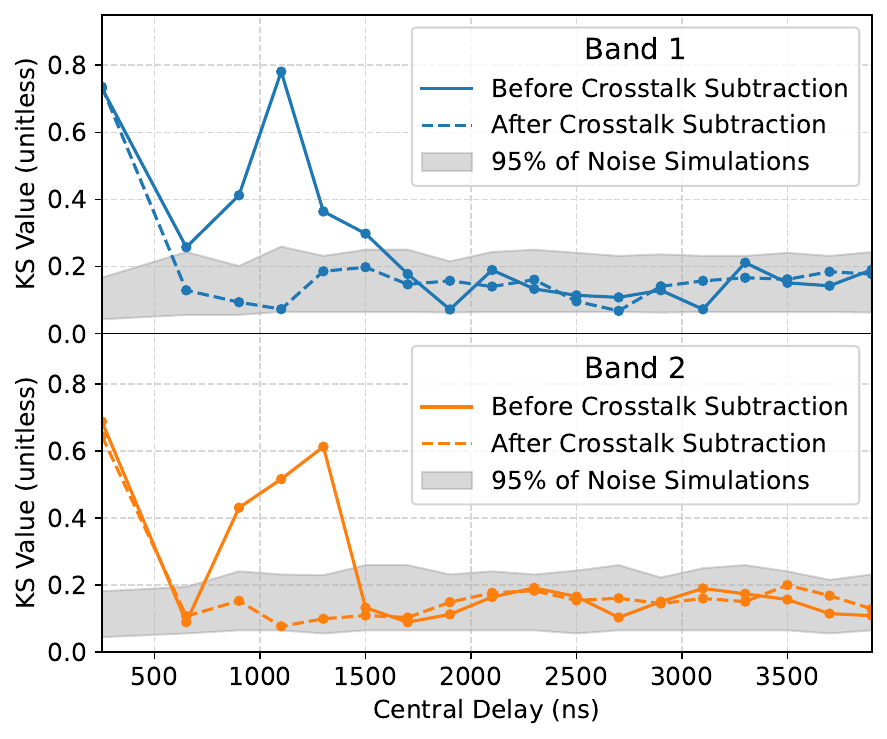}
    \caption{Here we quantify the impact of crosstalk subtraction with a two-sample Kolmogorov-Smirnov computed for various ranges of delays for 29\,m east-west baseline group. The KS test is performed by computing the cumulative distribution function of the bandpowers over different ranges of delays and LSTs in Field C and comparing it to the CDF of simulated noise. Grey regions mark $2\sigma$ confidence intervals for the delay range to be consistent with thermal noise expectations. Delay ranges are 200\,ns wide with the exception of the point, which encompasses delay ranges of 0--500\,ns. The upper and lower panels show the KS tests for Band 1 and Band 2 respectively. KS values after crosstalk subtraction, shown in dashed lines, are consistent with thermal noise expectations outside 500\,ns.}
    \label{fig:KS_delays}
\end{figure}

Here we iterate slightly on that test with a two-sample KS test comparing the CDF of measured bandpowers for the redundantly-averaged 29\,m east-west baseline group to that of the same 100 noise realizations as in our $z$-score test above. Each CDF is computed over LSTs in Field C and over a series of 200\,ns-wide delay bins (except the first bin, which spans 0--500\,ns). In \autoref{fig:KS_delays}, we show the results of that work for both bands and for data with (dashed lines) and without cross-talk subtraction (solid lines).

To infer the expected range of distances between CDFs, we also compute a distribution of two-sample KS statistics for each delay ranges by comparing many pairs of independent noise realizations. We show the 2$\sigma$ range of that distribution in the gray band of \autoref{fig:KS_delays}. We can see that while the distribution of bandpowers was highly inconsistent with noise before crosstalk subtraction, afterwards they are consistent beyond 500\,ns for this baseline group.

\needspace{2cm}\subsection{Jackknife Tests Across Epochs} \label{sec:tests:jk}

\begin{figure*}
    \centering
    \includegraphics[width=\textwidth]{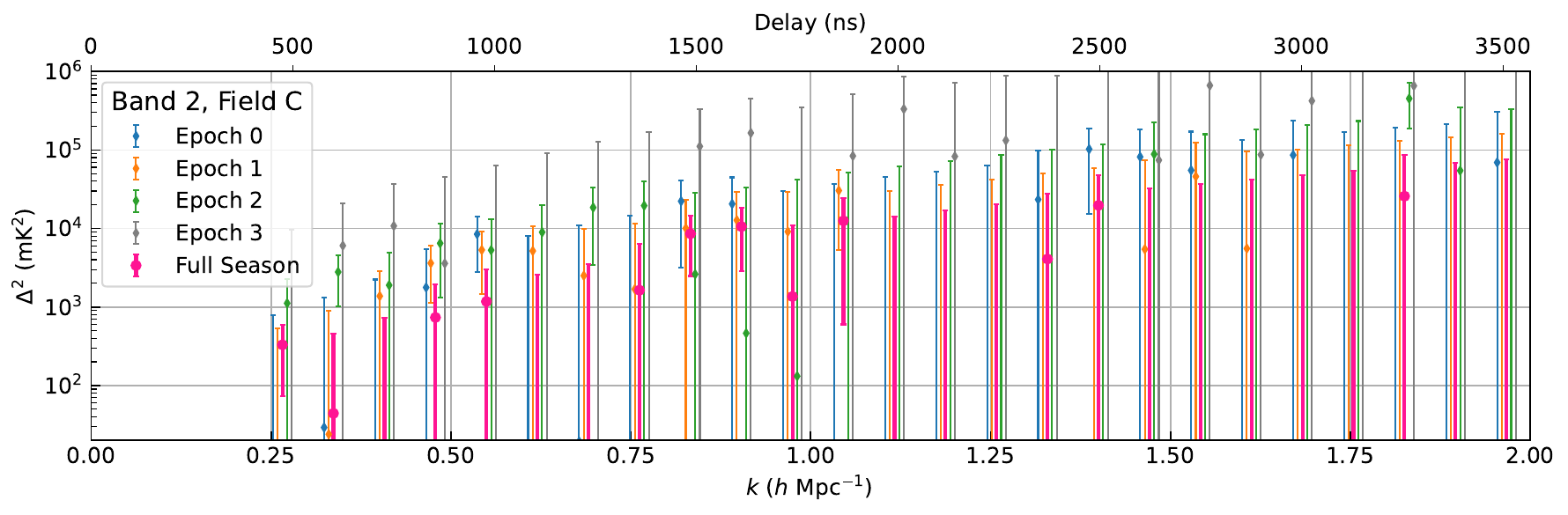}
    \vspace{-15pt}
    \caption{Here we show power spectra estimated for each epoch individually, as well as those for the full data set, for Band 2, Field C. Each individual epoch has higher noise levels than the full season combined, especially Epoch 3 which only partially overlaps with Field C. Vertical error bars are 2$\sigma$. For clarity, horizontal error bars have been omitted and individual epochs' power spectra are plotted a slightly displaced $k$s. Whether there is clear evidence for any particular epoch being a strong outlier, here or for other bands or fields, is a question we seek to answer in \autoref{fig:jackknife}.}
    \label{fig:multi_epoch_limits_b2}
\end{figure*}

In order to check the consistency of the epochs (defined in \autoref{sec:nights_and_epochs}) at the bandpower level and thus justify (post hoc) the decision to combine them to arrive at our final limits, it is useful to begin by looking at single-epoch power spectra. These are formed by simply going straight from crosstalk subtraction to forming pseudo-Stokes I and coherently averaging. In \autoref{fig:multi_epoch_limits_b2}, we show one such power spectrum from Band 2, Field C---our deepest limit---and compare measurements and vertical error bars from individual epochs to the full data set.
The full data set's power spectrum (which is identical to that in \autoref{fig:best_limit}), is generally lower than any individual epoch. Since most individual epochs are consistent with noise, this is not too surprising. Since the epochs are binned together with weights proportional to the number of visibility samples, it is not trivial to look at this figure and read off the impact of each epoch on any given $k$ mode. Therefore, while there are no obvious inconsistencies visible here, it behooves us to approach the question more quantitatively.

We answer that question by performing a Bayesian jackknife test across epochs. This test is described in detail in \citet{Wilensky2022}; we summarize our specific implementation of it here. For each band, field, and spherically-averaged $k$-mode, we consider $2^N$ hypothetical models for the $N\le 4$ epochs that contribute to the particular band and field's power spectrum at the specified $k$-mode (Epoch 0, for example, does not contain any observations in Field E). In each hypothetical model, we propose that a bandpower is either unbiased---i.e.\ consistent with zero-mean noise---or strongly biased---i.e.\ consistent with measuring a signal---by an unspecified amount within a relatively constrained prior range. To compute the posterior for each model, we marginalize over a multivariate Gaussian bias prior that is centered at $6\sigma_i$ with a width of $\sigma_i$ in each direction, where $\sigma_i$ is the square root of the estimated variance for the $i$th bandpower. This is intended to represent a strong but not excessive bias with an a priori SNR in the 4--8 range. We then use a series of Bayesian model comparisons to identify which epoch(s) is most likely to be biased for any given bandpower.

\begin{figure*}
    \centering
    \includegraphics[width=\textwidth]{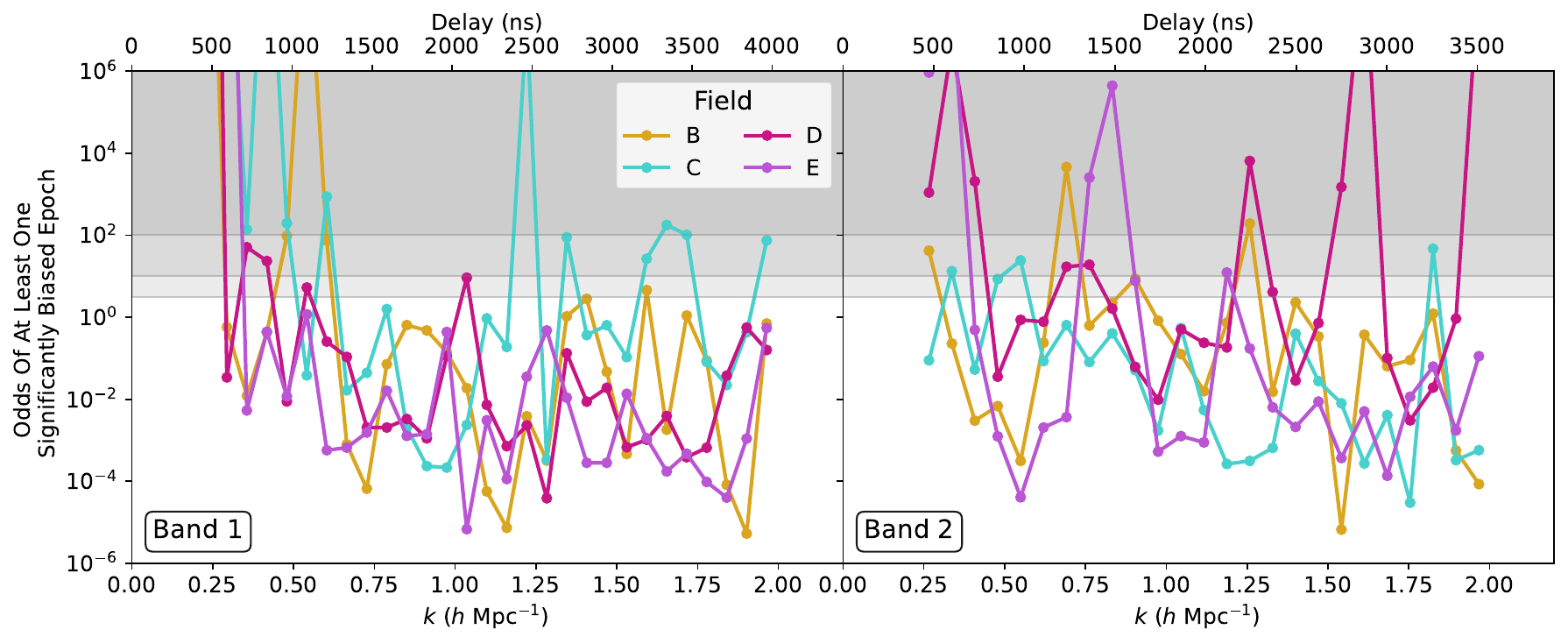}
    \vspace{-15pt}
    \caption{Here we show the odds of at least one epoch being significantly biased, as a function of $k$ for each band and field, as determined by our a Bayesian jackknife test. 
    We observe that for most bands, fields, and $k$ modes, there is only occasionally ``strong" evidence (odds $> 10^1$; medium gray region) or ``decisive'' (odds $> 10^2$; dark grey) for a significantly biased epoch, using the terminology of \citet{Kass1995}. The majority of points fall beneath the region of ``substantial" evidence (odds $> 3.2$; light gray) as well. Band 1, Field C shows consistent evidence for possessing a significantly biased epoch. Examination of the posterior over bias configurations \edited{(see \citealt{Wilensky2022})} suggests that Epoch 1 is most likely to be biased for a number of $k$ modes in Band 1, Field C. \edited{The biases observed in Band 2, Fields D and E are less clearly attributable to a single epoch.}
    Notably, Band 2, Field C (see \autoref{fig:multi_epoch_limits_b2}) shows only mild or occasionally strong evidence for the presence of a bias.}
    \label{fig:jackknife}
\end{figure*}

Figure~\ref{fig:jackknife} shows the results of this test for each band, field, and $k$ mode.
To calculate the inferred odds of there being at least one significantly biased epoch--- albeit only when the large biases reflected by the alternate hypotheses are considered---we have evaluated the posterior probability of each bias configuration, summed those in which at least one bias is present, and divided by the posterior probability of the null hypothesis that no biases were present. We use a flat prior for the bias configurations. We generally find that for the majority of bands and fields, most spherically-averaged Fourier modes are more consistent with pure noise than a scenario with an epoch-dependent bias, particularly for higher-$k$ modes (see the caption of \autoref{fig:jackknife} for a discussion of fields/epochs that potentially deviate from the null hypothesis). Since crosstalk subtraction is performed on a per-epoch basis, it is likely that residual crosstalk is a major source of bias that varies from epoch to epoch. This majority-null result justifies the decision to average the epochs together into one final power spectrum for each band and field.

This is not to say that we strongly suspect there is no underlying signal due to residual systematics or one of cosmological origin. Since only large biases are considered in the alternate hypotheses, our conclusion is that if biases are present, then they are sufficiently small to be difficult to distinguish from the expected statistical fluctuations in the bandpowers. Indeed, if we marginalize over bias priors centered at smaller biases (e.g. less than $\sigma_i$), the posterior probability of the bias configurations is more evenly diffused over the hypotheses, demonstrating a lack of certainty that arises from an inability to finely distinguish weak bias configurations with so few data.


\section{Constraints on the Astrophysics of Reionization and the Cosmic Dawn} \label{sec:theory}

Having established our new upper limits on the 21\,cm power spectrum at both $z=7.9$ and $z=10.4$, we now turn to their astrophysical implications. Just as much of this work so far has been the application of the techniques developed in \Ha{} and its supporting papers, in this section we directly apply the techniques laid out in \Hb{} to our updated power spectra. We begin by explaining how we compute model likelihoods in order to  perform astrophysical inference in a Bayesian framework (\autoref{sec:likelihoods}). In \autoref{sec:theory_overview}, we then briefly survey the four techniques employed in \Hb{} and compare their updated constraints on the ratio of the average spin temperature of the IGM to the temperature of the radio background, $\overline{T}_S / T_{\rm rad}$. After giving some background detail on the two simplest models in \autoref{sec:simpler_models}, we proceed with a more detailed report of updated techniques and results from both \texttt{21cmMC} (\autoref{sec:21cmMC}) and from models with an extra radio background generated by galaxies (\autoref{sec:radio_background}). Finally, we conclude in \autoref{sec:theory_summary} with a more comprehensive comparison of our models and the significance of their results in the broader context of 21\,cm cosmology.

\needspace{2cm}\subsection{Evaluating Model Likelihoods} \label{sec:likelihoods}

To interpret the power spectra and upper limits reported in \Ha{}, all four techniques examined in \Hb{} employ the same statistical approach to evaluating the posterior probability of model parameters in a Bayesian framework. Each theoretical model $\mathcal{M}$ with a set of parameters $\bm{\theta}$ compares the data $\mathbf{d}$ to modeled power spectra $\mathbf{m}(\bm{\theta})$ convolved with the window function $\mathbf{W}$. We write this difference as $\mathbf{t} \equiv \mathbf{d} - \mathbf{Wm}(\bm{\theta})$. We next assume that each measurement is due to some unknown combination of 21\,cm signal, noise, and systematics. Further we assume that systematics (typically residual foregrounds and crosstalk) can only add power. 
These assumptions are only appropriate when claiming upper limits on the 21\,cm power spectrum, as we do in this work. To claim a detection, one would have to impose additional constraints on the relative contribution of systematics to the measurement. To do that credibly, one likely needs high SNR detections at multiple $k$ modes, redshifts, and fields, which can be subjected to rigorous jackknives and other statistical tests for internal consistency, along with a theoretical framework able to match the data well.

Marginalizing over systematics, as we show in \Hb{}, yields a posterior probability of the form
\begin{align}
    p(\bm{\theta}|\bd,\mathcal{M}) \propto p(\bm{\theta}|\mathcal{M}) \prod_i^{N_d}\frac{1}{2}\left(1+{\rm erf}\left[\frac{t_i}{\sqrt{2}\sigma_i}\right]\right) \label{eq:final_marg_likelihood}
\end{align}
where the product on the right-hand side is the likelihood. Here $N_d$ is the number of data points considered and $\sigma_i$ is the standard deviation of $d_i$. This result is consistent with similar approaches in the literature \citep{Li2019, Ghara20}. In this derivation, we also assumed that measurements have uncorrelated errors.  While neighboring measurements in $k$ are in fact quite correlated, we follow \Hb{} and throw out every other measurement in $k$---keeping the absolute lowest limits---to largely eliminate that correlation. 

\begin{figure}
    \centering
    \includegraphics[width=.48\textwidth]{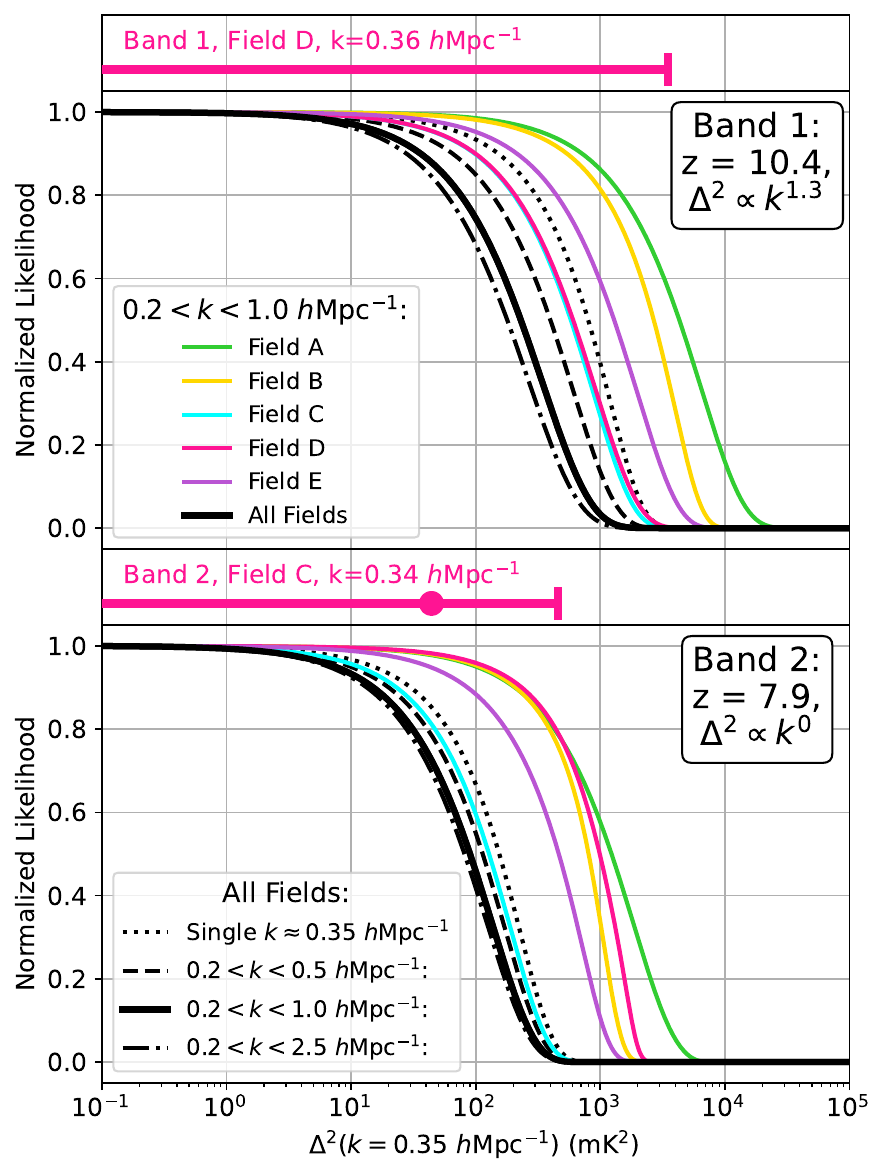}
    \caption{Here we show our marginalized likelihood from \autoref{eq:final_marg_likelihood} for each band and multiple fields and ranges of $k$. In solid colored lines, we show the likelihood for each field independently, but combine all $0.2 < k < 1.0$\,$h$ Mpc$^{-1}$ after throwing away every other power spectrum measurement to eliminate correlated errors. To combine multiple $k$ values, we have assumed a power law $\Delta^2 \propto k^\alpha$ where $\alpha=1.3$ at $z=10.4$ and $\alpha=0.0$ at $z=7.9$. In the various black curves, we combine together all fields but use different ranges of $k$ values, including just the $k$ of the best upper limits at $k \approx 0.35$\,$h$Mpc$^{-1}$. To help provide some intuition for how the upper limits yield these likelihoods, we also show the best upper limit from each band (pink error bars, taken from \autoref{fig:all_limits} and rotated 90$^\circ$). It is clear that combining together multiple fields has an effect, especially for Band 1 ($z=10.4$) where Fields C and D contribute roughly equally. Likewise, combining together multiple $k$ modes tightens the constraints significantly, especially at $z=10.4$ where power law is steeper. 
    When evaluating the posterior probability of our various models and parameters, we compare the models to the data at the proper $k$ values.
    }
    \label{fig:likelihoods}
\end{figure}

To understand the effect of the form of the likelihood in \autoref{eq:final_marg_likelihood}, we calculate it using a power law $\Delta^2 \propto k^\alpha$ over several orders of magnitude in $\Delta^2$ evaluated at $k=0.35$\,$h$Mpc$^{-1}$ in \autoref{fig:likelihoods}.
At $z=10.4$, we use $\alpha=1.3$; at $z=7.9$ we use $\alpha=0.0$. These values of $\alpha$ were derived by interpolating the \texttt{21cmMC} power spectrum realizations in our posterior without HERA (see \autoref{sec:21cmMC:pspec}) at the $k$ values of our two deepest measurements, fitting a power law, and then taking the median $\alpha$ over models that had not yet completely reionized. The flattening of $\Delta^2$ with decreasing redshift is a generic consequence of inside-out reionization, which suppresses small-scale power.

The first thing to note is that our likelihoods are essentially flat for $\Delta^2$ values much less than the upper limit; the signal could fall anywhere in that range because the systematics could fall anywhere in that range as well. Our results can therefore be heavily impacted by the prior, $p(\bm{\theta}|\mathcal{M})$, and what it considers the smallest viable power spectrum. 

\autoref{fig:likelihoods} also highlights the fact that, in contrast to \Hb{}, our result features fairly comparable limits from multiple fields and $k$ modes. Because our likelihoods are not Gaussian, the results do not scale like $1/\sqrt{N}$ as one might expect. Limits with larger measurements and smaller error bars look different in \autoref{eq:final_marg_likelihood} than measurements with the same $2\sigma$ upper limits but with smaller measured $\Delta^2$ and larger error bars. So, while it is not surprising that most of the information in Band 2 comes from Field C, it is interesting and not necessarily intuitive that Fields C and D contribute equal amounts of information in Band 1. Likewise, we can see that just using a single $k$ mode (dotted black lines)---even though it is the single best upper limit and combines together all fields---yields somewhat worse constraints than combining multiple $k$s. The vast majority of the constraining power comes form $k < 1$\,$h$Mpc$^{-1}$ (solid black line). Of course, combining $k$ modes together in this way is only technically correct if we are evaluating the likelihood in \autoref{eq:final_marg_likelihood} for a model with $\Delta^2 \propto k^\alpha$. In general, our model posteriors use all fields and all $k$-values, but compare them to the data at the proper $k$ values.


\needspace{2cm}\subsection{Overview of Theoretical Models and their IGM Spin Temperature Constraints} \label{sec:theory_overview}

In \Hb{}, we interpreted the results of \Ha{} using a suite of four different theoretical models to infer constraints on the IGM and high-$z$ galaxies:
\begin{enumerate}
    \item a simple ``density-driven" linear bias model, in which the 21\,cm fluctuations are assumed to trace the density fluctuations, multiplied by a bias factor that depends on the average thermal and ionization state of the IGM;
    \item  a phenomenological ``reionization-driven" model, which parameterizes the ionized bubble distribution and IGM spin temperature directly without making any explicit assumptions about galaxies \citep{Mirocha2022};
    \item a semi-numeric model of the 21\,cm signal, \texttt{21cmFAST} \citep{Mesinger2011}, along with the inference engine \texttt{21cmMC} \citep{Greig2017b} that uses it to explore and constrain a range of parameterized galaxy properties (e.g. \citealt{park19}) with multi-wavelength probes; and
    \item an independent semi-numeric model that models the galaxies differently and allows for a radio background in excess of the CMB at high redshifts \citep{Reis:2020}.
\end{enumerate}

Our models are constructed in two qualitatively different ways. The density-driven and reionization-driven models interpret the 21\,cm signal directly as a function of IGM properties. They avoid making explicit assumptions about the sources generating the radiation fields that determine the 21\,cm signal (though their priors on IGM properties carry some implicit assumptions), and as such have a lighter implementation. The semi-numeric models, on the other hand, start with a model of galaxy evolution and simulate a cosmological volume in order to predict the 21\,cm signal. The latter are significantly more complicated but have the distinct advantage of a parameter set that is rooted in our current understanding of galaxy properties, which allows us to combine 21\,cm measurements with other constraints on the galaxy population (such as UV luminosity functions and the X-ray background). The disadvantage of the semi-numeric approach is that the resulting  constraints on galaxy physics are only sensible if their parameterization is flexible enough to include all the relevant physics. In \Hb{} we considered two independent semi-numeric models to mitigate this problem.

The key result of \Hb{}, across all these models, was that the IGM must have been heated above the expectation for an adiabatically cooling IGM from recombination to $z=7.9$ at $>$95\% credibility.\footnote{When we say, e.g., ``at 95\% credibility" or present a ``95\% credible limit'' we mean that the particular parameter value bounds the 95\% credible interval---the region of the posterior on that parameter that contains 95\% of the integrated probability density. This is different from a frequentist's 95\% confidence interval, which is by definition bounded by a pair of random variables that should contain the true parameter value in 95\% of repeated trials. The credible interval depends directly on the theoretical model, its parameterization, and its priors. To say that a specific model with a given set of parameters is ruled out at 95\% confidence, by contrast, one would have to compare the model's prediction for $\Delta^2(k,z)$ directly to the measurements and their error bars (see \autoref{tab:band1_limits} and \autoref{tab:band2_limits}).} If this heating is interpreted in the conventional way---as the result of X-rays generated by accretion onto black hole remnants of early star formation and then depositing that thermal energy in the IGM---this level of heating suggested that early galaxies were more efficient X-ray emitters than their local counterparts. If an excess radio background were allowed, the observations jointly constrained the efficiencies of radio and X-ray emission from galaxies, also restricting the parameter space of otherwise viable models of these mechanisms during the EoR.

\begin{figure*}
    \centering
    \includegraphics[width=\textwidth]{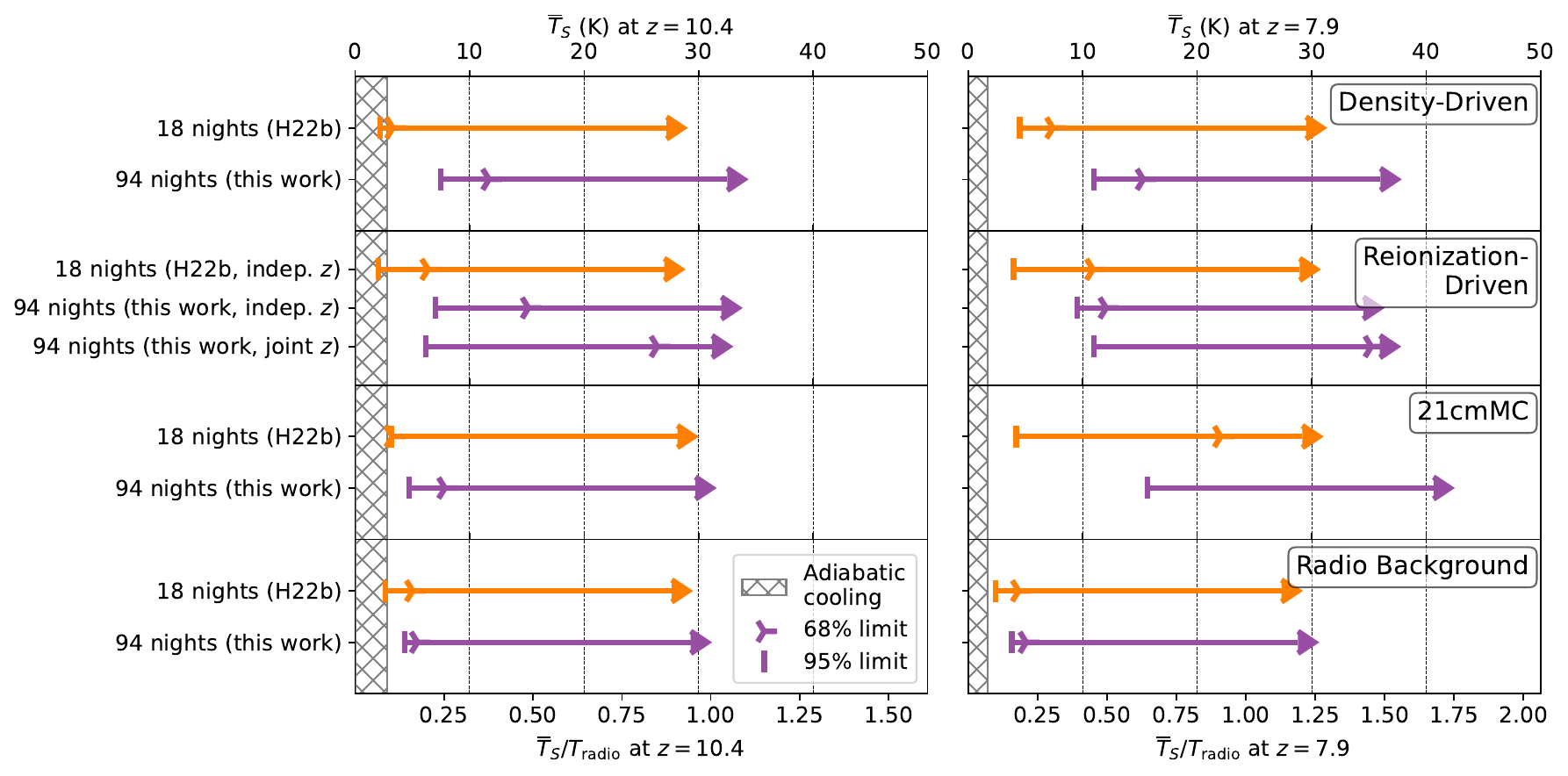}
    \caption{Here we summarize HERA's constraints on the IGM spin temperature $T_S$ and contrast $T_S / T_{\rm radio}$ at $z = 10.4$ (left) and $z = 7.9$ (right). Each row shows the HPD results obtained with a different model for 21-cm fluctuations (described in \autoref{sec:theory}). These include a linear bias model with density fluctuations only (top row; see \autoref{sec:simpler_models}); a phenomenological model that parametrizes the ionized bubble size distribution and assumes an IGM of uniform temperature (second row; also see \autoref{sec:simpler_models}); \texttt{21cmMC}, a Bayesian technique for fitting parameters of semi-numeric \texttt{21cmFAST} simulations (third row; see \autoref{sec:21cmMC}); and another semi-numerical model that includes a prescription for radio emission generated by galaxies (bottom row; see \autoref{sec:radio_background}). In each panel, we compare results obtained in this work with the previous set of upper limits published in \Hb{}, where we only saw evidence for heating above the adiabatic limit (gray hashed region) at $z=7.9$. In this work, we see consistent evidence across all our models for an IGM heated above the adiabatic at $z=10.4$ as well. 
    The more dramatic rise of the $z=7.9$ spin temperature in \texttt{21cmMC} relative to the other models is driven by the $z=10.4$ constraints combined with independent constraints on galaxy luminosity functions (as we discuss in detail in \autoref{sec:theory_summary}).
    Note that for the first three models, $T_{\rm{radio}} = T_{\rm{CMB}}$, which enables a conversion between the top and bottom axes. For the model in the bottom row with an excess radio background, $T_{\rm{radio}} \neq T_{\rm{CMB}}$ in general, and so the $T_{S}$ tick marks along the top axis should be ignored.
    }
    \label{fig:theory_summary}
\end{figure*}

In this work, we apply those same four techniques---with some minor adjustments, as we will discuss below---to our data from both bands. As \Hb{} argued, the easiest point of comparison for the models is the inferred average spin temperature of the neutral IGM, $\overline{T}_S$. In \autoref{fig:theory_summary} we 
compare the 68\% and 95\% highest posterior density (HPD) credible interval on $\overline{T}_S$ in all four models and show how those constraints have improved since \Hb{}.\footnote{In this work, we generally compute HPD credible intervals in the space sampled in, either logarithmic or linear. For $\overline{T}_S$, this is done in by minimizing the logarithmic interval in most models. The exception is when we compute posteriors in $\Delta^2$ (see \autoref{fig:ps_posteriors_with_limits} and \autoref{fig:full_ps_posterior}), a derived quantity in our models, which can often be spread over a wide dynamic range, multi-modal, and include zeros where the universe is fully reionized. There we instead used the equal-tailed credible interval, which has equal integrated probabilities above and below the interval.} 
The improvements are modest but consequential, especially at $z=10.4$: all four models now independently require the IGM to be heated above the adiabatic cooling limit at $>$95\% credibility. The precise values of lower limits vary from model to model, reflecting their different assumptions and approximations. In particular, \texttt{21cmMC} shows the most dramatic increase in the spin temperature constraint at $z=7.9$, a fact that deserves further examination.\footnote{In fact, its 68\% credible lower bound at $z=7.9$ is 79.0\,K, off the right edge of the plot (see \autoref{fig:TKS_posterior}).} As we discuss the specific models and their detailed results throughout this section, we will look to understand what precisely drives these differences. In \autoref{sec:theory_summary}, we will summarize our takeaways and put these spin temperature constraints in the broader cosmological context.


\needspace{2cm}\subsection{Phenomenological Models of the 21\,cm Power Spectrum} \label{sec:simpler_models}

Before we discuss the detailed results of our more complex techniques for astrophysical inference based on semi-numerical simulations, here we briefly detail the two simpler methods used in \Hb{} to infer constraints on the spin temperature and other IGM properties.

In the first, which we refer to as the ``density-driven'' model, the 21\,cm power spectrum is assumed to follow that of matter, multiplied by a bias parameter squared (as is standard in perturbation theory, see e.g.\ \citealt{McQuinn:2018zwa,Georgiev2022,Qin:2022xho,Sailer:2022vqx} for more complex approaches). Limits on \edited{this} bias can then be translated into lower bounds on $\overline{T}_S/T_{\rm CMB}$ under some assumptions, \edited{including redshift-space distortions as a function of the line-of-sight cosine $\mu$, see~\Hb{}) and} that the IGM properties are roughly homogeneous (as the model ignores ionization and temperature fluctuations beyond adiabatic, see \Hb{} for details).
Our limits on the bias $b_m$ using 94 nights (and all fields and every other $k$, as discussed above) are $|b_m| < 60$\,mK at $z=7.9$ and $|b_m| < 160$\,mK at $z=10.4$ at 95\% credibility, which assuming $x_{\rm HI}=1$ translate into the IGM limits shown in \autoref{fig:theory_summary} and in \autoref{fig:TsoverTrad}.
Note that these limits are on the absolute value $|b_m|$. This is unimportant for cases where $\overline{T}_S \ll T_{\rm radio}$, but as the limits tighten we will obtain contours around $\overline{T}_S = T_{\rm radio}$ (which yields no 21-cm signal and thus $b_m=0$).
In fact, the 68\% credible limit on the $z=7.9$ bias is $|b_m| < 30$ mK, which translates into the double-sided region from  $\overline{T}_S=15$\,K to $\overline{T}_S=60$\,K (assuming $T_{\rm radio} = T_{\rm CMB}$, as is standard).

In the second, which we refer to as the ``reionization-driven'' model, the IGM is modeled as a two-phase medium, with fully ionized bubbles drawn from a log-normal size distribution and the ``bulk'' IGM outside bubbles assumed to be of uniform temperature \cite[as described in][]{Mirocha2022}. The advantage of this approach\footnote{\url{https://github.com/mirochaj/micro21cm}} is that it works directly with IGM quantities, which makes it easy to interpret, and avoids making explicit assumptions about galaxies\footnote{However, there are latent assumptions. For example, given that flat priors on astrophysical parameters in \texttt{21cmMC} do not correspond to flat priors on properties of the IGM (see, e.g., Figures \ref{fig:LX_posterior} and \ref{fig:TKS_posterior}), flat priors on IGM properties in this phenomenological model will not yield flat priors in astrophysical parameter space.}. In its simplest form, it requires four parameters: the volume filling fraction of ionized gas, $Q \equiv (1 - x_{\rm HI})$, the IGM spin temperature, $\overline{T}_S$, the characteristic bubble size, $R_b$, and log-normal dispersion, $\sigma_b$. In this work, when jointly fitting both HERA bands, we also use a 7-parameter version of the model in which the ionized fraction and characteristic bubble size are allowed to evolve with redshift as power laws. We require only that $Q$ and $R_b$ increase with time, that reionization completes at $z \geq 5.3$, and that all parameters are positive. We perform our Markov Chain Monte Carlo (MCMC) inference using \textsc{emcee} \citep{Foreman-Mackey2013}.

Now, with 94 nights of data, we infer spin temperatures in excess of the adiabatic cooling limit at $z=7.9$ and $10.4$, both for fits that consider each band separately and the joint fit to both bands using a 7-parameter version of the model (see \autoref{fig:theory_summary}). At 95\% (68\%) credibility we obtain $\overline{T}_S > 11.0$ (35.2)\,K at $z=7.9$, and $\overline{T}_S > 6.2$ (26.4)\,K at $z=10.4$, when fitting both bands simultaneously. Note that, in this joint fit, lower limits on $\overline{T}_S$ grow at $z=7.9$, as expected, but actually slightly decrease at $z=10.4$ (at 95\% credibility) relative to the results of single-band fits. This is a result of using an HPD estimate of the credible interval, combined with change in the shape of the $\overline{T}_S$ posterior, which goes from being roughly flat as $\overline{T}_S \rightarrow 10^3 \ \rm{K}$ to a more ``peaked'' distribution, slightly away from the maximal value of $\overline{T}_S$. This is consistent with the results of the density-driven model, suggesting that our limits are beginning to disfavor scenarios with saturated $\overline{T}_S$ at $z=7.9$. 

The differences between these approaches explains some of the differing conclusions in \autoref{fig:theory_summary}. For example, the density-driven model yields higher $\overline{T}_S$ limits than the reionization-driven model, because the latter assumes ionization and density are positively correlated. As a result, the IGM must be made colder to compensate for the loss of the densest regions to ionization. We see also in the reionization-driven model the power of jointly fitting multiple bands; the $z=7.9$ limit increases by roughly a factor of 2 in the joint fit because a given temperature at $z=7.9$ is only viable if the temperature evolution is consistent with the $z=10.4$ data.

\needspace{2cm}\subsection{Updated Constraints on Reionization and X-ray Heating of the IGM with \texttt{21cmMC}} \label{sec:21cmMC}

\texttt{21cmFAST} \citep{Mesinger2011, Murray_2020} is a semi-numerical simulator for computing the evolution of the 21\,cm signal across cosmic time by assuming that star-forming galaxies, hosted by dark matter halos, drive the cosmic radiation fields that heat and ionize the IGM. It uses empirical scaling relations to assign galaxy properties to dark matter halos based on their halo masses. These include the stellar mass to halo mass ratio, the UV ionizing escape fraction, the star formation rate, and the X-ray luminosity. \texttt{21cmFAST} simulates reionization using the excursion-set formalism \citep{Furlanetto2004}. It accounts for inhomogeneous recombinations \citep{Sobacchi2014} and includes a numerical correction for photon conservation \citep{Park2021}.

In Section~5.1 of \Hb{}, we discussed the nine physically-meaningful free parameters that go into the scaling relations describing galaxy properties (also see \citealt{park19}). We adopted either flat linear or logarithmic priors---depending on the parameter's dynamic range and how it enters into the model---within our 9-dimensional hypercubic parameter space. The ranges on these priors were picked to allow a broad range of physically plausible values while not strongly constraining the parameters most constrained by other high-redshift probes. These include UV luminosity functions \citep{Bouwens2015, Bouwens16, Oesch18} and measurements of the IGM opacities during the EoR such as the Ly-$\alpha$ forest \citep{McGreer2015, BB15, Qin21} and CMB polarization and optical depth \citep{Planck18, Qin20}. The Ly-$\alpha$ forest and the CMB's $\tau$ provide important clues as to the timing and duration of reionization, which then constrain the ionizing escape fraction, given the observed UV luminosity function.
We use \texttt{21cmMC} \citep{Greig2015a, Greig2017b} and its MultiNest sampler \citep{FHB09, Qin21b} to perform Bayesian inference in this work. The posterior probability distribution without HERA serves as the starting point for comparing models against HERA data.  

One key result from these other high-redshift probes is the strong constraint on the star formation efficiency of dark matter halos at high redshift, which is determined by the UV luminosity functions at $z \sim 6$--8; existing observations constrain both the peak efficiency and show that it declines toward small halo masses (e.g., \citealt{tacchella13, mason15, Mirocha2017, park19,Sabti:2021xvh}). The version of \texttt{21cmFAST} used here assumes that this behavior can be extrapolated to higher redshifts and smaller haloes. As a result, the range of galaxy formation models that are allowed by \texttt{21cmMC} is relatively restricted, and all display a rapid increase in the stellar mass density between the two redshifts measured by HERA. The resulting constraints must be interpreted with this in mind, as more complex galaxy evolution histories 
(which break the assumed extrapolation by, for example, introducing a new population of sources, see e.g.\ \citealt{Munoz:2019rhi,Qin:2020xyh,Munoz:2021psm} for implementations in \texttt{21cmFAST}) are not included in the prior and are left for future work.

\subsubsection{\texttt{21cmMC} Constraints on X-Ray Luminosity} \label{sec:21cmMC:xray}

In \Hb{} we explored how adding HERA affected the full posterior parameter covariance. In this work, we focus on the parameter most constrained by HERA, the ratio of the integrated soft-band (${<}2$keV) X-ray luminosity to the star formation rate. Since \texttt{21cmFAST} assumes that X-ray photons govern the thermal history of the neutral IGM, this $L_{\rm X < 2keV}$/SFR parameter essentially describes the heating power of EoR galaxies per unit of star formation. In \Hb{}, we obtained the first observational evidence for an enhanced X-ray luminosity of high-redshift ($z>6$) galaxies, with a 68\% HPD credible interval of $L_{\rm X < 2keV}/{\rm SFR} \sim 10^{39.9}$--$10^{41.6}$\,erg\,s$^{-1}$\,${\rm M}_\odot^{-1}$\,yr. This disfavors a relationship between star formation and soft X-ray luminosity at or below the one seen in local, metal-enriched galaxies at $>$68\% credibility.

\begin{figure}
    \centering
    \includegraphics[width=.48\textwidth]{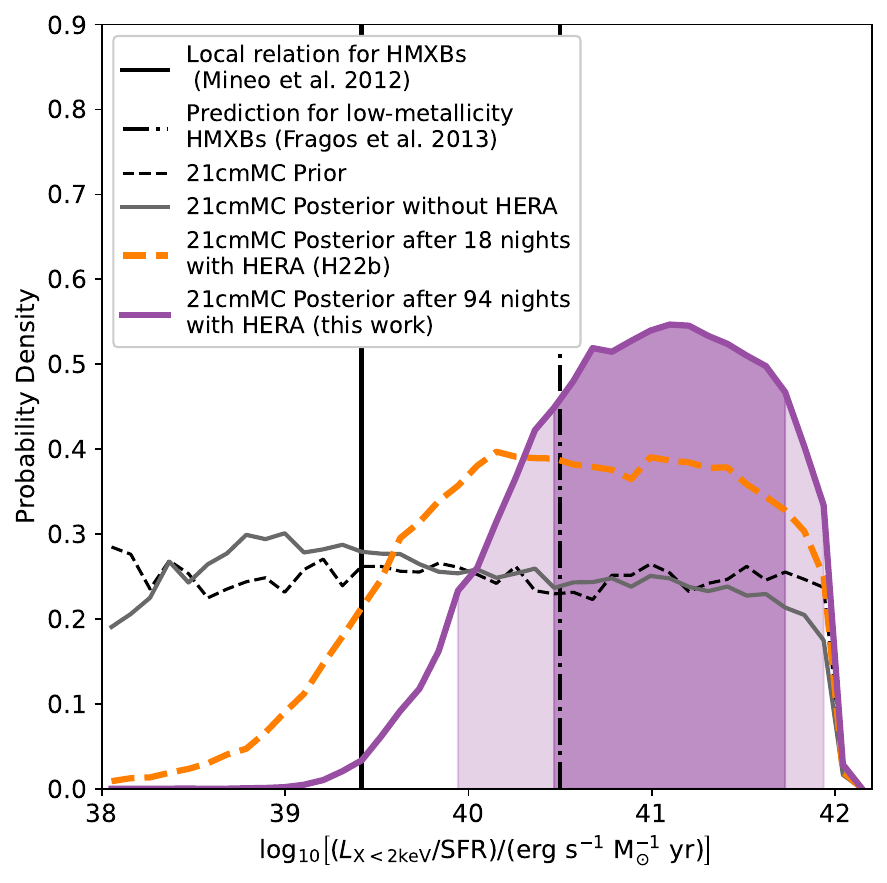}
    \caption{Here we show how our marginalized \texttt{21cmMC} posterior PDF of the ratio of soft X-ray luminosity to SFR, $L_{\rm X < 2keV}$/SFR, tightens with a full season of HERA data. The shaded regions show the 68\% and 95\% credible intervals of the posterior. Probability densities are plotted per logarithmic interval. Our results are consistent with theoretical expectations for X-rays produced during the cosmic dawn by a population of low-metallicity high-mass X-ray binaries (HMXB) \citep{Fragos2013}, likely a more representative model of the first galaxies (dash-dotted black vertical line). Compared to \Hb{} (orange dashed line), our result's 99\% credible interval excludes models where the local relation for X-ray efficiency (solid black vertical line; \citealt{Mineo2012}) continues to hold at high redshift.}
    \label{fig:LX_posterior}
\end{figure}

As \autoref{fig:LX_posterior} shows, we find that the full season of HERA observing constrains the 95\% credible interval on $L_{\rm X < 2keV}/{\rm SFR}$ to the range $10^{40.4}$--$10^{41.8}$\,erg\,s$^{-1}$\,${\rm M}_\odot^{-1}$\,yr. 
This result assumes as a prior that $L_{\rm X < 2keV}/{\rm SFR} < 10^{42}$\,erg\,s$^{-1}$\,${\rm M}_\odot^{-1}$\,yr, beyond which 
X-rays reionize the universe too quickly \citep{Mesinger2013}.
More than 99\% of the \texttt{21cmMC} posterior volume excludes the possibility of the local relation for HMXBs \citep{Mineo2012} continuing to hold at high redshift. It is consistent, however, with models of extremely low-metallicity galaxies, where high mass stars have less mass-loss from line-driven winds than their solar-metallicity counterparts \citep{Fragos2013}.

\needspace{2cm}\subsubsection{\texttt{21cmMC} Constraints on the IGM's Thermal History} \label{sec:21cmMC:temp}

\begin{figure*}
    \centering
    \includegraphics[width=\textwidth]{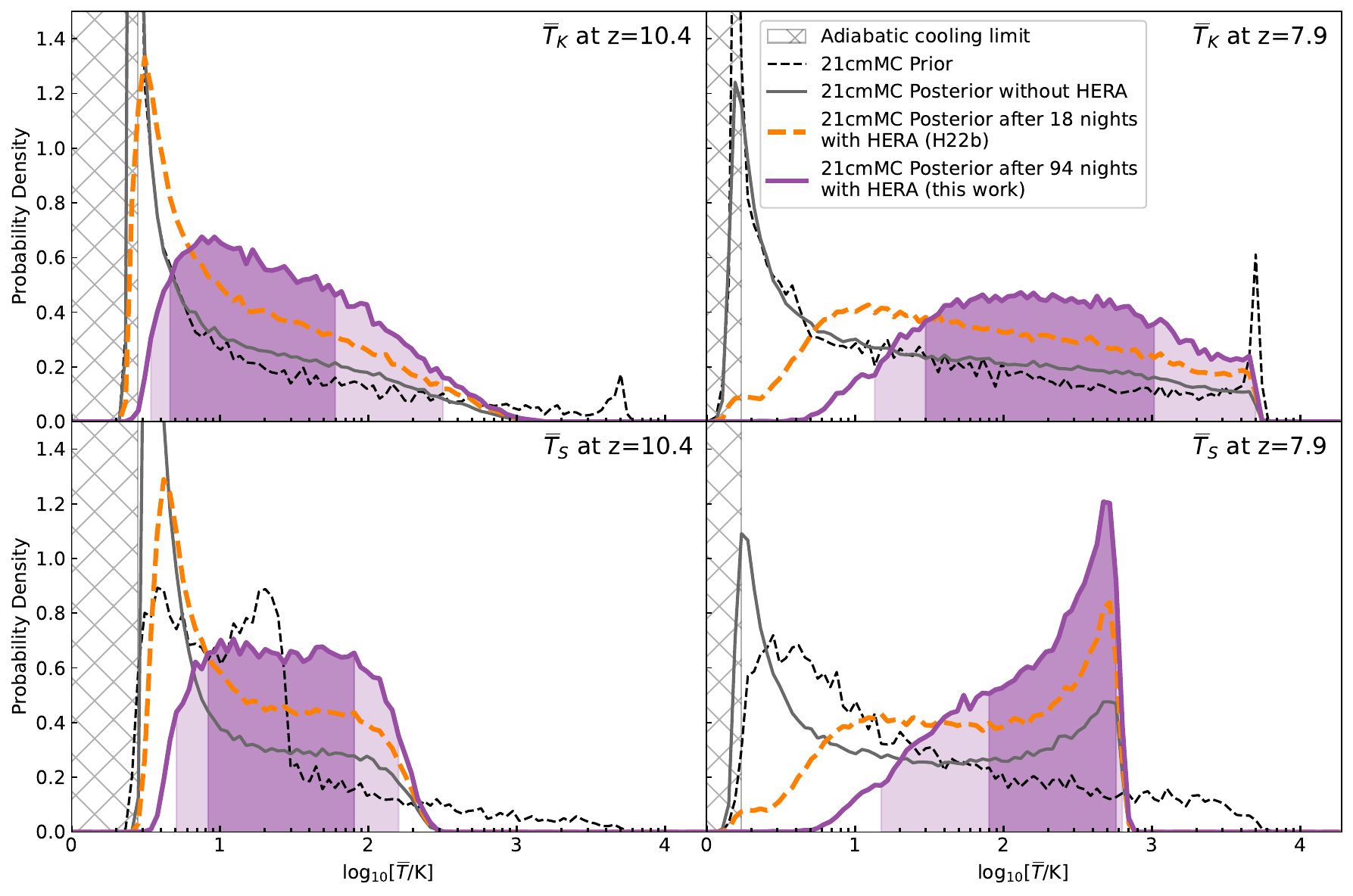}
    \caption{Here we show marginalized 1D PDFs from \texttt{21cmMC} per logarithmic interval in the kinetic temperature (top row) and spin temperature (bottom row), averaged over the neutral IGM, at both $z=10.4$ (left column)  and $z=7.9$ (right column). We compare the prior on these quantities to the posterior from \Hb{} data (orange dashed) and the posterior after a full season (purple, with 68\% and 95\% credible intervals shaded). We also show our prior (black dashed) and the posterior after including non-HERA astrophysical constraints on reionization (gray). The averaging is performed over neutral cells with $x_{\rm HI}>0.95$; for models completely reionized at $z=7.9$, we take the average temperature from the last time-step with neutral cells. The hashed region indicates temperatures below the adiabatic cooling limit. Our observations rule out an unheated IGM at $>$99\% credibility at both $z=10.4$ and $z=7.9$, placing new constraints on the population of X-ray emitting compact objects during the cosmic dawn.}
    \label{fig:TKS_posterior}
\end{figure*}

Our constraints on the soft X-ray efficiency are themselves a consequence of our ability to use our upper limits to exclude a range of scenarios with low levels of IGM heating. In \autoref{fig:TKS_posterior} we show our updated marginalized posteriors on the predicted average IGM temperatures---both the spin temperature, $\overline{T}_{S}$, and the kinetic temperature, $\overline{T}_{K}$---along with results from \Hb{} for comparison. 
As demonstrated in \Hb{}, current EoR constraints from Planck and quasar spectra already disfavor a large number of models in the prior volume which predict either highly ionized IGM at $z \ge 10.4$ or completely neutral one at $z \le 10.4$. These constraints also have a slight impact on the average IGM temperature, excluding models with high $\overline{T}_{\rm K}$ or $\overline{T}_{\rm S}$ at these redshifts. However, because a decently-sized fraction of parameter space with an unheated IGM at these redshifts is not ruled by these probes, and since \texttt{21cmMC} cannot produce spin temperatures below the adiabatic limit,  our posterior without HERA shows a pileup of probability right around that limit.

When we incorporate the HERA limits, a significant range of models with low IGM temperatures can be further excluded. We showed in \Hb{} how HERA observations substantially improve our understanding of the neutral IGM at $z=7.9$. However, there was still a small fraction of the total posterior with low values of $\overline{T}_{\rm S}$, so \Hb{} could not completely rule out an unheated IGM the observed redshifts. With the improved limits presented in this work, we now find that an unheated IGM is disfavored at greater than 99\% credibility at both $z=10.4$ and 7.9. The new HPD 95\% credible intervals on the spin and kinetic temperatures are 
$4.7$\,${\rm K} < \overline{T}_{S} < 171.2$\,K and 
$3.2$\,${\rm K} < \overline{T}_{K} <  313.2$\,K at $z=10.4$ and
$15.6$\,${\rm K} < \overline{T}_{S} < 656.7$\,K and
$13.0$\,${\rm K} < \overline{T}_{K} < 4768$\,K at $z=7.9$.\footnote{Our upper limits on $T_S$ are very similar to those in \Hb{}, in large part because at high $T_K$, $T_S$ asymptotes to value set by the Ly-$\alpha$ coupling, which cannot be too high without reionzing the universe too early (see Section~5.4.2 of \Hb{} for details).}
 
Since the 21\,cm brightness temperature is proportional
to $(1 - T_\text{CMB} / T_\text{S})$, any model where $T_\text{S} < 0.5 T_\text{CMB}$ will show an enhanced power spectrum amplitude relative to one where $T_\text{S} \gg T_\text{CMB}$. Since the CMB is 24.3\,K at $z=7.9$, our results rule out the range of cold reionization scenarios with enhanced power spectra at $z=7.9$ with greater than 95\% credibility. While we cannot definitively rule out any power spectrum enhancement due to a cold IGM $z=10.4$, we have ruled out the large swath of parameter space that produces the largest power spectra.

\autoref{fig:TKS_posterior} brings into sharper relief the question raised by \autoref{fig:theory_summary} of why the \texttt{21cmMC} results show such a large change in $\overline{T}_{S}$. Once we reach the regime where $\overline{T}_{S} > T_{\rm CMB}$, large increases in spin temperature have only a modest impact on the power spectrum. It follows then that the updated power spectrum limits at $z=7.9$ are not primarily driving the spin temperature constraint at $z=7.9$. Just as in the reionization-driven phenomenological model where jointly fitting the two redshifts produced tighter $\overline{T}_S$ constraints, the combination of our two redshift bands is crucial. In particular, the star formation histories inferred by \texttt{21cmMC} from UV luminosity functions (in which the stellar mass density increases significantly from $z=10.4$ to $z=7.9$) causes substantial IGM heating between the two bands. Thus the modest heating demanded by HERA at $z=10.4$ translates to a strong inference about the temperature at the later times. 
We will see quite a similar effect when we examine the inferred power spectrum posteriors in \autoref{fig:full_ps_posterior}.

\needspace{2cm}\subsubsection{\texttt{21cmMC} Derived Constraints on the 21\,cm Power Spectrum} \label{sec:21cmMC:pspec}

\begin{figure*}[!ht]
    \centering
    \includegraphics[width=\textwidth]{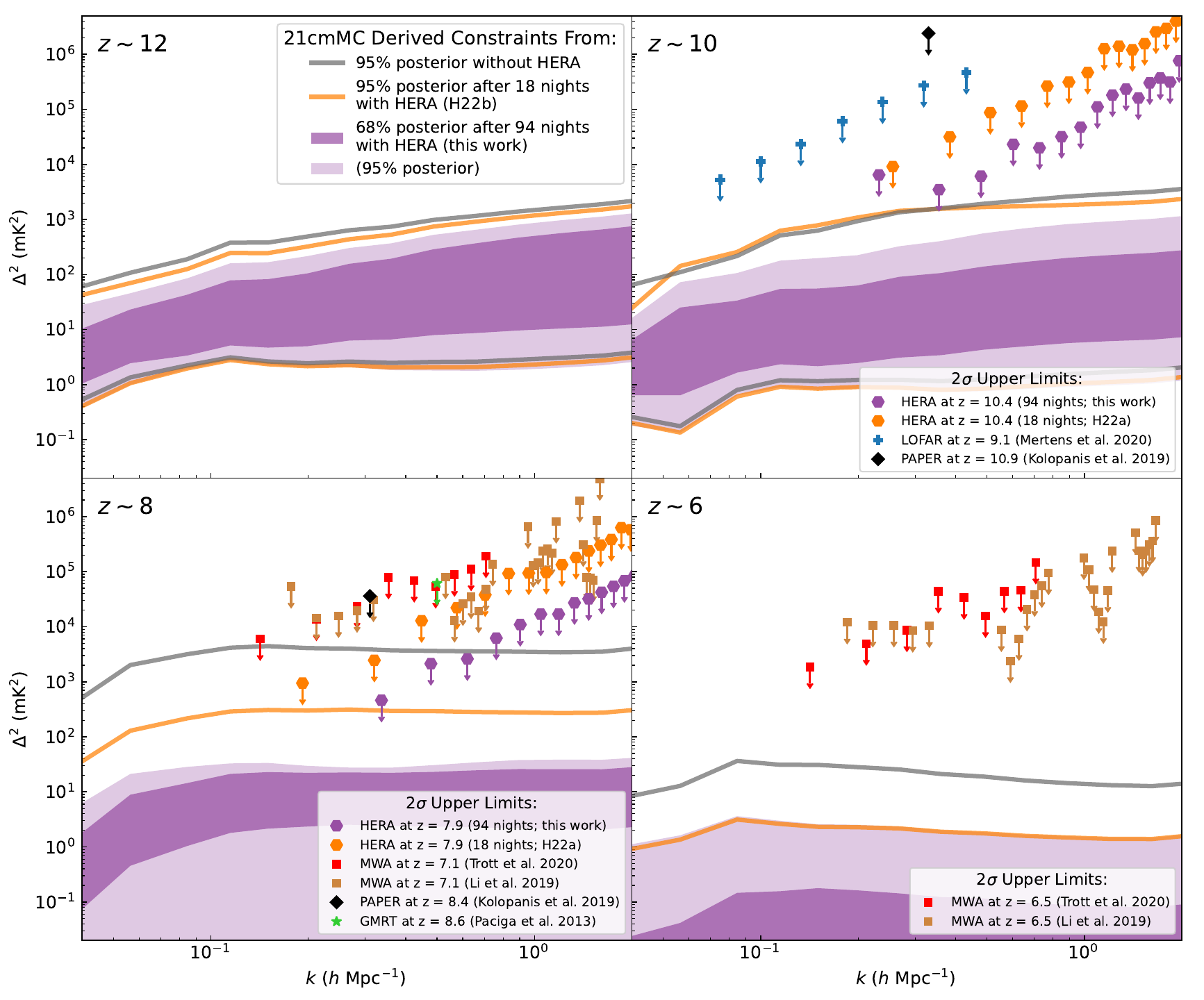}
    \caption{The new upper limits on the 21\,cm power spectrum presented here further constrain the possible range of derived power spectra in our \texttt{21cmMC} posteriors. Here we show the 68\% and 95\% equal-tailed credible intervals for the inferred power spectra at redshifts 12, 10, 8, and 6, after a full season of HERA (purple). We also show 95\% credible intervals after 18 nights of HERA (orange, reproduced from \Hb{}) and without HERA (gray).
    We also include the HERA $2\sigma$ limits from \Ha{} and this work that most strongly constrain the likelihood, namely the single deepest limit over all fields at each $k$. When evaluating likelihoods, data is compared to models using all fields and $k$, which can make a big difference (see \autoref{sec:likelihoods}).
    Note that, due to the form of the likelihood in  \autoref{eq:final_marg_likelihood}, which depends on both our measurements and our error bars, models with power spectra just below our limits are more disfavored than models with power spectra well below them.
    To better understand how the shapes of these likelihoods were updated at $z\sim 10$ and $z\sim 8$ and why the contours are often surprisingly far from the deepest limits, see the full posteriors at the $k$ values of our deepest limits in \autoref{fig:full_ps_posterior}.
    Following \Hb{}, we use every other $k$ to avoid unmodeled correlations between measurements at different $k$ values.
    For context, we also show the most competitive $2\sigma$ limits from other telescopes at similar redshifts, including GMRT \citep{Paciga2013}, PAPER \citep{Kolopanis2019}, MWA \citep{Li2019, Trott2020}, and LOFAR \citep{Mertens2020}.}
    \label{fig:ps_posteriors_with_limits}
\end{figure*}

Because we have ruled out very large negative 21\,cm brightness temperatures arising from cold reionization, we can significantly constrain the range of possible values of the 21\,cm power spectrum in our model. \autoref{fig:ps_posteriors_with_limits} shows the inferred 68\% and 95\% posterior equal-tailed credible intervals on the power spectrum at redshifts 12, 10, 8 and 6 after incorporating results of a full season of HERA observation. 
One can think of these contours as the testable---and therefore falsifiable---predictions of \texttt{21cmMC} given the HERA data, the other astrophysical constraints, and the assumptions of the model. In \autoref{fig:ps_posteriors_with_limits}, we also show the 95\% equal-tailed credible interval from \Hb{} and from our posterior without HERA. Upper limits from \Ha{} and this work as well as a number of other previously published measurements from GMRT \citep{Paciga2013}, PAPER \citep{Kolopanis2019}, MWA \citep{Li2019, Trott2020}, and LOFAR \citep{Mertens2020} are also presented in \autoref{fig:ps_posteriors_with_limits}. 

It is evident that the HERA limits have been significantly improved since \Ha{} and that this work represents our best constraint so far on the neutral IGM during the EoR.  As a result, the posterior distribution of 21\,cm power spectra in our model has also become much tighter, which is consistent with our exclusion of an unheated IGM at these redshifts. That said, one might wonder why the 95\% posteriors are well below our 2$\sigma$ power spectrum upper limits in \autoref{fig:ps_posteriors_with_limits}. The reason is threefold. First, recall that our likelihood model is not Gaussian; it is an error function that asymptotes to a flat probability at small model $\Delta^2$ (\autoref{eq:final_marg_likelihood}). As \autoref{fig:likelihoods} shows, power spectra just below our limits are far less likely than power spectra well below them. This concentrates posterior probability at lower power spectrum values. The effect is especially important when multiple fields and $k$ modes contribute significantly to the likelihood, instead of the two measurements from a single field that dominated the results in \Hb{}.

\begin{figure*}[!ht]
    \centering
    \includegraphics[width=\textwidth]{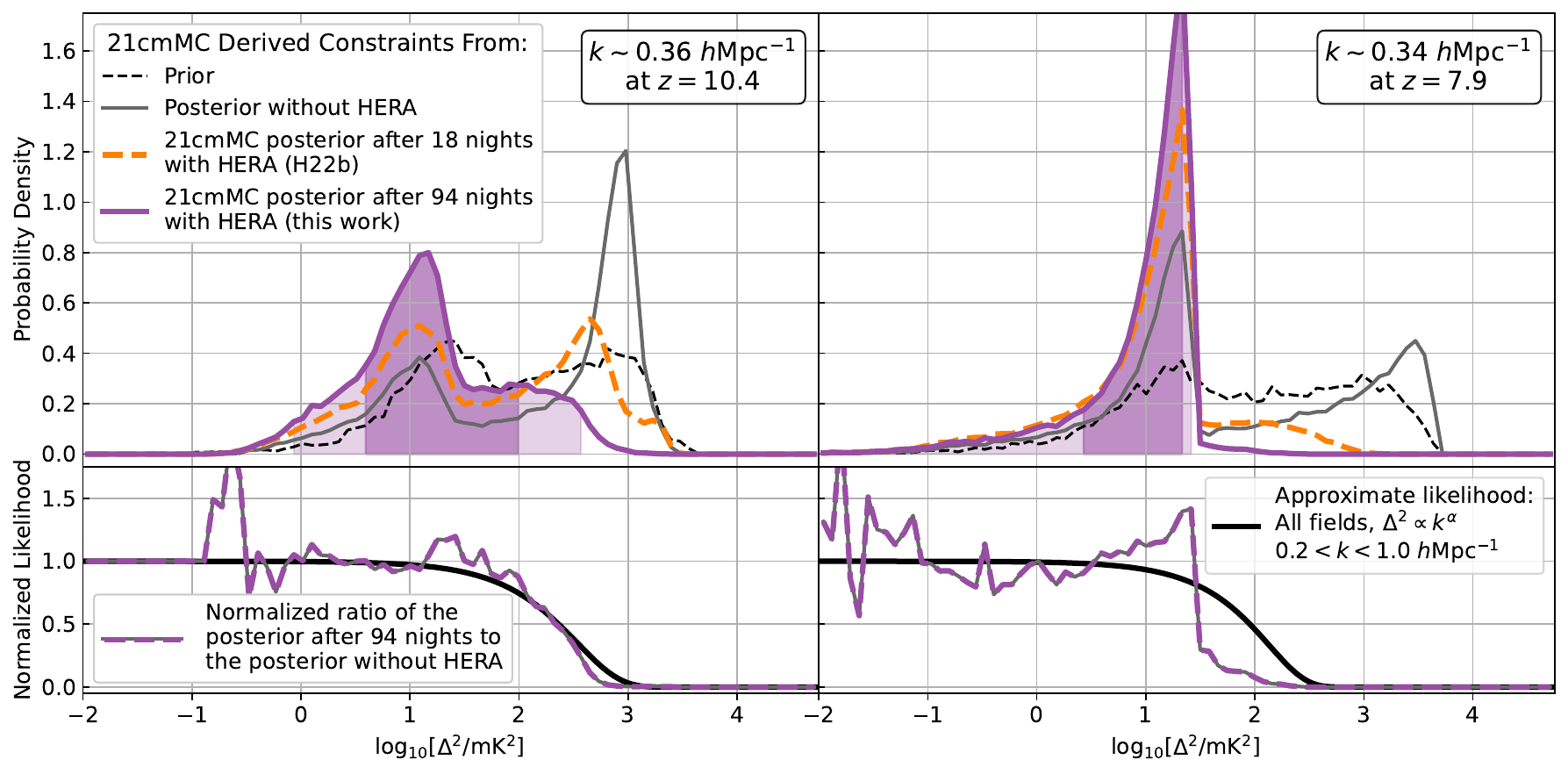}
    \vspace{-10pt}
    \caption{
    Here we show the full \texttt{21cmMC} derived prior and posterior PDFs per logarithmic interval in $\Delta^2$ at the redshifts and $k$ modes roughly corresponding to our best upper limits (top row). Just as in \autoref{fig:TKS_posterior}, we show our priori and three posteriors: one with other astrophysics constrains but no HERA (black), one from \Hb{} (orange dashed), and one from this work (purple, with 68\% and 95\% equal-tailed credible intervals shaded).  In the bottom row, we show the ratio of this work's \texttt{21cmMC} posterior to the posterior without HERA data.
    After renormalizing, this ratio is effectively the likelihood that went into the Bayesian update. We compare those effective likelihoods to those calculated in \autoref{fig:likelihoods} using all fields and all $0.2 < k < 1.0$\,$h$Mpc$^{-1}$. It is clear that the update at $z=10.4$ is largely (though not entirely) attributable to the measures at $z=10.4$.  However, the update at $z=7.9$ is sharper, indicating that much of the information about the inferred $\Delta^2$ comes from ruling out models with inefficient X-ray heating of the IGM, which are better constrained by the $z=10.4$ measurement despite the larger upper limit on $\Delta^2$. This in turn helps us understand this work's 95\% contours in \autoref{fig:ps_posteriors_with_limits} relative to \Hb{}, especially at $z=7.9$. Because we have eliminated a broad range of cold reionization scenarios---the right-hand peaks in the pre-HERA posteriors---the \texttt{21cmMC} posteriors have shifted substantially toward the peaks associated with hot reionization.
    }
    \label{fig:full_ps_posterior}
\end{figure*}

The second reason is that, as already discussed, our inference is heavily informed by other high-redshift observations and the galaxy model assumed by \texttt{21cmFAST}. This is particularly relevant in the context of the third reason, which is the influence of constraints from the two bands on each other.  
To better understand the impact of the model, we show in the top row of \autoref{fig:full_ps_posterior} the prior and full posterior probability distributions from the three inferences, i.e.\ without HERA and with HERA after 18 (\Hb{}) and 94 nights (this work). 
We only show the distributions at the $k$ values of our deepest limits, roughly 0.35\,$h$Mpc$^{-1}$. 
First consider the PDF without incorporating HERA. It has two clear peaks: the one at smaller $\Delta^2$ corresponds to models with $\overline{T}_S \gg T_{\rm CMB}$ (with abundant X-ray heating) while the other is mostly ``cold reionization" with little heating. In between, there is a valley in the distribution because one must fine-tune the heating and ionization to get a signal between these two extremes, which is comparatively unlikely given our priors and the other high-redshift constraints.

The \Hb{} results ruled out the extreme end of the cold reionization peak $z=7.9$, but at $z=10.4$ much of that stronger peak was still viable. With our new limits, an IGM near the adiabatic limit is essentially excluded at $z=10.4$, which clearly favors models with considerable X-ray heating. Because that heating is assumed by the models to continue through $z=7.9$ (when the stellar mass density, and hence density of X-ray sources, has increased by a factor of a few), the higher-redshift measurement helps to constrain the lower-redshift one as well. In particular, it eliminates 
the last little tail of large amplitude power spectra in the posterior, which is why the 95\% contour moves so much between \Hb{} and this work in \autoref{fig:ps_posteriors_with_limits}, though the 68\% contour (not shown here for \Hb{}) does not change very much. 

As further evidence for the impact of the $z=10.4$ measurements on the $z=7.9$ posterior, we show two additional sets of curves in the bottom row of \autoref{fig:full_ps_posterior}. The first are the ratios of the posterior after 94 nights with HERA to the pre-HERA posterior. The second are the likelihoods that we get from \autoref{eq:final_marg_likelihood} by combining together measurements from all fields and $k$ modes from $0.2 < k < 1.0$\,$h$Mpc$^{-1}$ (dropping every other $k$, as discussed in \autoref{sec:likelihoods}). The likelihoods are evaluated assuming power law $\Delta^2 \propto k^{1.3}$ at $z=10.4$ and $\Delta^2 \propto k^{0.0}$ at $z=7.9$ (see \autoref{sec:likelihoods} for details). Both curves are normalized to plateau at 1. Since the pre-HERA posterior is treated as the prior for the post-HERA inference, we should expect this ratio to match the likelihood by Bayes' theorem. 
It should be noted that using a power-law power spectrum is only an approximation. To understand the precise disagreements between between the likelihood and the ratio of the posteriors, we would have to account for the detailed dependence of $\Delta^2$ on $k$, the ways in which the $z=7.9$ measurement constrains the $z=10.4$ posterior (and not just vice versa), and sampling noise.

With those caveats the match looks reasonable at $z=10.4$, but not nearly as good at $z=7.9$. It follows then that the inferred constraints on both $\overline{T}_{S}$ and $\Delta^2$ at $z=7.9$ are driven externally to the data at $z=7.9$, which is all that goes into the black likelihood curve. Therefore, the $z=7.9$ results must be a consequence of the $z=10.4$ limits and the way \texttt{21cmFAST} models the evolution the X-ray luminosity by tying it to star formation and then constraining that star formation rate with other probes, most notably the UV luminosity function. In \autoref{sec:theory_summary}, we will return to the question of the impact of the specific modeling choices of \texttt{21cmMC} and how they compare to the other three techniques.

\needspace{2cm}\subsection{Updated Constraints on Astrophysical Models with Excess Radio Background} \label{sec:radio_background}

In \Hb{}, we reported parameter constraints on an alternate semi-numerical model that allows for a significant excess radio background. EoR scenarios with high levels of radio background at rest-frame 21\,cm wavelengths can potentially produce strong 21\,cm absorption signals, since the brightness temperature is proportional to $(1 - T_\text{radio} / T_\text{S})$ \citep{Feng_2018}. We know there exists today a radio background well in excess of the CMB from observations with ARCADE 2 \citep{fixsen11, seiffert11} and the LWA \citep{dowell18}. It has been theorized that if this excess is sourced by a population of unresolved high-redshift, potentially exotic sources \citep{Ewall_Wice_2018, Fraser:2018, jana19,  Pospelov:2018, Brandenberger:2019, Superstrings2021}, it could explain the anomalously strong absorption signal seen by EDGES \citep{Bowman2018}. Such explanations are not without difficulty; they would have to feature sources far stronger than what would be expected from local observations \citep{Ewall_Wice_2018, mirocha19, Mebane2020, ewall20} and would need to include rapid X-ray heating at $16 \gtrsim z \gtrsim 14$ to explain the high-frequency side of the EDGES trough (e.g.\ \citealt{Mittal_2022}).

Since a radio background can also increase the amplitude of 21\,cm fluctuations \citep{Ewall_Wice_2014,FB19,Reis:2020}, limits from HERA can constrain astrophysical parameters describing models with excess radio background. In \Hb{}, we used a semi-numerical simulation \citep{Visbal12, Fialkov2014b, FB19, Cohen:2020, Reis:2020, Reis:2021} in which the key radiation fields are all driven by the cosmic star formation rate. This is set (in part) by the star formation efficiency $f_{\rm *}$
with which collapsed gas in halos turns into stars, and the circular velocity $V_c$ which determines the minimum mass for star forming halos.
Just as in \Hb{}, the efficiencies of X-ray and radio background luminosity relative to the star formation rate are parameterized by $f_X$ and $f_r$,
and reionization is parameterized by the CMB optical depth $\tau$.
For more details on the implementation of the model and its parameterization, see Section~8 of \Hb{}.\footnote{Unlike in \Hb{} we only consider here the more realistic radio galaxy model, where we expect to have greater constraining power, and do not reconsider more exotic models with a homogeneous synchrotron radio background.}

There are a few differences between the precise inference procedure used here and the one used in \Hb{}.  More parameters are now allowed to vary rather than fixing them to specific values. Most significantly, we no longer limit ourselves to  the X-ray spectral energy distribution (SED) of X-ray binaries \citep{Fragos2013, Fialkov2014b} and instead parameterize the SED by a power law spectral index $\alpha_X$ (either 1.0, 1.3, or 1.5) and a minimum frequency cutoff $\nu_{\rm min}$ which we vary from 0.1 to 3\,keV with log-uniform prior. Additionally, we no longer fix the mean free path to $R_{\rm mfp}=40$\,cMpc but vary it between 10 and 70\,cMpc.\footnote{The other priors have been widened as well: $f_{\rm *}$ now has a log-uniform prior between $10^{-4}$ and $0.5$; $V_c$ has a log-uniform prior between 4.2 and $10^2$; $f_X$ has a log-uniform prior between $10^{-5}$ and $10^3$; $f_r$ has a log-uniform prior between $10^{-1}$ and $10^6$; and $\tau$ has a uniform prior between 0.02 and 0.1.}
The new parameterization of the X-Ray SED implies that more varieties of X-ray heating are captured. For the same value of $f_X$, a softer SED (with a larger number of low-energy photons, e.g.\ lower  $\nu_{\rm min}$) leads to a more efficient heating with shorter X-ray mean free path, while a harder SED (lower number of low-energy photons, e.g. higher values of  $\nu_{\rm min}$) results in less heating with a larger fraction of photons remaining unabsorbed.  

Another improvement over the analysis performed in \Hb{} is the application of more accurate emulators. Our new emulator, based on the \texttt{GLOBALEMU} formalism \citep{Bevins2021} and implemented using \texttt{scikit-learn} \citep{Pedregosa2011}, uses a
more general neural network reaching an accuracy of $\substack{+11\%\\-7\%}$ (at 68\% confidence) at $z=7.9$ and $k=\text{\lowzk}$.
Additionally, instead of an MCMC, we now use nested sampling with \texttt{PolyChord} \citep{Handley_2015,Handley_2015b}.

The updated analysis and the improved HERA upper limit together result in tightened constraints, two of which are of particular interest. The biggest impact is on the X-ray efficiency $f_X$, which is consistent with the \texttt{21cmMC} results on $L_X$/SFR (see \autoref{sec:21cmMC:xray}).
In general, HERA excludes models with high $f_r$ and low $f_X$, since they would produce the brightest amplitude of 21\,cm fluctuations.
The lower bound on the 68\% HPD credible interval of $f_X$ increases from $f_X > 0.03$, using the data from \Ha{}, to $f_X > 0.18$ using the limits presented here. However, the upper bound on the 68\% HPD credible interval of $f_r$
decreases only slightly, from $f_r < 586$ to $f_r < 575$. Note that these numbers differ from the values quoted in \Hb{} due to the changed X-ray SED and priors.

\begin{figure}
    \centering
    \includegraphics[width=.48\textwidth]{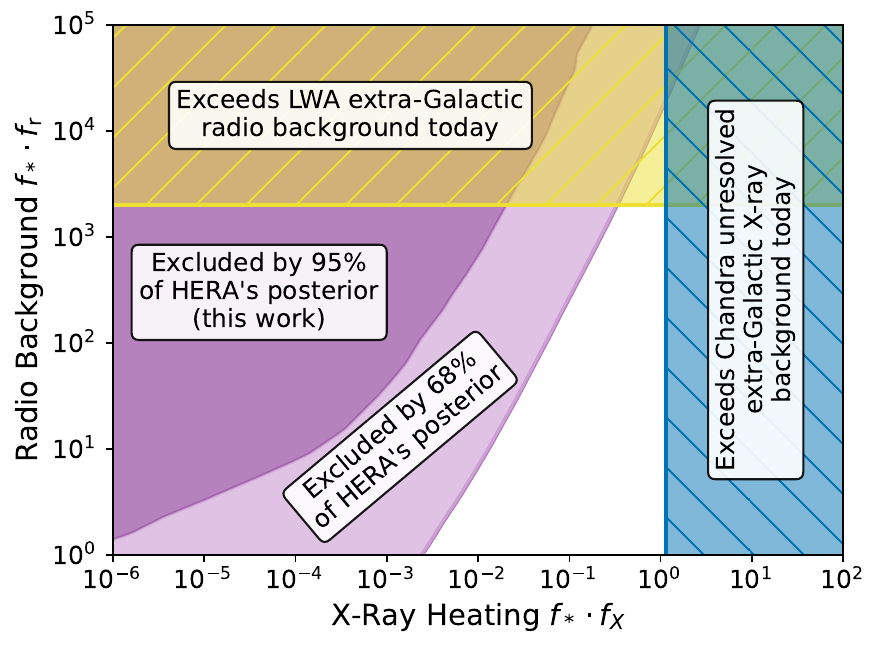}
    \caption{Models with excess high-redshift radio background can produce much larger power spectra than the standard scenario where the 21\,cm brightness temperature is seen in contrast to the CMB. However, such models without accompanying X-ray heating of the IGM are excluded by HERA. Here we show the region of parameter space disfavored by 68\% and 95\% of HERA's posterior, as well as regions inconsistent with either the LWA's radio background measurements \citep{dowell18} or Chandra's X-ray background measurements \citep{Lehmer_2012}. Between HERA's constraints and Chandra's, models where LWA's extra-Galactic radio background as entirely explained by $z > 8$ emission (i.e.\ models of the radio background at the bottom of the yellow region) are disfavored but not entirely excluded.}
    \label{fig:xray_vs_rad}
\end{figure}

In \autoref{fig:xray_vs_rad}, we show the region of the parameter space disfavored by HERA in this model, where we include the impact of $f_{\rm *}$ on both X-ray heating and the total radio background.
This space was already constrained by other measurements of astrophysical backgrounds. \citet{Reis:2020} showed that models with strong $f_{\rm *} \cdot f_{r}$ are inconsistent with the radio background from LWA \citep{dowell18} and  ARCADE 2 \citep{fixsen11, seiffert11}, where the lower frequency measurements from LWA are the more constraining of the two \citep{Reis:2020}. Likewise, \citet{Fialkov:2017} show that large values of $f_{\rm *} \cdot f_{X}$ are generally ruled out by Chandra X-ray background measurements in the 0.5--2\,keV band \citep{Lehmer_2012}. We show those limits in \autoref{fig:xray_vs_rad} in yellow and blue, respectively. Compared to the results in \Hb{}, our new 21\,cm constraints leave little room for the LWA radio background to be explained entirely as a cosmological signal originating from $z > 8$. Such models are not entirely excluded, but they are mostly disfavored at 68\% credibility or greater.

\begin{figure}
    \centering
    \includegraphics[width=.48\textwidth]{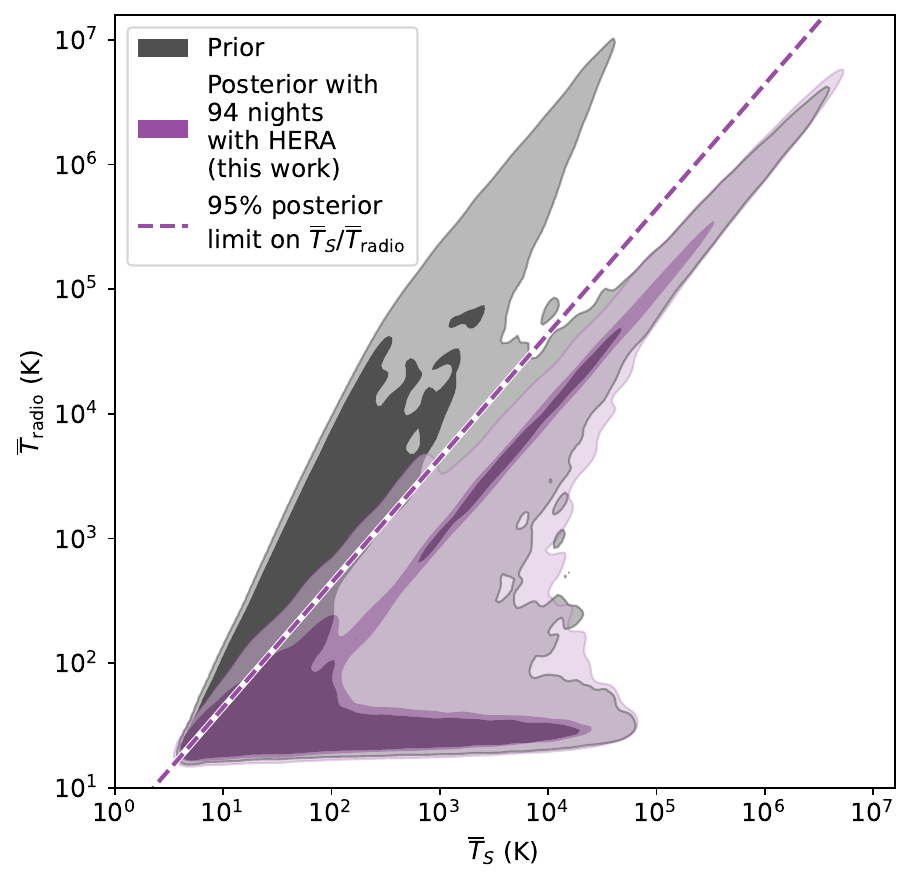}
    \caption{The derived constraints from our model with excess radio emission from high-redshift galaxies allows for a large range of both radio background and IGM spin temperatures at $z=7.9$. HERA generally favors models with lower radio background, though large radio backgrounds counteracted by large spin temperatures are still possible. Both temperatures are averaged over the neutral IGM where 21\,cm emission or absorption might be observed. 95\% of the posterior volume with HERA falls below the line where $\overline{T}_S / \overline{T}_\text{radio} = 0.21$. Performing the same calculation with the $z=10.4$ posterior (not shown) we can constrain $\overline{T}_S / \overline{T}_\text{radio} > 0.18$ at 95\% credibility.}
    \label{fig:ts_vs_trad}
\end{figure}

The second result, which is also an effect of the higher X-ray efficiency, can be seen via our derived constraints on the spin temperature of the IGM. In \autoref{fig:ts_vs_trad}, we show how HERA data updates our model's prior on $\overline{T}_S$ and $\overline{T}_\text{radio}$. 
Both of these quantities are averaged over the neutral IGM at $z=7.9$.

In general, our results exclude models with large radio background temperatures without also featuring large spin temperatures. Specifically, our 95\% credible interval requires that $\overline{T}_S / \overline{T}_\text{radio} > 0.21$ at $z=7.9$. By performing the same calculation for Band 1, we can also show that at $z=10.4$, $\overline{T}_S / \overline{T}_\text{radio} > 0.18$ with 95\% credibility. While the posterior largely prefers relatively low radio background temperatures, there remain a set of models where a large radio background becomes coupled to the spin temperature, yielding $\overline{T}_S / \overline{T}_\text{radio} \approx 1$ and thus small power spectra.

\needspace{2cm}\subsection{Comparison of Astrophysical Models and Constraints} \label{sec:theory_summary}

Finally, we can now more fully compare the results of our four models and their implications for the thermal history of the IGM. As we saw in \autoref{fig:theory_summary}, the results from \texttt{21cmMC} exhibit a more dramatic improvement at $z=7.9$ in this work relative to \Hb{} than either of the phenomenological models (see \autoref{sec:simpler_models}) or the radio background model (see \autoref{sec:radio_background}). In the discussion around \autoref{fig:full_ps_posterior}, we saw clear evidence that \texttt{21cmMC}'s change in $\overline{T}_S$ and thus in the inferred posterior probability distribution of $\Delta^2$ was driven by the power spectrum constraints at $z=10.4$. This tight connection between redshifts requires rather rapid thermal evolution of the IGM, which is generally the case for the range of \texttt{21cmMC} models we consider.

In the other models, the spin temperature history can evolve quite gradually, and thus yield similarly gradual 21\,cm signal evolution that will be more difficult to disfavor with improved $z=10.4$ limits. For the reionization-driven model, this is because $\overline{T}_S$ is parameterized as a power-law in redshift with a uniform prior on the power-law index. In the radio background model, there are two reasons for gradual evolution. First, the stellar-mass-halo-mass relation is constant in the radio background model, which gives rise to a much more gradual cosmic star formation history than, e.g., models in which the relation is itself a function of halo mass. Second, because the radio and X-ray backgrounds are both sourced by galaxies, both $\overline{T}_{\rm radio}$ and $\overline{T}_S$ will grow similarly as the cosmic star formation rate density rises, resulting in relatively gradual evolution in the 21\,cm signal. By contrast, in our \texttt{21cmMC} simulations, $T_{\rm radio}=T_\text{CMB}$, which of course declines as $(1+z)$. This means $T_{\rm radio} / \overline{T}_S$ will evolve more rapidly than in than in the radio background models, even if both share an identical $\overline{T}_S$ history.

Another reason for the difference between models is the way \texttt{21cmMC} directly incorporates data from other wavelengths (especially the galaxy luminosity function)---a major advantage of the approach. Models that fit the luminosity function at all redshifts require the star formation efficiency to increase with galaxy mass (at least up to galaxy masses comparable to the Milky Way, see e.g., \citealt{Mirocha2017,park19}). The \texttt{21cmMC} models therefore also favor massive galaxies, whose abundance evolves quite rapidly at high redshifts. Because these galaxies are also the sources of IGM heating in this model, their rapidly evolving abundance also leads to rapid IGM heating. Because the luminosity function fits ``bake in'' this kind of behavior, \texttt{21cmMC} models that feature modest heating at $z=10.4$ require stronger heating at $z=7.9$. The expected increase is roughly the ratio of the stellar mass density at the two redshifts, which is a factor of $\sim$3. This explains why the strong $\overline{T}_S$ limit in \texttt{21cmMC} at $z=7.9$ is actually informed in large part by the ostensibly weaker temperature limits at $z=10.4$. 

Of course, this conclusion is reliant on the parameterization of the galaxy luminosity function within \texttt{21cmMC}. While a wide range of models agree that a physically-motivated extrapolation of the observed luminosity function to higher redshifts behaves similarly, it is of course reasonable to suppose that the extrapolation breaks down---perhaps because of processes unique to high-redshifts, or perhaps simply because young dwarf galaxies behave differently than their larger cousins (see e.g., \citealt{Qin:2020xyh}). One could always broaden the allowed set of galaxy models to accommodate such exotic astrophysics. We have, of course, explored one class of such models by examining the effect of an excess radio background in \autoref{sec:radio_background}. One could also imagine adding another source of early heating like Pop~III remnants that could inject sufficient heat into the IGM at $z=10.4$ without necessarily affecting the constraints at $z=7.9$, though this might require some degree of fine tuning---a question left for future work. Regardless, it is useful to understand how we should interpret the HERA limits in the context of ``vanilla" galaxy models. Here, the \texttt{21cmMC} results demonstrate that the ``known'' galaxy population requires substantial IGM heating by $z \sim 8$. 

\begin{figure}
    \centering
    \includegraphics[width=.48\textwidth]{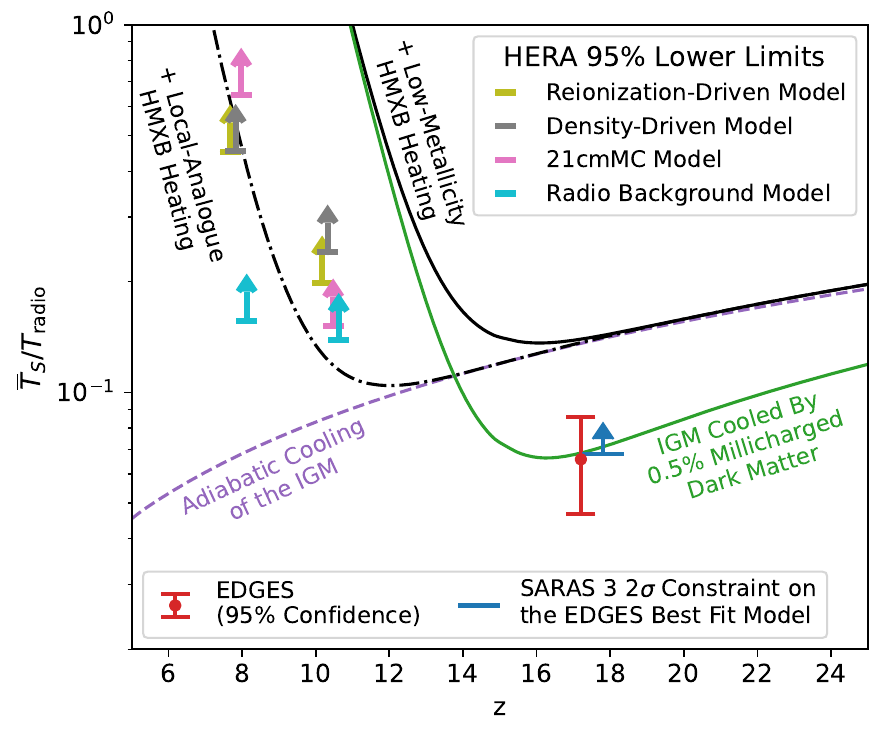}
    \caption{To put the results from our four models in \autoref{fig:theory_summary} into their cosmological context, we show each model's 95\% credible limits on $\overline{T}_S / T_\text{radio}$. Each model's results are shown offset from their proper redshifts, 7.9 or 10.4, for visual clarity. They differ in their conclusions, both due to their different physical assumptions as well as their different priors and incorporation of other high-redshift probes.  However, they all conclude that the IGM spin temperature at both $z=7.9$ and $z=10.4$ is in excess of the temperature one would expect from adiabatic cooling since recombination, assuming $T_\text{radio} = T_\text{CMB}$ (purple dashed line). They are also generally inconsistent with an IGM weakly heated by X-rays (black dot-dashed line), as it would be if the local relation for $L_X/{\rm SFR}$ for HMXBs held at high redshift (see \autoref{fig:LX_posterior})---although that statement is somewhat model dependent. However, all our models are consistent with a more rapidly heated IGM, such as the one shown as a solid black line where $L_{\rm X < 2keV}/{\rm SFR} = 10^{41}$\,erg\,s$^{-1}$\,${\rm M}_\odot^{-1}$\,yr, which is closer to the expectation for low-metallicity HMXBs. To put HERA's result in the context of high-redshift global signal measurements, we also show results from EDGES \citep{Bowman2018} and SARAS 3 \citep{Singh_2022}. For EDGES, we show the implied $\overline{T}_S / T_\text{radio}$ from the depth of their best fit model, as well as the $2\sigma$ range of model amplitudes. For SARAS, which disfavors the EDGES best fit at $\sim$95\% confidence when marginalizing over only the amplitude, we show the lower limit on models with the same shape as the EDGES signal. The SARAS central redshift is also offset from EDGES for visual clarity. Finally, we show a model with 0.5\% millicharged dark matter, as in Figure~11 of \Hb{}, which could explain EDGES and still be consistent with HERA if there is sufficient X-ray heating between $z \sim 17$ and $z \sim 10$.
    }
    \label{fig:TsoverTrad}
\end{figure}

Regardless of the differences between our models, we should not lose sight of the fact that they are all pointing in fundamentally the same direction: a heated IGM during the EoR. To put them in the broader cosmological context of other 21\,cm experiments, we compare them to a few illustrative scenarios for the thermal history of the IGM in \autoref{fig:TsoverTrad}.
The arrows show the lower bounds of the 95\% credible regions of $\overline{T}_S / T_\text{radio}$ at $z=7.9$ and $z=10.4$ inferred by each of the four models. We also show the expectation for an IGM temperature set by adiabatic cooling since recombination with no additional heating from astrophysical sources. Just as \autoref{fig:theory_summary} showed, the lower limits at both redshifts are above this curve, requiring substantial X-ray heating (as illustrated by the solid black line where $L_{\rm X < 2keV}/{\rm SFR} = 10^{41}$\,erg\,s$^{-1}$\,${\rm M}_\odot^{-1}$\,yr).  This is a significant improvement from \Hb{}, which could not demonstrate any heating beyond the adiabatic cooling curve at $z=10.4$. More interestingly, they are all inconsistent with an IGM thermal history that one would predict by extrapolating the local relation between X-ray luminosity and star formation rate to high redshift (black dot-dashed line, see \autoref{fig:LX_posterior}). The accordance of multiple models here strengthens the argument in \autoref{sec:21cmMC:xray} that our results broadly favor an IGM heated by low-metallicity HMXBs.

\autoref{fig:TsoverTrad} also shows measurements at higher redshifts. The EDGES result \citep{Bowman2018} found a surprisingly large absorption signal at $z \sim 17$, requiring either substantial cooling of the gas below the adiabatic limit or a larger radio background than provided by the CMB. The implied range of $\overline{T}_{\rm S} / T_{\rm radio}$ is shown by the red error bars, but note that EDGES found this absorption trough to be very narrow, implying that the Universe was heated shortly afterward. However, the recent SARAS 3 measurement \citep{Singh_2022} did not detect such a signal; the upper limit is subtle to express quantitatively, but here we show the limit their measurement places on a feature with the shape of the best-fit EDGES signal with an unknown amplitude. Though we show the SARAS 3 limit offset in redshift for clarity, the measurement spanned a wide frequency range corresponding to $15.8 < z < 24.6$. 

Because HERA observed at much smaller redshifts than EDGES and SARAS 3, it is difficult to compare directly---even if the deep EDGES trough is real, any heating between $z \sim 15$ and $z \sim 10$ (as indeed the EDGES best-fit model requires!) would make the two measurements consistent. We show an example of such a model (green solid line) where the IGM is cooled by interactions with a fraction (0.5\%) of millicharged dark matter with mass 10\,MeV and charge $10^{-5}$ times that of the electron, following \citet{Munoz:2018pzp}.
The gas is subsequently heated by HMXBs. 
This model is designed to explain EDGES and still be consistent with lower redshift observations like HERA's, though it is in mild tension with SARAS 3. More independent measurements like SARAS 3, as well as low-frequency power spectra from HERA Phase II and other interferometers, will be extremely valuable in understanding this early era and any new physics that may have impacted it. 


\section{Conclusion} \label{sec:conclusion}
In this work we have presented improved upper limits on the 21\,cm power spectrum using 94 nights of observing with Phase I of the Hydrogen Epoch of Reionization Array (HERA), as well as their astrophysical implications for the X-ray heating of the IGM. We have found with 95\% confidence that $\Delta^2(k=\text{\lowzk{}}) \leq \text{\lowzlimit{}}$ at $z=7.9$ and that $\Delta^2(k=\text{\highzk{}}) \leq \text{\highzlimit{}}$ at $z=10.4$, an improvement by a factor of 2.1 and 2.6 respectively over previous HERA limits in \Ha{} with 18 nights of data and roughly the same number of antennas. Our full set of upper limits across $k$ are detailed in \autoref{sec:upper_limits}.

Our results rely heavily on the application of existing techniques to this new data set. In particular, we adapted most of the techniques used in \Ha{} to our larger data volume, noting where and why we made different analysis choices. We replicated many of the statistical tests developed in \Ha{}---and some new ones---to show that our data largely integrate down as expected and that our techniques for systematics mitigation do not introduce any new biases. Likewise, we have performed a number of the simulation-based tests developed in \citet{Aguirre_2022} using our updated techniques, including an end-to-end test of our analysis pipeline that simulates the full data volume before reducing it to power spectra. While many small adjustments were made to accommodate the larger data set considered here, the fundamental philosophical approach remains the same. Instead of subtracting or filtering foregrounds, we focused on maintaining spectral smoothness and systematics control while minimizing and rigorously quantifying potential sources of signal loss. 

We also revisit four independent theoretical models for inferring constraints on IGM and galaxy properties during the EoR and the cosmic dawn. These techniques were applied to the \Ha{} data set in \Hb{} where we showed that the IGM had to heated above the adiabatic cooling limit by at least $z=7.9$. Using the improved upper limits presented in this work, the four techniques broadly agree with at least 95\% of their posterior volumes (and in some cases greater than 99\%): the IGM had to be heated above the expectation from adiabatic cooling by at $z=10.4$ the latest.

There are two key consequences of this result. The first follows from our current understanding from existing probes of reionization---especially the integrated optical depth to reionization, galaxy UV luminosity functions, and quasar
spectroscopy---that the bulk of reionization happens after $z=10.4$. 
If that is the case, then we have ruled out the ``cold reionization'' scenarios in which the IGM continues to adiabatically cool until it reionizes, creating a bright 21\,cm power spectrum is boosted by a strong contrast between the CMB and the IGM spin temperature.
It is still possible that the IGM is slightly heated at $z=10.4$ but still colder than the CMB. However, a broad class of models where the IGM remains very cold until reionization are no longer viable. 

Second, if high mass X-ray binaries dominate the soft X-ray budget of high redshift galaxies and thus were responsible for the heating of the IGM, as is generally expected \citep{Fragos2013}, then more than 99\% of HERA's posterior excludes the local relationship between star formation and soft X-ray production \citep{Mineo2012} extrapolated to high redshift. We instead favor models with low-metallicity HMXBs, which is clear evidence for the impact on the IGM of some of the first compact objects to form during the cosmic dawn.

We also used a semi-numerical model that allows galaxies to create radio backgrounds brighter than the CMB to jointly constrain those radio backgrounds and the X-ray luminosity of those galaxies. Specifically, we combined HERA limits (which disfavor strong radio backgrounds with weak X-ray heating) and Chandra X-ray background constraints (which rule out strong X-ray heating) to exclude most models that would explain the radio backgrounds observed by LWA and ARCARDE 2 as originating at $z \gtrsim 8$.

Looking forward, we see a number of ways HERA might continue to improve the constraints on the 21\,cm signal as we continue the steady march to greater and greater sensitivity. One approach would be to move beyond cautious foreground-avoidance and attempt to apply more aggressive and more nearly optimal filters, removing foregrounds \citep{Ewall-Wice2021} in delay space and integrating down coherently for longer with better-tailored fringe-rate filters \citep{Parsons2016}. These techniques might help us explore more frequency bands and claw back some of the most sensitive baselines that we had to excise in this work. However, they will likely incur higher levels of signal loss that will have to be rigorously quantified and taken into account.

More importantly, this analysis only used a small fraction of HERA's final size and bandwidth. HERA Phase II, now being commissioned, will have 350 antennas observing from 50--250\,MHz which corresponds to $4.7 < z < 27.4$. We now have a well-tested analysis pipeline taking us all the way to power spectra and astrophysical inference, including a suite of statistical tests and end-to-end simulations with which to validate our results. This work, along with \Ha{}, \Hb{}, and their supporting papers, will serve as a foundation for future HERA analysis. Of course, HERA Phase II has an entirely new signal chain---from feeds to correlator---and will likely have to contend with a somewhat different set of systematics. However, if these can be overcome, HERA could be the instrument to detect and characterize the 21\,cm power spectrum from the epoch of reionization and push our knowledge of early stars, black holes, and galaxies into the cosmic dawn.


\section*{Acknowledgements}

Work similar to that presented here appeared previously in unrefereed formats in public HERA Team Memos \#97 \citep{H1C_IDR3_2_Memo}, \#104 \citep{Crosstalk_Model_Memo}, and \#107 \citep{H1C_IDR3_2_Pspec_Memo}. 

This analysis utilized custom-built, publicly-accessible software by the HERA Collaboration (\url{https://github.com/Hera-Team}) in addition to software built by both HERA members and collaborators (\url{https://github.com/RadioAstronomySoftwareGroup}), especially \texttt{pyuvdata} \citep{Hazelton2017}. 
This analysis also relied on number of public, open-source software packages, including \texttt{numpy} \citep{2020NumPy-Array}, \texttt{scipy} \citep{scipy2020}, \texttt{scikit-learn} \citep{Pedregosa2011}, \texttt{matplotlib} \citep{Hunter:2007}, and \texttt{astropy} \citep{astropy:2018}.

This material is based upon work supported by the National Science Foundation under grants \#1636646 and \#1836019 and institutional support from the HERA collaboration partners. 
This research is funded in part by the Gordon and Betty Moore Foundation through Grant GBMF5212 to the Massachusetts Institute of Technology.
HERA is hosted by the South African Radio Astronomy Observatory, which is a facility of the National Research Foundation, an agency of the Department of Science and Innovation. 

This work used the Extreme Science and Engineering Discovery Environment (XSEDE; \citealt{XSEDE}), which is supported by National Science Foundation grant number ACI-1548562.
We acknowledge the use of the Ilifu cloud computing facility (\url{www.ilifu.ac.za}) and the support from the Inter-University Institute for Data Intensive Astronomy (IDIA; \url{https://www.idia.ac.za}).

\edited{The authors wish to thank the anonymous referee for their insightful feedback. }
J.S.~Dillon gratefully acknowledges the support of the NSF AAPF award \#1701536.
N.~Kern gratefully acknowledges support from the MIT Pappalardo fellowship.
P.~Kittiwisit and M.G.~Santos acknowledge support from the South African Radio Astronomy Observatory (SARAO; \url{www.sarao.ac.za}) and the National Research Foundation (Grant No.\ 84156).
This result is part of a project that has received funding from the European Research Council (ERC) under the European Union's Horizon 2020 research and innovation programme (Grant agreement No.\ 948764; P.~Bull and M.J.~Wilensky). P.~Bull and H.~Garsden acknowledge support from STFC Grant ST/T000341/1.
Y.~Qin would like to acknowledge the support from the High Performance Computing centers of the Scuola Normale Superiore (Italy), the Council for Scientific and Industrial
Research (South Africa), the OzSTAR national facility (Australia) and XSEDE (U.S.) for computational resources.
J.~Mirocha acknowledges computational resources and support on the supercomputer Cedar at Simon Fraser University, which is managed by Compute Canada and funded by the Canada Foundation for Innovation (CFI).
Parts of this research were supported by the Australian Research Council Centre of Excellence for All Sky Astrophysics in 3 Dimensions (ASTRO 3D), through project number CE170100013.
S.~Heimersheim acknowledges support from STFC studentship 2277533 in project  ST/T505985/1.
J.B.~Mu\~noz is supported by a Clay Fellowship at the Smithsonian Astrophysical Observatory.
G.~Bernardi acknowledges funding from the INAF PRIN-SKA 2017 project 1.05.01.88.04 (FORECaST), support from the Ministero degli Affari Esteri della Cooperazione Internazionale -- Direzione Generale per la Promozione del Sistema Paese Progetto di Grande Rilevanza ZA18GR02 and the National Research Foundation of South Africa (Grant Number 113121) as part of the ISARP RADIOSKY2020 Joint Research Scheme, from the Royal Society and the Newton Fund under grant NA150184 and from the National Research Foundation of South Africa (grant No.\ 103424).
E.~de Lera Acedo acknowledges the funding support of the UKRI Science and Technology Facilities Council SKA grant.
A.~Liu acknowledges support from the New Frontiers in Research Fund Exploration grant program, the Canadian Institute for Advanced Research (CIFAR) Azrieli Global Scholars program, a Natural Sciences and Engineering Research Council of Canada (NSERC) Discovery Grant and a Discovery Launch Supplement, the Sloan Research Fellowship, and the William Dawson Scholarship at McGill.


\appendix

\section{A Physical Model for HERA Phase I Crosstalk Systematics}
\label{appendix:crosstalk}

One of the most important systematics that we needed to mitigate before estimating power spectra is been crosstalk. The strategy developed in \citet{Kern2019}, demonstrated in \citet{Kern2020a}, and applied in \Ha{} is quite effective at removing that crosstalk. However, as we discovered in \autoref{sec:xtalk_updates}, our crosstalk was significantly less time-stable over the entire season that it was during the 18 nights analyzed in \Ha{}. This motivated LST-binning over epochs and new cuts on the data (see \autoref{sec:xtalk_residual}). 

This discovery also motivated a renewed attempt to understand the physical origin of the crosstalk. We began with the basic mathematical model of crosstalk presented in \citet{Kern2019}.  If we postulate that the voltage measured on an antenna $i$ is due to both the incident sky signal $v_i$ absorbed by antenna $i$ and contributions from other antennas, we can write that voltage (ignoring noise) as
\begin{equation}
    v_i' = v_i + \sum_{n\neq i} \varepsilon_{ni} v_n,
\end{equation}
where $\varepsilon_{ni}$ is the coupling between the $n$th antenna and antenna $i$. Since visibilities are formed by cross-correlating voltages, a visibility with cross-coupling takes the form
\begin{equation}
    V'_{ij} = \langle (v_i + \sum_{n\neq i}\varepsilon_{ni} v_n)(v_j + \sum_{n \neq j}\varepsilon_{nj} v_n)^* \rangle
\end{equation}
where angle brackets indicate a time average. Assuming the coupling is small and dropping all terms that are second order in $\varepsilon$ or first order in $\varepsilon$ but only involve cross-correlations, we get
\begin{align}
    V'_{ij} &\approx \langle v_i v_j* + \varepsilon_{ij}^* v_i v_i^* + \varepsilon_{ji} v_j  v_j^* \rangle \nonumber \\
    & \approx V_{ij} + \varepsilon_{ij}^* V_{ii} +  \varepsilon_{ji} V_{jj}. \label{eq:coupling}
\end{align}
Thus, to leading order, the main cross-coupling contributions appear as autocorrelations of each antenna involved leaking into a visibility. It should be noted that we are ignoring antenna-to-antenna coupling effects due to reflections off antennas, which \citet{Josaitis_2022} showed would be an important systematic for HERA Phase II. At this point, we have seen no evidence that this is a dominant systematic for HERA Phase I.

\citet{Kern2019} was agnostic as to the origin of the coupling. They instead focused on how autocorrealtions, which evolve much more slowly in time than cross-correlations with any appreciable east-west projected length, can be distinguished from the primary sky signal in fringe-rate space. As long as $\varepsilon$ was time-stable, its structure was of secondary concern. Our observation of discontinuities in the crosstalk in delay space (see \autoref{fig:xtalk_discon}) meant that $\varepsilon$ was effectively no longer stable over a full season of LST-binned data.

Explaining the range of delays over which crosstalk was observed in e.g.\ \autoref{fig:xtalk_discon} and \autoref{fig:xtalk_demonstration}---generally 800 to $\sim$2000\,ns---required solving a puzzle. Those delays are too long to be explained by a broadcasting antenna, which would created correlated signals at delays explicable by the light travel time across the array. They are also too short to be explained by invoking cable reflections, which require two traversals of the $\sim$150\,m cables connecting antennas to the receivers and thus take $\sim$1200\,ns. \citet{Kern2019d} explored and rejected several possible explanations. No model could explain that range of delays and also the asymmetric structure we see in a single visibility, whose positive and negative delays can exhibit completely different crosstalk structure.

\begin{figure*}
    \centering
    \includegraphics[width=1\textwidth]{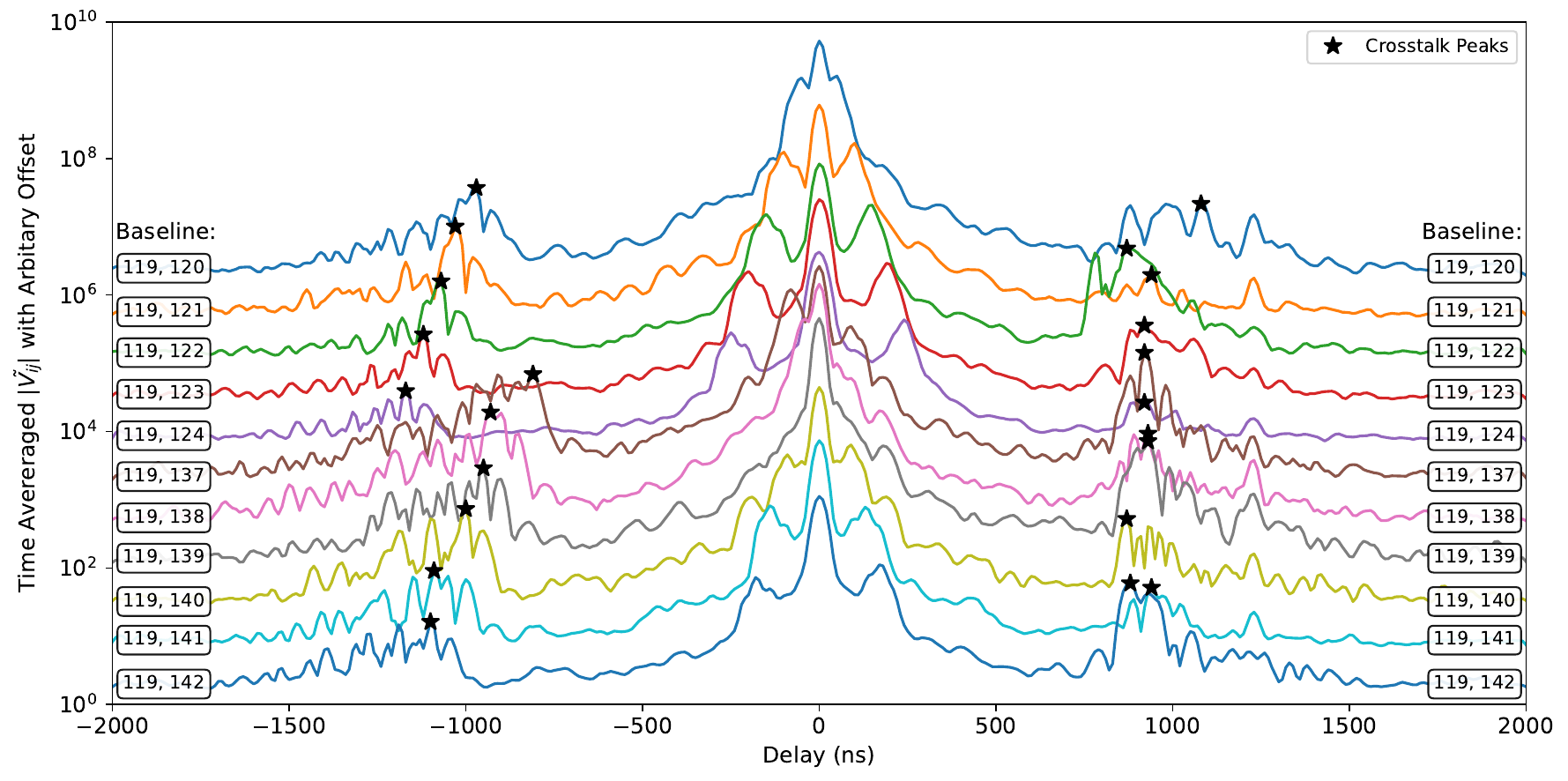}
    \caption{Examining the time-averaged delay spectrum of several baselines all sharing a single antenna reveals a clear pattern in the delay structure of the crosstalk feature between $800 \lesssim |\tau| \lesssim 1500$\,ns, but only one on side. Here we show eleven baselines all sharing antenna 119 (and all north-polarized). All the antennas are part of two rows (see \autoref{fig:layout}). We plot their time-averaged amplitudes in delay space, each arbitrarily offset for readability. We also mark the peak delays of each baseline's crosstalk with black stars, which our model must explain. Crosstalk at negative delay shows similar structure within rows of the array (120--124 as compared to 137--142) with diminishing amplitude as we move eastward. By contrast, the crosstalk feature at positive delay shows no clear pattern. This supports the argument that the negative delay feature is associated with antenna 119. It follows from symmetry that, since this effect is not unique to antenna 119, the positive delay feature must be associated with the other antenna. At low delay we can also see the foregrounds peak in the main lobe of the primary beam, as well as a widening ``pitchfork'' feature associated with foregrounds on the horizon \citep{Thyagarajan2015a, Thyagarajan2015b, Pober2016}, which is at higher delay for longer baselines.}
    \label{fig:xtalk_offset_delay_spec}
\end{figure*}

However, \citet{Kern2019d} did note that attributing positive or negative delays to the first and second antennas in a baseline reveals interesting patterns in the dependence of the delay and amplitude of the cross-talk peak on position in the array. That proved to be a key insight. Taking the time-averaged amplitude of the Fourier transform of a several inpainted visibilities from a single epoch (in this case Epoch 2) which share one antenna in common illustrates that pattern. In \autoref{fig:xtalk_offset_delay_spec} we show this for a subset of the baselines that share antenna 119, all of them in the same two rows (see \autoref{fig:layout}).
What we see is a remarkable asymmetry between positive and negative delays. At negative delay, we see a delay structure that looks similar within rows, but with decreasing amplitude and at more negative delays as the second antenna moves east. Meanwhile, at the crosstalk systematics at positive delay show up at largely similar delays but with wildly ranging amplitudes and structures. Holding a single antenna fixed and seeing such similarities at negative delays indicates a clear association between antenna $i$ and negative delays in $V_{ij}$---and thus by symmetry, an association between antenna $j$ and positive delays. A model for this crosstalk must explain how the voltage signal from e.g.\ antenna 119 was received by every other antenna with a position-dependent peak delay and amplitude. 

To understand these delays, recall that HERA Phase I reused PAPER feeds and signal chains. These included $\sim$150\,m coaxial cables from each feed to a set of eight ``receiverators,'' RF-shielded mini-fridges each containing 16 receiver and post-amplification modules \citep{Bradley2017}, located just west of the array. These were then connected by $\sim$20\,m coaxial cables to a shipping container housing the analog to digital converters and the correlator \citep{DeBoer2015}. We hypothesize that at one of those connection points, likely after the receiverators, was leaking and broadcast virtually every antenna's voltage signal. These signals were then picked up by every other antenna, leading to autocorrelations appearing in cross-correlation visibilities at high delay. We thus explain the crosstalk peak delays and amplitudes in \autoref{fig:xtalk_offset_delay_spec} with a model for $\varepsilon_{ij}$ where
\begin{equation}
    \varepsilon_{ij} \approx A_i \left(\frac{d_{*j}}{\rm 1\,m}\right)^{-\alpha} \exp\left[2\pi i \nu \left(\tau_{i,{\rm cable}} + d_{*j} / c + \tau_{\rm offset}\right)\right]. \label{eq:xtalk_model}
\end{equation}
Here $\tau_{i,{\rm cable}}$ is the light travel time along the $\sim$150\,m cable connecting antenna $i$ to the receiverators; it can be easily measured by examining the delay spectrum of autocorrelations since cable reflections appear at $2\tau_{i,{\rm cable}}$ \citep{Kern2020a}. The rest of the terms are free parameters. The first two are the positions of the emitter, $x_*$ and $y_*$, from which we can calculate $d_{*j}$, the distance from the emitter to antenna $j$. The next is $\tau_{\rm offset}$ which can account for the possibility that emission occurred after traversing another $\sim$20\,m cable between the receiverators and the correlation. No per-antenna variation is allowed in $\tau_{\rm offset}$. Finally, the amplitude of the crosstalk seen on each baseline depends on the ``leakiness" of antenna $i$---$A_i$, each a free parameter---and the distance it must travel to the power $-\alpha$. Since emitted voltages go down as $1/r$, we should expect $\alpha=1$. However, to attempt to account for the effects of complicated mutli-path propagation through a lattice of antennas, we leave $\alpha$ as a free parameter. 

\begin{figure*}
    \centering
    \includegraphics[width=1\textwidth]{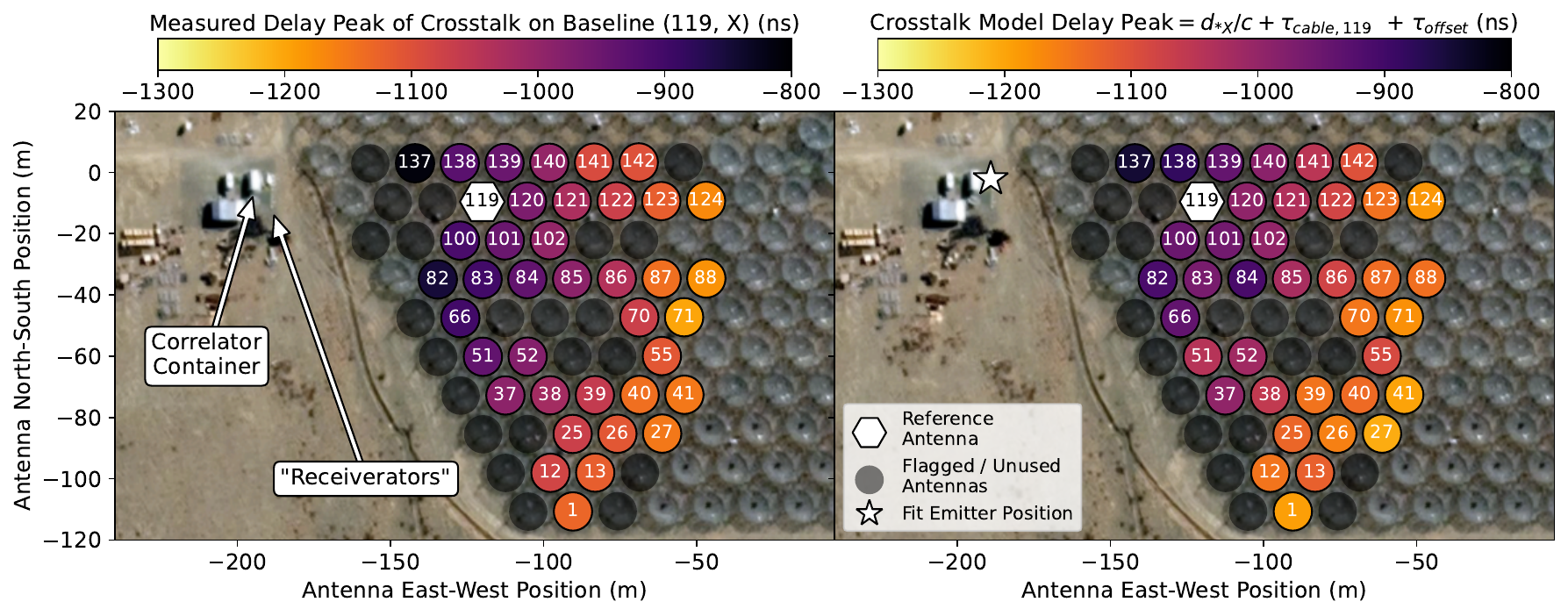}
    \caption{We can predict the measured peak delays in the crosstalk in each visibility $V_{ij}$ using a model (\autoref{eq:xtalk_model}) where antenna $i$'s signal travels down a $\sim$150\,m cable, gets amplified in the ``receiverators," and then is emitted shortly thereafter. That voltage signal then travels over the air and is picked up by antenna $j$, producing a contribution of $V_{ii}$ to $V_{ij}$ at large negative delays (\autoref{eq:coupling}). In the left-hand panel, we show all such measurements involving a single antenna, 119, and the prediction of our best fit model (right-hand panel) using all baselines. We overlay our data and fit over recent satellite imagery of HERA from Google Maps (accessed in June 2022), which shows that our best fit for the position of the emitter is spatially consistent with our receiverators and correlator container. This explanation of the physical origin of the crosstalk validates our approach to removing it.}
    \label{fig:xtalk_dly_fit}
\end{figure*}

Since this model does not predict the full delay spectrum of the crosstalk, our goal here is not to solve for every parameter optimally. Instead, we want to test its physical plausibility. Therefore, we fit both parts---the phase and the amplitude---separately with potentially different emitter positions. First we fit the delays, since that is a simpler model with far fewer free parameters. In \autoref{fig:xtalk_dly_fit}, we show in the left panel the measured crosstalk peak positions at negative delay (black stars in \autoref{fig:xtalk_offset_delay_spec}) for all baselines $V_{ij}$ where $i = 119$. 
In the right panel, we show our best fit source position and the prediction for the total delay term in \autoref{eq:xtalk_model}. Both are overlaid over recent satellite photography of HERA. This fit is performed over all unflagged baselines, with the exception of a few baseline that had too little crosstalk to measure the peak reliably. The fit is quite good; the average delay error is only 47\,ns which is only a few times larger than the 12.8\,ns delay resolution of the Fourier transform after removing flagged channels. More tellingly, the emitter position is quite consistent with the suspect signal chain elements, namely the connections between the receiverators and the correlator. The fit $\tau_{\rm offset} = 99$\,ns which consistent with expectations for the for the $\sim$20\,m cable \citep{Kern2020a}, given the speed of light in the medium. While this result lends significant credence for our model, we cannot definitively state that it locates the emitter as the input ports on the correlator container. The emitter position and $\tau_{\rm offset}$ are correlated and, as \autoref{fig:xtalk_offset_delay_spec} shows, quantifying the error in measuring peak delays is challenging. 

\begin{figure*}
    \centering
    \includegraphics[width=1\textwidth]{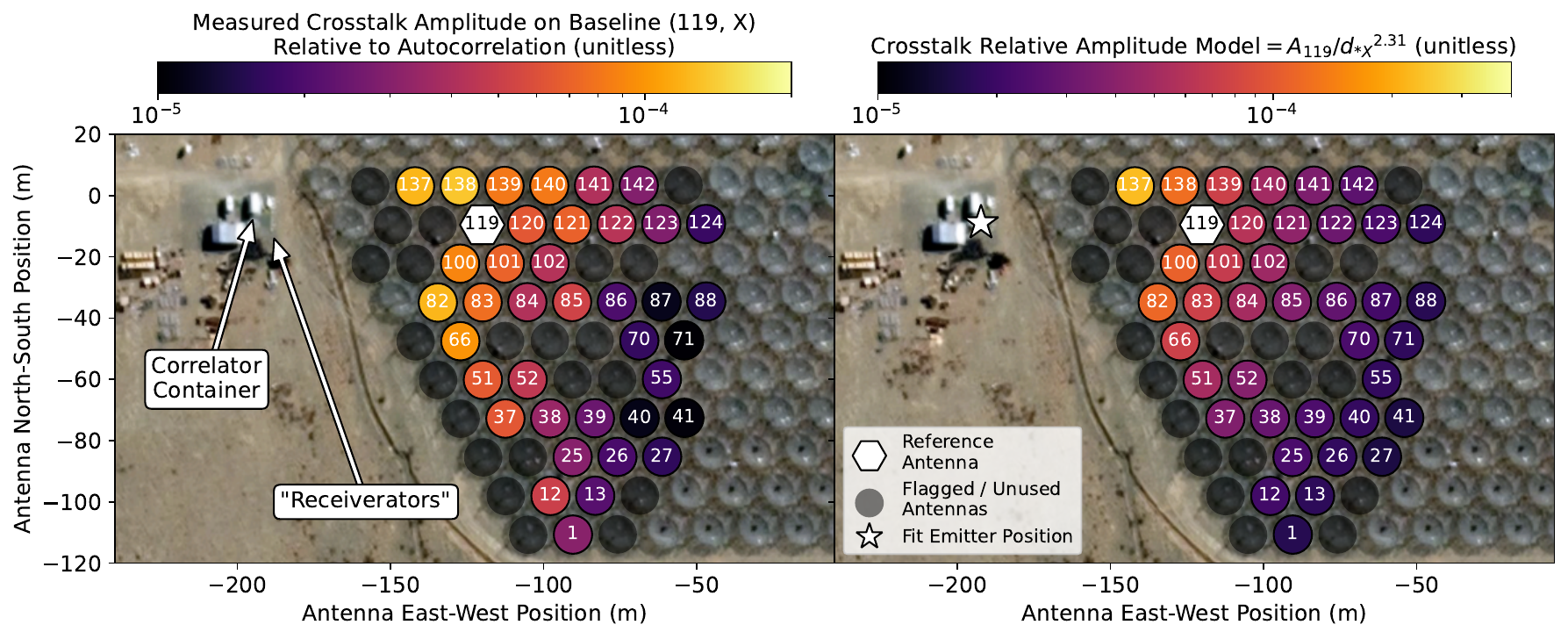}
    \caption{Analogously to \autoref{fig:xtalk_dly_fit}, we can also predict the amplitudes of the crosstalk as they appear in each baseline using the same model (\autoref{eq:xtalk_model}). In the left-hand panel we show measured peak crosstalk amplitudes, relative to the autocorrelations that source them, for all baselines involving antenna 119. On the right, we show the best-fit model prediction for that same data set using all baselines. The relatively good agreement here, as well as the fact that we find a consistent emitter position despite fitting for it independently fit, lend credence to our physical model for the crosstalk.}
    \label{fig:xtalk_amp_fit}
\end{figure*}

As a second test, we also fit the amplitude of each crosstalk signal using \autoref{eq:xtalk_model}. This test is a bit less straightforward, since it requires a free parameter $A_i$ for every antenna. That said, an independent fit of the emitter position yields a quite consistent result, as we can see in \autoref{fig:xtalk_amp_fit}.
Again, we show the crosstalk amplitudes associated with antenna 119 (left-hand panel) but perform our fit over all baselines. The result is a fairly good fit, with a mean amplitude error of $1.2 \times 10^{-5}$ relative to the autocorrelations. Oddly, the best fit power law is not $\alpha=1$ as we had expected, but $\alpha=2.31$. We do not have an explanation for why it should follow that power law, though we note that the problem is a fair bit more complicated than simple free-space propagation. There are also some parameter degeneracies to consider; one can still get a decent fit by fixing $\alpha=1$ and moving the emitter position much closer to the array. 

Without a proper error analysis, it is difficult to validate the precise parameterization of our model. What should $\alpha$ be? What about $\tau_{\rm offset}$? Can we explain the full delay structure of the crosstalk signal as some sort of multi-path propagation effect? Can construction of new antennas explain the epoch-to-epoch change in the crosstalk we saw in \autoref{fig:xtalk_discon}? Unfortunately we can only speculate. The system in question has long-since been disassembled and HERA Phase II does not use receiverators nor does this systematic appear in more recent data. That said, having a plausible physical mechanism that allows autocorrelations to leak into cross-correlations in a relatively time-stable way adds significant support to our strategy for removing the systematic.


\bibliographystyle{aasjournal}
\bibliography{biblio}

\end{document}